\documentclass{mn2e}
\input{psfig.sty}
\usepackage{mnras_cite}
\newcommand{\apj}{Astrophys. J.}
\newcommand{\aj}{Astron. J.}
\newcommand{\apjs}{Astrophys. J. Supp.}

\newcommand{\mnras}{Mon. Not. R. Astron. Soc.}

\newcommand{\gal}{{\rm gal}}

\newcommand{\lin}{{}}
\def\sun{\hbox{$\odot$}}

\newdimen\hssize
\newdimen\hdsize 
\begin{document}

\title[Observational Constraints on Galaxy Properties]{
Halo Model at Its Best:  Constraints on Conditional Luminosity Functions from Measured Galaxy Statistics} 
\author[Cooray]{Asantha Cooray\\ 
Center for Cosmology, Department of Physics and Astronomy, University of California, Irvine, CA 92697\\
E-mail:acooray@uci.edu}

\maketitle

\begin{abstract}
Using the conditional luminosity function (CLF) --- the luminosity distribution of
 galaxies in a dark matter halo --- as the fundamental
building block, we present an empirical model for the galaxy distribution.
The model predictions are compared with the published luminosity function and clustering statistics from 
Sloan Digital Sky Survey (SDSS) at low redshifts, and galaxy correlation functions from
COMBO-17 survey at a redshift of 0.6, DEEP2 survey at a redshift of unity, 
Great Observatories Deep Origins Survey (GOODS)
at a redshift around 3, and Subaru/XMM-Newton Deep Field data at a redshift of 4. 
The comparison with statistical measurements allows us to constrain certain parameters related to analytical descriptions
on the relation between a dark matter halo and
its central galaxy luminosity,  its satellite  galaxy luminosity,  and the fraction  of early- and late-type galaxies
of that halo. With the SDSS r-band LF at $M_r <-17$, the log-normal scatter in the
central galaxy luminosity at a given halo mass in the central galaxy--halo mass, $L_c(M)$, relation is constrained to be
$0.17^{+0.02}_{-0.01}$, with 1 $\sigma$ errors here and below. 
For the same galaxy sample, we find no evidence for a
low-mass cut off in the appearance of a single central galaxy in dark matter halos, 
with the 68\% confidence level upper limit on the minimum mass of dark matter halos to host a central galaxy, with
luminosity  $M_r <-17$, is $2 \times 10^{10}$ $h^{-1}$ M$_{\sun}$. On the other hand,
the appearance of satellites with luminosities $M_r < -17$ at $z < 0.1$, using
a total  luminosity-halo mass relation of the form $L_c(M)(M/M_{\rm sat})^\beta_s$,
is constrained with SDSS to be at a halo mass of
$M_{\rm sat} = (1.2_{-1.1}^{+2.9}) \times 10^{13}$ $h^{-1}$ M$_{\sun}$ with a 
power-law slope $\beta_s$ of $(0.56^{+0.19}_{-0.17})$.
At $z \sim 0.6$, COMBO-17 data allows these parameters for  $M_B < -18$  galaxies
to be constrained as $(3.3_{-3.0}^{+4.9}) \times 10^{13}$ $h^{-1}$ M$_{\sun}$  and
 $(0.62^{+0.33}_{-0.27})$, respectively.  
At $z \sim 4$, Subaru measurements constrain these parameters for $M_B < -18.5$ galaxies
as $(4.12_{-4.08}^{+5.90}) \times 10^{12}$ $h^{-1}$ M$_{\sun}$  and
 $(0.55^{+0.32}_{-0.35})$, respectively. The redshift evolution associated with these
parameters can be described as a combination of the evolution associated with the halo mass function and
the luminosity--halo mass relation. The single parameter well constrained
by clustering measurements is the average of total satellite galaxy luminosity 
corresponding to the dark matter halo distribution probed by the galaxy sample.
For SDSS, $\langle L_{\rm sat}\rangle =(2.1^{+0.8}_{-0.4}) \times 10^{10}$ h$^{-2}$ L$_{\sun}$,  
while for GOODS at $z \sim 3$, $\langle L_{\rm sat}\rangle < 2 \times 10^{11}$ h$^{-2}$ L$_{\sun}$.
For SDSS, the fraction of galaxies that appear as satellites is $0.13^{+0.03}_{-0.03}$,
$0.11^{+0.05}_{-0.02}$, $0.11^{+0.12}_{-0.03}$, and $0.12^{+0.33}_{-0.05}$    for galaxies with luminosities
in the $r'$-band between -22 to -21, -21 to -20, -20 to -19, and -19 to -18, respectively.
In addition to constraints on central and satellite CLFs, we  also determine model parameters of the analytical relations 
that describe the fraction of
early- and late-type galaxies in dark matter halos. 
We use our CLFs to establish probability distribution of halo mass
in which galaxies of a given luminosity could be found either at halo centers or as satellites.
Finally, to help establish further properties of the galaxy distribution, we propose the measurement of
cross-clustering between galaxies divided into two distinctly different luminosity bins.
Our analysis show how CLFs provide
a stronger foundation to built up analytical models of the galaxy distribution when compared to models based on the
halo occupation number alone.

\end{abstract}


\section{Introduction}

The conditional luminosity function  (CLF; Yang et al. 2003b, 2005), or the
luminosity and color distribution of galaxies within a dark matter halo of mass $M$, $\Phi(L,c|M)$,  
captures important astrophysical information that determines how the large scale structure
galaxy distribution is related to that of the dark matter. As shown in Cooray \& Milosavljevi\'c (2005b; Cooray 2005a),
a simple empirical model for the CLF, when combined with the halo mass function,
describes the galaxy luminosity function (LF) accurately;
This empirical model recovers the Schechter (1976) form of the galaxy LF given by 
$\Phi(L) \propto (L/L_\star)^\alpha \exp(-L/L_\star)$
with a characteristic luminosity $L_\star$ and a power-law slope at the faint end of $\alpha$.
A basic ingredient in this model is the relation between
central galaxy luminosity and the mass of the halo in which the galaxy is found --- the $L_c(M)$ relation of
Cooray \& Milosavljevi\'c (2005a). The characteristic luminosity of the Schechter function
is the luminosity when the scatter in the $L_c(M)$ relation, at a given halo mass,
 dominates over the increase in the luminosity with mass or when $d \ln L_c/d \ln M \approx \ln(10) \sigma$ where
$\sigma$ is the log-normal dispersion in the $L_c(M)$ relation. Given the observed dispersion, we find
$M_\star \approx 2 \times 10^{13}$ M$_{\sun}$ and $L_\star$, corresponding to $L_c(M_\star)$, agrees with
estimates from observations. The faint-end power-law slope of the LF
is a combination of the power-law slope of the $L_c(M)$ relation at $M \ll M_\star$ and the slope
of the dark matter halo mass function. The combination puts a strict bound that  $\alpha < -1.25$, 
consistent with observations that  indicate  $\alpha \approx -1.3$ (Blanton et al. 2004; Huang et al. 2003). 

The $L_c(M)$ relation, as appropriate for galaxies at low-redshifts and in the K-band, was established in 
Cooray \& Milosavljevi\'c (2005a)  from a combination of weak galaxy-galaxy lensing data (e.g., Sheldon et al. 2004; Yang et al. 2003a)
and direct measurements of galaxy luminosity and mass in groups and clusters (e.g., Lin et al. 2004; Lin \& Mohr 2004).
The same relation, as appropriate for lower wavelengths, 
has been established with a statistical analysis of the 2dF Galaxy Redshift Survey (2dFGRS; Colless et al. 2001)
 $b_J$-band LF  (e.g., Norberg et al. 2002) by Vale \& Ostriker (2004) and, 
independently, by Yang et al. (2005) based on the 2dFGRS galaxy
group catalog.  The shape of the $L_c(M)$ relation, where luminosities grow rapidly with increasing mass but
flattens at a mass scale around $\sim$ 10$^{13}$ M$_{\sun}$ is best explained through dissipationless 
merging history of central galaxies (Cooray \& Milosavljevi\'c 2005a). A large fraction of these
bright galaxies, in centers of groups and clusters, are early-type and observational evidences for 
dry mergers as a dominant process in the formation and evolution of massive, luminous early-type galaxies are now beginning to appear
(Bell et al. 2005). In Cooray (2005b), the approach based on CLFs was extended to higher redshifts and a comparison with
galaxy LFs observed  out to redshifts 2 and higher, with surveys such as DEEP2 (Davis et al. 2003; Willmer et al. 2005;
Faber et all. 2005)  and COMBO-17 (Wolf et al. 2001, 2003; Bell et al. 2004),
allowed constraints on the redshift evolution of the $L_c(M)$ relation.

Beyond the total galaxy LF, 
the empirical modeling approach based on CLFs can easily be extended to consider statistics of galaxy types as well.
For example, in Cooray (2005a), we studied the environmental dependence of galaxy colors, broadly
divided into blue- and red-galaxies given the bimodal nature of the color distribution 
(e.g., Baldry et al. 2004; Balogh et al. 2004). There, we described early- and late-type
conditional  LFs measured with 2dFGRS  as a function of the galaxy overdensity (Croton et al. 2004) 
based on an empirical description of the fraction of late- and early-type galaxies
relative to the total number of galaxies in dark matter halos as a function of the halo mass. With an increasing fraction  of
early-type galaxies as the halo mass is increased, the simple analytical model
considered in Cooray (2005a) explains why the LF of galaxies in
dense environments is dominated by red galaxies. 

While LFs are well produced by this analytical model, with statistics related to early- and late-type
galaxies parameterized by analytical functions, it is timely to consider how this model
compares with higher order statistics of the galaxy distribution such as those related to the clustering
pattern of galaxies. A comparison to data could also
provide additional constraints on ingredients related to this model and especially those related to
CLFs of central and satellite galaxies.
While numerous predictions and limited comparisons to data 
exist in the literature on how well clustering measurements can constrain galaxy properties,
 these are mostly considered in terms of the simple halo occupation model involving
the number of galaxies in a dark matter halo as a function of the halo mass, $N_g(M)$
(e.g., Seljak 2000; Scoccimarro et al. 2001; Berlind \& Weinberg 2002; Cooray 2002; Berlind et al. 2003; Zehavi et al. 2004; Zheng et al. 2004). 

The same approach of constraining the halo occupation number  based on galaxy clustering data
has been applied at redshifts $\sim$ 0.6 with COMBO-17 (Phleps et al. 2005)
and, more frequently, at $\sim$ 3 and higher using Lyman-break galaxies (LBGs) and similar drop-out samples
(e.g., Bullock et al. 2002; Lee et al. 2005). Since the 
halo occupation number is an integral function and treats all galaxies the same, 
regardless of the color or the galaxy luminosity, meaningful models that account
for differences in galaxy physical parameters cannot easily be considered. 
It is also no surprise that the halo model, based on the halo occupation number alone, cannot
be used to model the LF of galaxies. Even with clustering statistics, 
due to differences involving luminosities of galaxies in different
samples and potential variations with redshift when defining galaxy samples,
constraints on the simple halo occupation number at different
redshifts cannot be compared with each other easily. 

The best approach to overcome these drawbacks 
is to make use of the conditional occupation number or, more appropriately, CLF
as the fundamental quantity when describing galaxy statistics (Yang et al. 2003b, 2005). 
The CLFs extend the analytical halo model  (see, Cooray \& Sheth 2002 for a review)
by dividing the mean number of galaxies to a distribution in galaxy luminosity such that $\Phi(L|M)=dN_g(M)/dL$
and using $\Phi(L|M)$ to model observed statistics rather than $N_g(M) $ (Yang et al. 2003b, 2005).
Similarly, the same approach can be extended for galaxy color or any other property as
one can easily define the subsample of galaxies related to that property, but with the restriction
that the whole sample be contained within the total LF.
Thus, the approach based on CLFs is useful when comparing with measurements conditioned in terms
of galaxy properties such as the luminosity or the color. With wide-field surveys, where statistics of a
few hundred thousand galaxies or more are easily available, the division of measurable statistics to
galaxy properties is now common. Given the existence of measurements already, for
example clustering properties as a function of the galaxy luminosity (Zehavi et al. 2004)
or galaxy-mass cross-correlation through galaxy-galaxy lensing studies as a function of the galaxy luminosity 
(Mandelbaum et al. 2004; Sheldon et al. 2004),  the need for an improved halo model is clear.

Here, we extend our previous discussions related to CLFs (Cooray 2005a, 2005b), where we modeled the LF, to study
galaxy clustering and make predictions for
 clustering  statistics at the two-point level involving projected correlation functions as a function of the redshift.
These models are then compared with
existing results from surveys such as Sloan Digital Sky Survey (SDSS; York et al, 2000; Zehavi et al. 2004) at redshifts
less than 0.1, COMBO-17 survey at redshifts between 0.4 and 0.8 (Wolf et al. 2001, 2003; Phleps et al. 2005),
DEEP2 survey (Davis et al. 2003; Coil et al. 2004) at a redshift around unity,  Great Observatories Deep Origins
Survey (GOODS; Lee et al. 2005) at a redshift around 3, and Subaru/XMM-Newton Deep Field LBG clustering at
a redshift of 4 (Ouchi et al. 2005)  to derive general constraints  on the underlying
CLF. 

A previous attempt related to extracting properties of the galaxy sample in 2dFGRS through CLFs is described in
Yang et al. (2005). In this analysis, authors made use of a priori assumed Schechter function shape for the
CLF (Yang et al. 2003b), though we make no such assumptions here.
In fact, galaxy cluster or group LF data suggest that Schechter function shapes are not the appropriate form to
describe their luminosity distribution, given the central galaxy. A combination of a log-normal component and
a power-law fits the data best, consistent with the CLF models we have suggested.
Motivated by Yang et al. (2003b), Yan et al. (2003) used the same CLF description involving Schechter function shapes
to compare galaxy clustering between
2dFGRS and DEEP2 and suggested that the CLF does not strongly evolve with redshift. 

While we make use of SDSS clustering measurements in our
analysis here,  we also note that an attempt has been made to establish $\Phi(L|M)$  based on
differences in halo occupation models, as a function of luminosity, when describing clustering statistics as a function of the galaxy
luminosity (Zehavi et al. 2004). The modeling approach we use here involves the CLF as the
basic parameter to be extracted from the data and
provides a consistent way to model both the galaxy LF and clustering statistics of the same galaxy sample
and a mechanism to extend the same underlying CLF model to describe galaxy statistics at higher redshifts. Since CLFs are recovered, 
we can easily integrate over the luminosity to calculate halo occupation numbers allowing an easy comparison to previous analyses.
 Our approach also demonstrates why and how certain
parameters related to CLFs are sensitive to LFs, such as those related to central galaxies, while others, especially
those involving satellite galaxies, can be determined better with the non-linear 
1-halo part of the correlation function.
This is consistent with suggestions in the literature, that, for example,
halo occupation statistics --- which are dominated by satellite galaxies --- are better constrained with
clustering of galaxies within groups (e.g., Coil et al. 2005; Collister \& Lahav 2005).

The paper is organized as follows: In the next  section, we will outline basic ingredients in the empirical model  for CLFs
and how galaxy clustering statistics can be derived from CLFs. We refer the reader to Cooray
\& Milosavljevi\'c (2005b) and Cooray (2005a) 
for initial discussions related to this empirical modeling approach and
to Cooray (2005b) for details on the extension to higher redshifts. Previous studies related to the CLF, involving
mostly the 2dFGRS catalog, are described in Yang et al. (2003b, 2005). In Section~3,  we present a
comparison of our model with the observed LF, and LFs of galaxy types, 
from SDSS and in Section~4, we extract parameters related to CLFs as a function of the redshift
from SDSS clustering at redshifts below 0.1 to Subaru data at redshifts $\sim$ 4.
In Section~5, we conclude with a summary of our main results and implications related to the galaxy distribution and propose
several new statistics that can help constrain CLFs better.
Throughout the paper we assume cosmological parameters consistent with most observational analyses of
measurements modeled here and take $\Omega_m=0.3$, $\Omega_\Lambda=0.7$, and  a scaled Hubble constant of  $h=1$ 
in units of 100 km s$^{-1}$ Mpc$^{-1}$, unless otherwise stated explicitly.

\begin{figure}
\centerline{\psfig{file=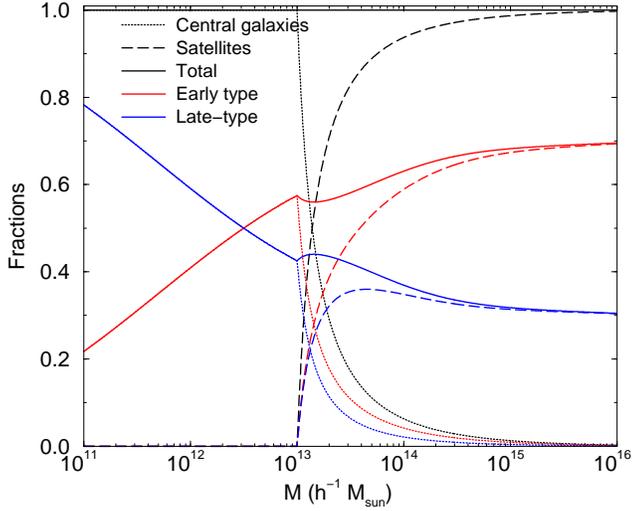,width=\hssize,angle=-90}}
\caption{The fraction of early- (red lines) and late-type (blue lines)
galaxies, both appearing as  central and satellite galaxies,
relative to the total number of galaxies in dark matter halos, as a function of the halo mass. 
For reference, we also show the fraction of total central (dotted lines) and
satellite galaxies (dashed lines). These fractions assume fiducial
values for various  model parameters, appropriate for SDSS galaxies with $M_r <-17$, 
as discussed in the text. At halo masses below 10$^{13}$ M$_{\sun}$ fractions are 
determined by central galaxies; For low-mass halos the fraction of late-type galaxies
is close to 0.8, while the same fraction for high mass halos, dominated by satellites, is
$\sim$ 0.3. In addition to the halo mass,
the early- and late-type fraction of satellite galaxies also depends on the
galaxy luminosity. Here, we show the fractions for satellites with luminosity of $10^{10}$ L$_{\sun}$.
Later, based on parameter constraints, we will highlight the satellite galaxy fraction at a given galaxy luminosity
and show that, while some parameters such as the total satellite luminosity is well constrained,
the fraction is not.}
\end{figure}

\begin{figure}
\centerline{\psfig{file=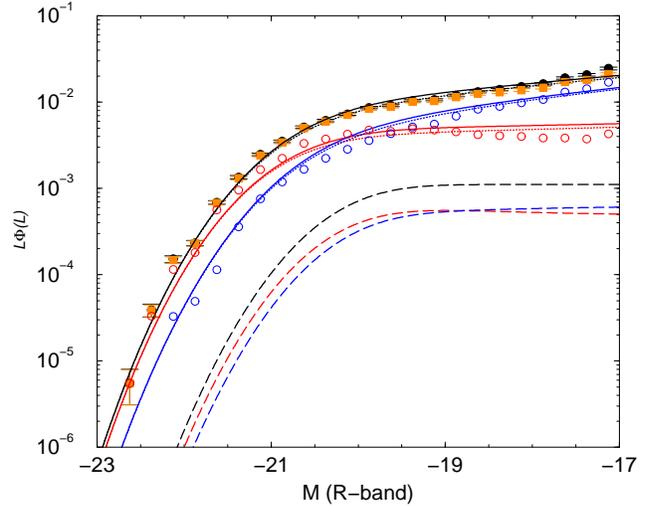,width=\hssize,angle=-90}}
\caption{The LF of SDSS galaxies in the $r$-band from Blanton et al. (2003; 2004). We concentrate
here only on galaxies with $M_r < -17$ as these form the sample used by Zehavi et al. (2004)
for galaxy clustering measurements. In addition to the total luminosity function  ---
 we show both the uncorrected and corrected estimates as filled symbols with error
bars (see, Blanton et al. 2004 for details) ---
we also show the LF of early- and late-type galaxies (open symbols). For clarity, we do not
plot the error bars for the LF of galaxy types, but they are at the same level as that of the total
sample. The curves show the predictions based on CLFs, with fiducial best-fit parameters as described in the text. 
The dotted lines show contribution from central galaxies, while the dashed lines
show satellites. The solid lines show the total galaxy LF as predicted in this model. 
As shown, and discussed in Cooray (2005a), central galaxies dominate LF statistics; as we find later, parameters related to
central galaxies are better determined with LFs when compared to the information provided in clustering measurements.}
\end{figure}

\section{Conditional Luminosity Functions: An Overview}

In order to construct galaxy clustering statistics as a function of redshift, we follow Cooray \& Milosavljevi\'c (2005b) and Cooray (2005) 
to define the redshift-dependent conditional luminosity function (CLF; Yang et al. 2003b, 2005), denoted by $\Phi(L|M,z)$,
giving the average number of galaxies with luminosities between $L$ and $L+dL$ that 
resides in halos of mass $M$ at a redshift of $z$. As in our previous applications, the CLF is separated 
into terms associated with central and satellite galaxies,  such that
\begin{eqnarray}
\Phi(L|M,z)&=&\Phi^{\rm cen}(L|M,z)+\Phi^{\rm sat}(L|M,z) \nonumber \\
\Phi^{\rm cen}(L|M,z)  &=& \frac{f_{\rm cen}(M,z)}{\sqrt{2 \pi} \ln(10)\sigma_{\rm cen} L} \times \nonumber \\
&& \quad \quad \exp \left\{-\frac{\log_{10} [L /L_{\rm c}(M,z)]^2}{2 \sigma_{\rm cen}}\right\}  \nonumber \\
\Phi^{\rm sat}(L|M,z) &=& A(M,z) L^{\gamma(M)} g_{\rm s}(L)\, .
\label{eqn:clf}
\end{eqnarray}
Here, $f_{\rm cen}(M,z)$ is a selection function  introduced to
account for the efficiency for galaxy formation as a function of the halo mass given the
fact that at low mass halos galaxy formation may be inefficient and not all dark matter halos may host a galaxy:
\begin{equation}
f_{\rm cen}(M,z) = \frac{1}{2}\left[1+{\rm erf}\left(\frac{\log(M)-\log(M_{\rm cen-cut}(z))}{\sigma}\right)\right] \, .
\label{eqn:fcm}
\end{equation}

The motivation for the separation of
galaxies to central and satellite galaxies is numerous: from theory,
a better description of the galaxy occupation statistics is obtained when one separates to central and
satellite galaxies (Kravtsov et al. 2004), while from observations, central and satellites galaxies
are known to show different properties, such as color and luminosity (e.g., Berlind et al. 2004).
In our fiducial description, we will take numerical values of  $M_{\rm cen-cut}= 10^{10}$ M$_{\rm sun}$ and $\sigma=0.5$.

We introduced the selection function $f_{\rm cen}(M,z)$  in Eq.~(2)
in Cooray (2005a) to explain the faint-end slope of the 2dFGRS LF with a value of $\sim$ -1.05 (Norberg et al. 2002).
When considering model fits to the data with $M_{\rm cen-cut}$ as a free parameter, we find that this parameter
can only be constrained as an upper limit with SDSS data. As we discuss later, 
the lack of a clear constraint on $M_{\rm cen-cut}$ in our model fits 
differs from analysis based on halo occupation numbers where a minimum mass for the presence of galaxies
in halos is usually suggested. The minimum mass in halo occupation number generally corresponds to the minimum mass for halos
that host galaxies at the low-end of the galaxy luminosity distribution probed with the data and such a cut-off
is naturally present in models related to CLFs.

\begin{figure*}
\centerline{\psfig{file=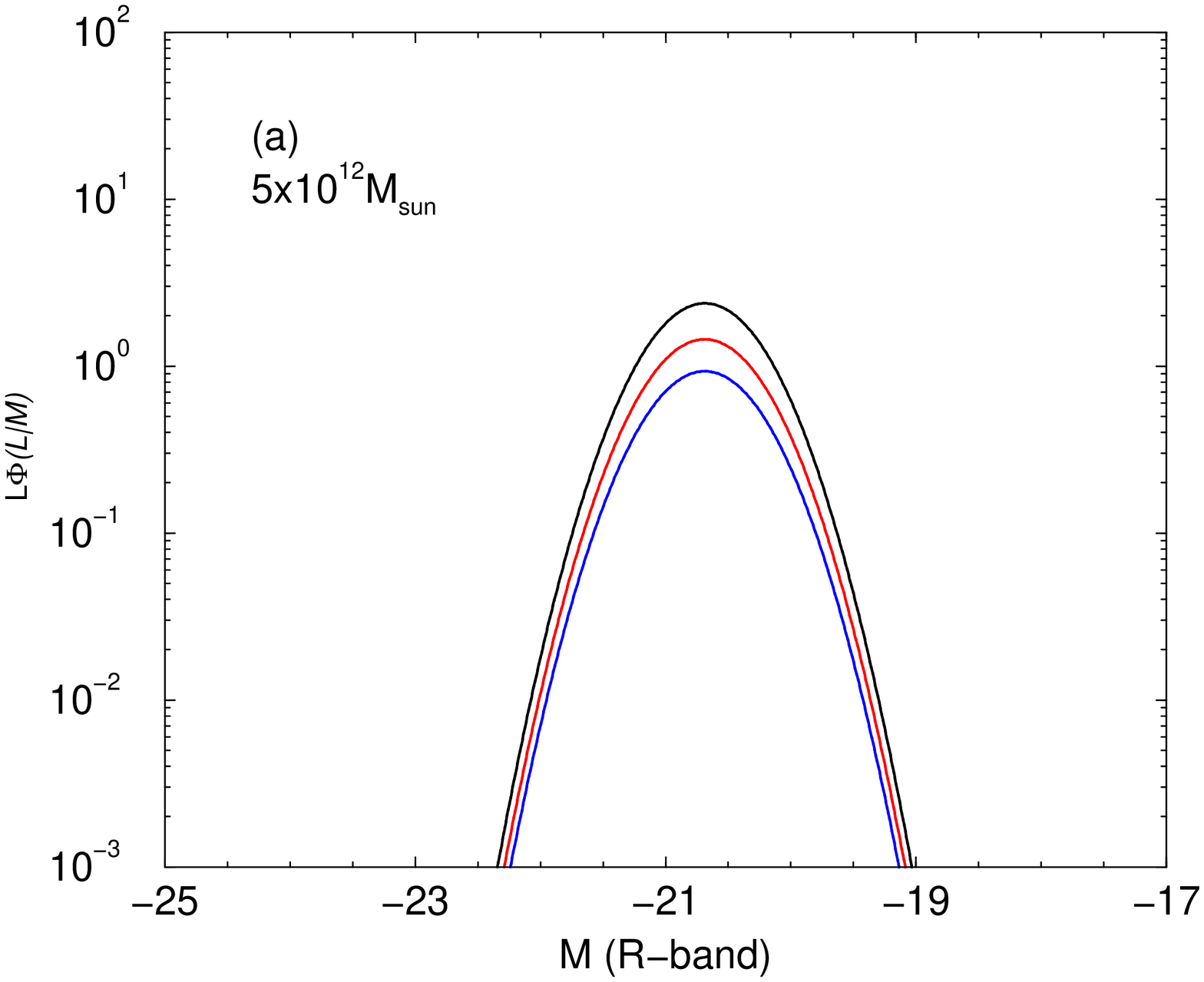,width=\hssize,angle=-90}
\psfig{file=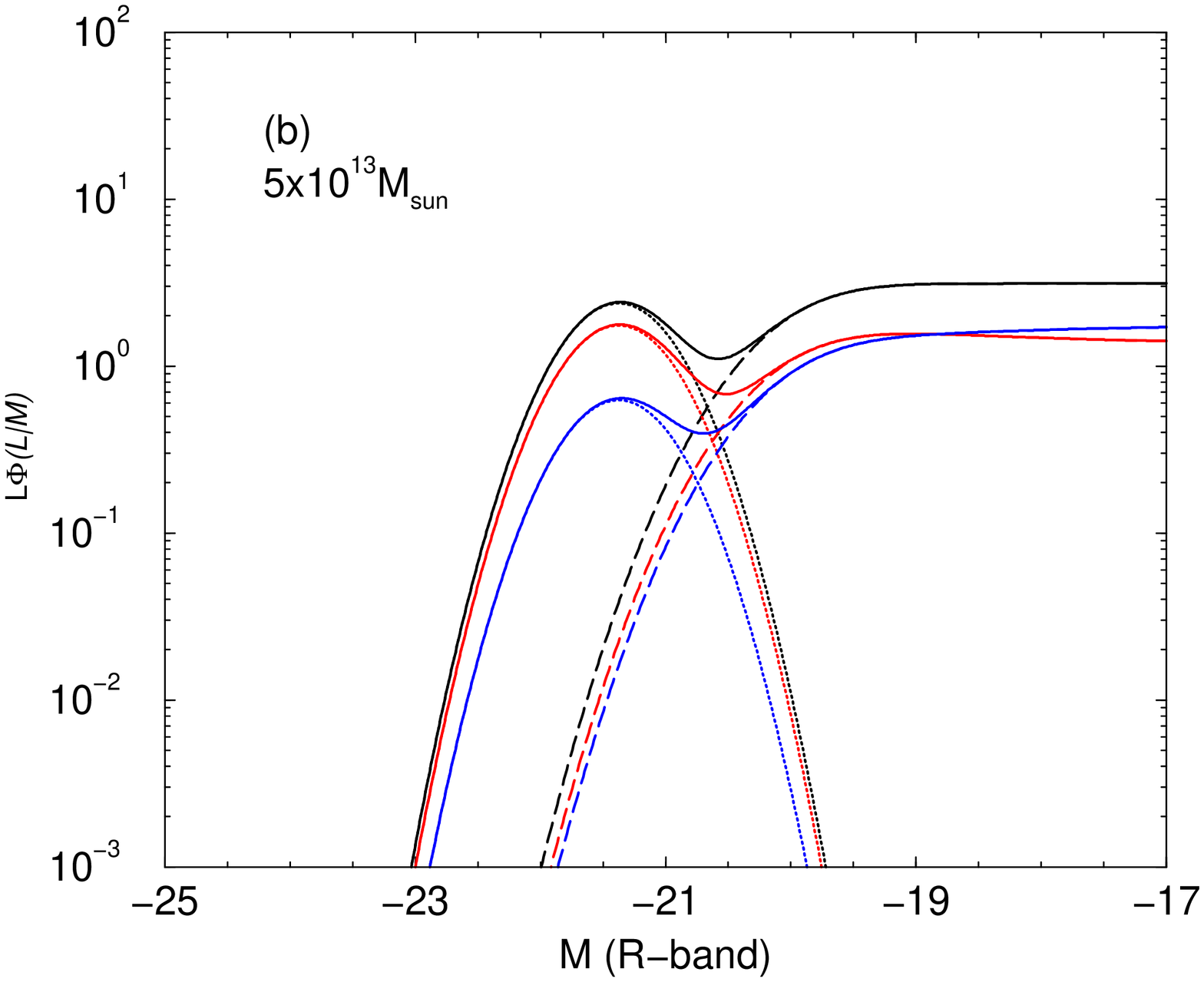,width=\hssize,angle=-90}}
\centerline{\psfig{file=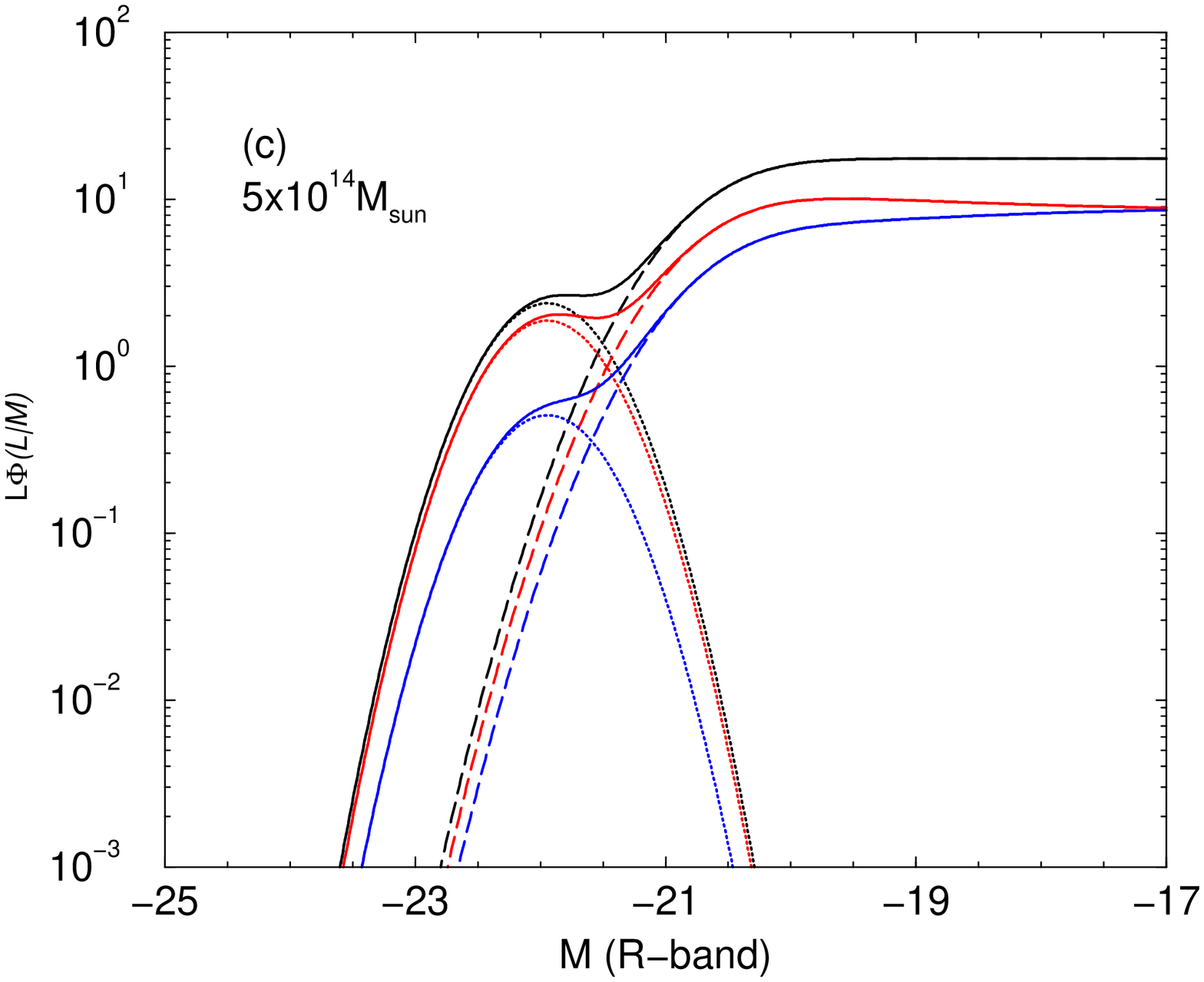,width=\hssize,angle=-90}
\psfig{file=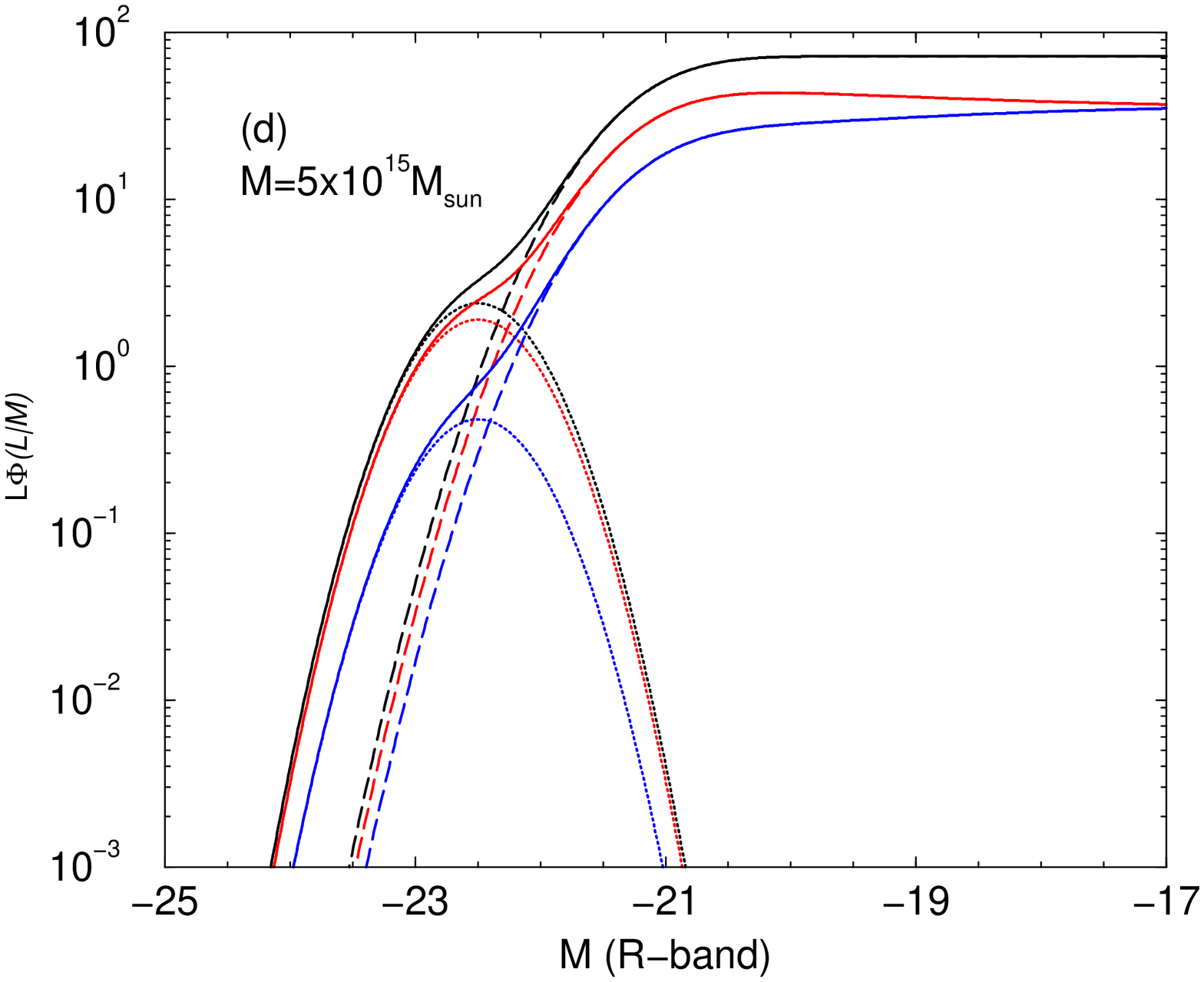,width=\hssize,angle=-90}}
\caption{Conditional luminosity functions today for a variety of masses as labeled on each of the four panels.
The CLFs are divided to early- (red lines) and late-type (blue lines) galaxies, while for reference,
we also show the total galaxy sample (black lines) with central and satellite galaxies
shown with dotted and dashed lines, respectively.  The CLF of high-mass halos are in good agreement
with galaxy cluster LF, such as from Coma (Trentham \& Tully 2002), that are neither fitted with
a single Schechter function nor a simple power-law at the faint-end of the LF (Cooray \& Cen 2005), though
due to the $M_r <-17$ cut-off in the measurements considered here, we ignore the latter complication.
}
\end{figure*}

\begin{figure*}
\centerline{\psfig{file=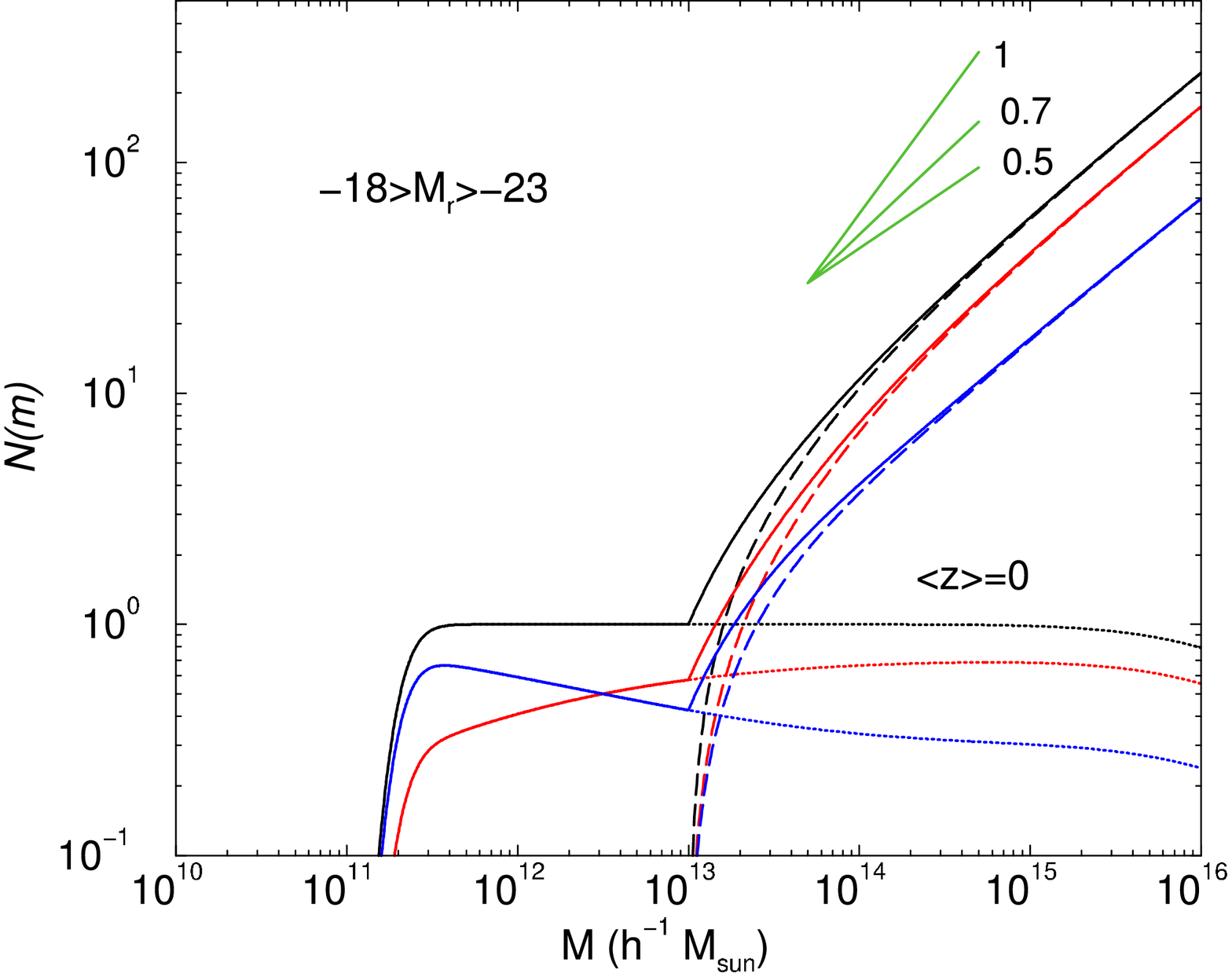,width=\hssize,angle=-90}
\psfig{file=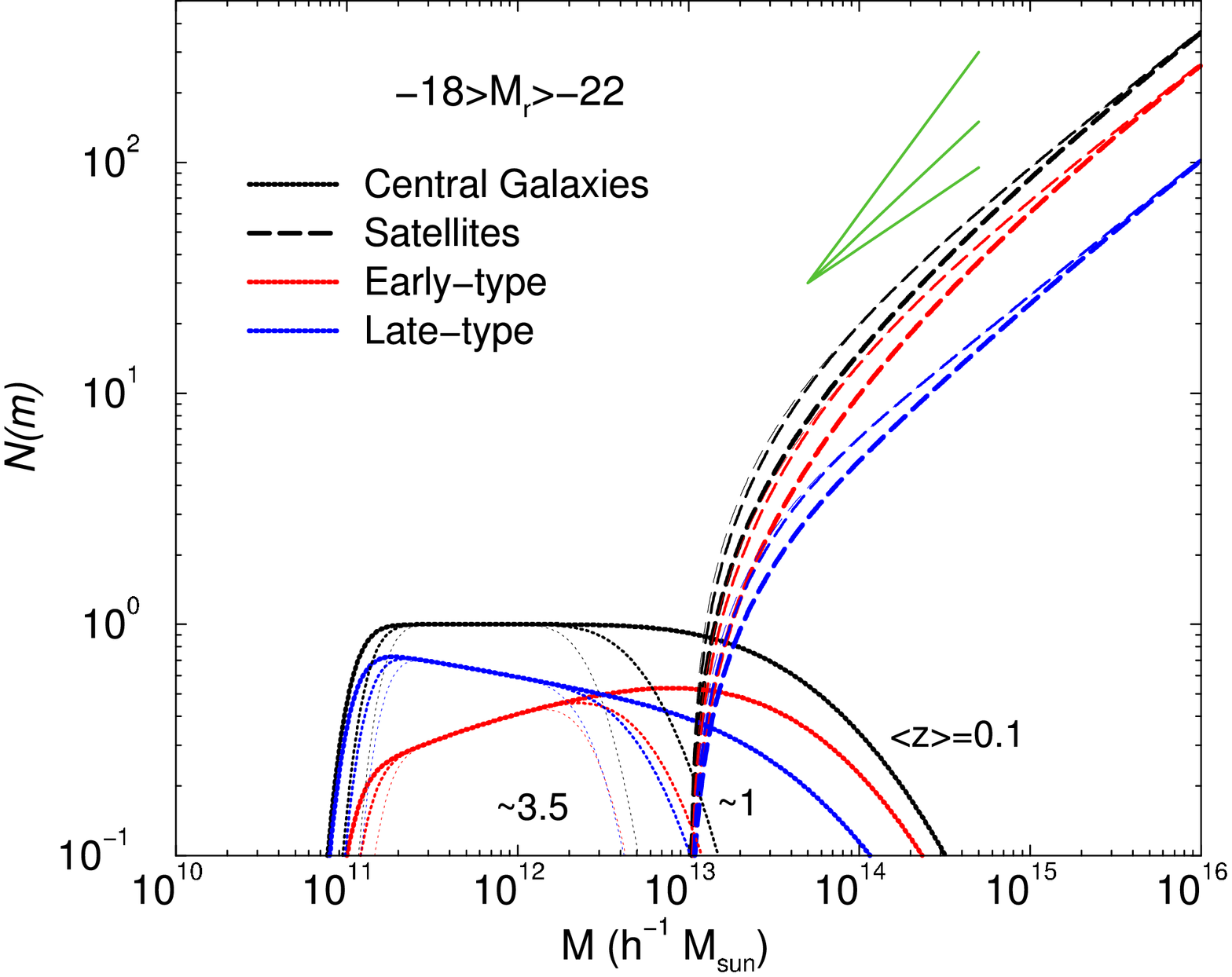,width=\hssize,angle=-90}}
\caption{Halo occupation numbers. {\it Left panel:} For galaxies with
absolute magnitudes between -18 and -23 in the SDSS $r$-band at redshifts $< 0.1$.
Dotted lines show the central galaxy occupation number, dashed lines show  the satellite occupation
number, and the solid line show the total occupation number. {\it Right panel:}
Redshift dependence of the halo occupation number, based on fiducial parameters for the $L_c(M,z)$
relation and $L_{\rm tot}(M,z)$ relation, as well as the redshift-dependent halo mass
function, for galaxies with -18 and -22 in the $r$-band (as appropriate for SDSS at a redshift of 0.1)
or B-band (for comparison with DEEP2 at a redshift of unity and GOODS at a redshift of 3.5). 
The occupation numbers suggest a power-law of $\sim 0.75$ for early-type galaxies
and $> 0.5$ for late-type galaxies at the high mass end when dominated by satellites.
Note also the transition from a dominant late-type fraction in
central galaxies to a dominant early-type fraction at a halo mass around
$\sim 3 \times 10^{12}$ M$_{\sun}$, regardless of the redshift.
When comparing halo occupation numbers determined from other data to the ones shown
above, luminosity ranges of galaxies between different samples should be taken into account.
Note that these occupation numbers are based on the fiducial model. When model fitting data,
we find large degeneracies suggesting that the halo occupation slope is not well constrained by the observations.
}
\end{figure*}

\begin{figure*}
\centerline{\psfig{file=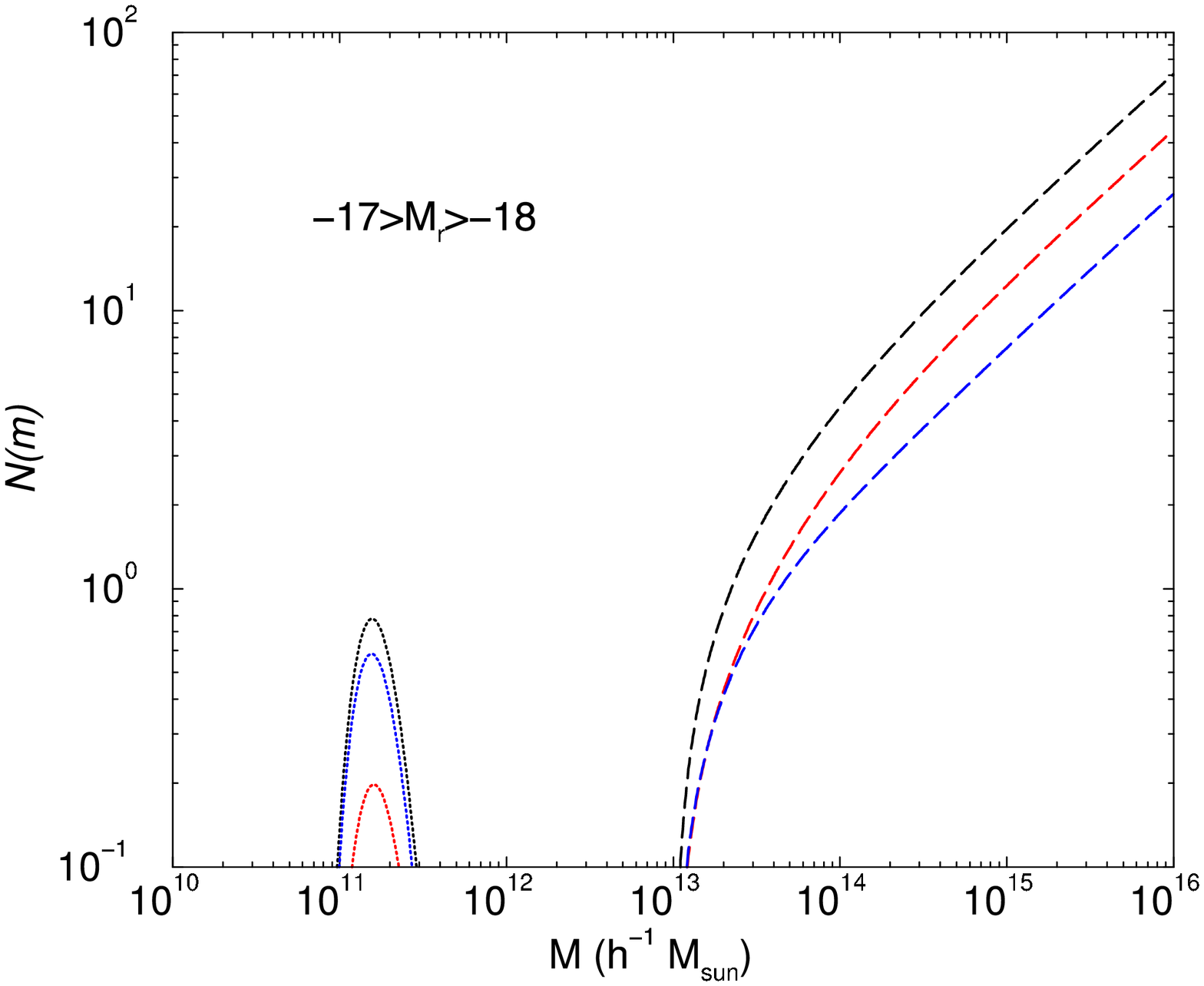,width=\hssize,angle=-90}
\psfig{file=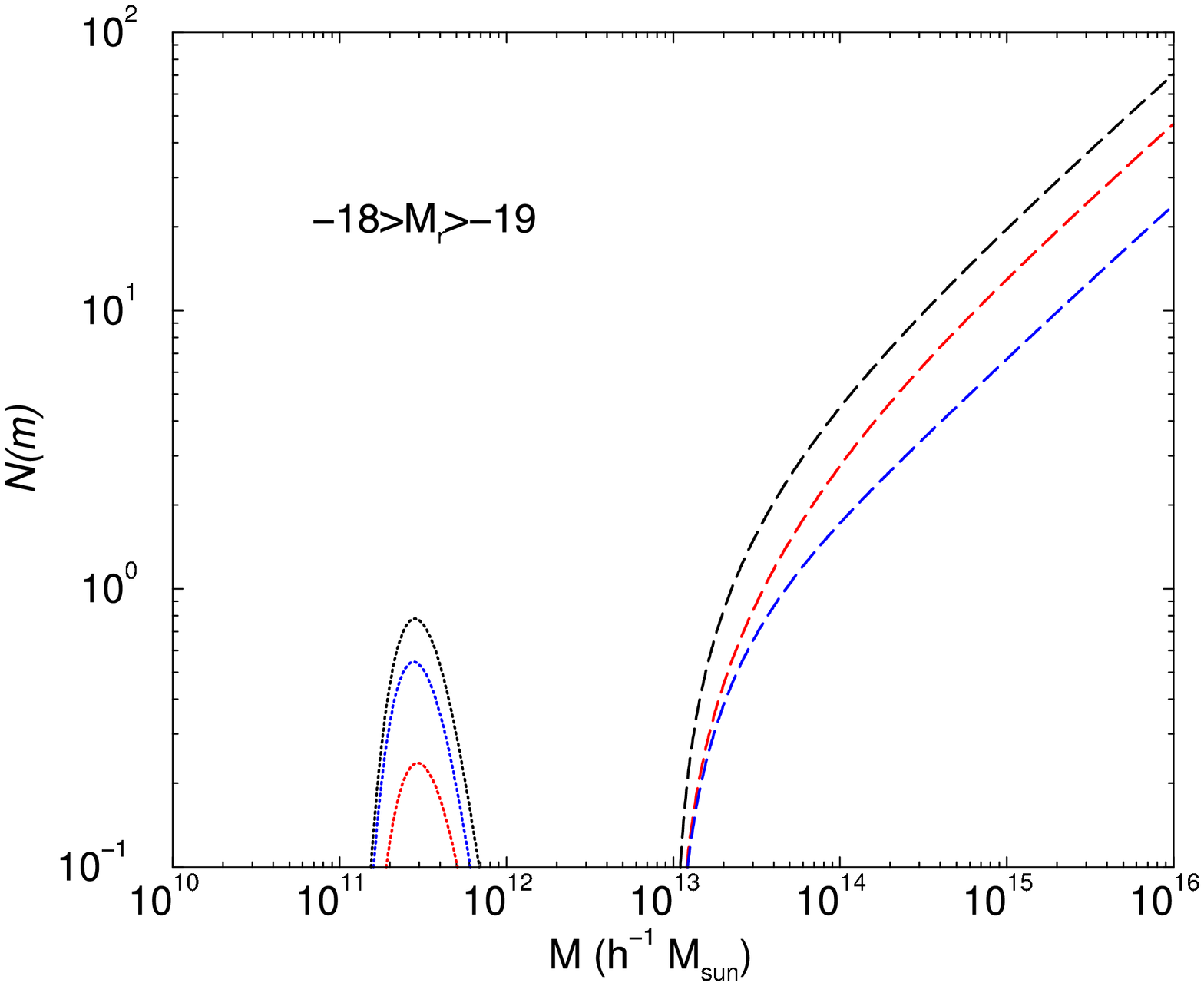,width=\hssize,angle=-90}}
\centerline{\psfig{file=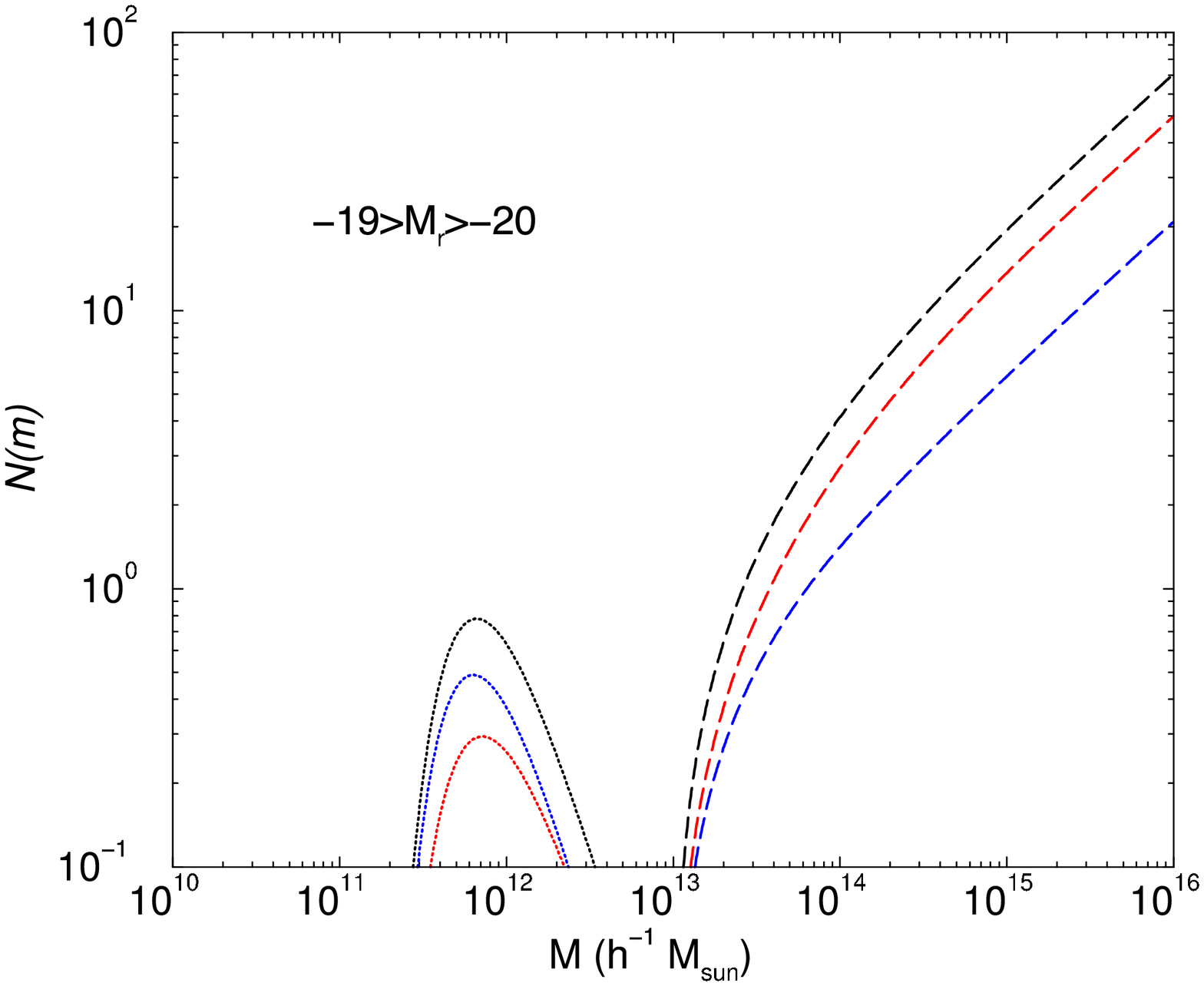,width=\hssize,angle=-90}
\psfig{file=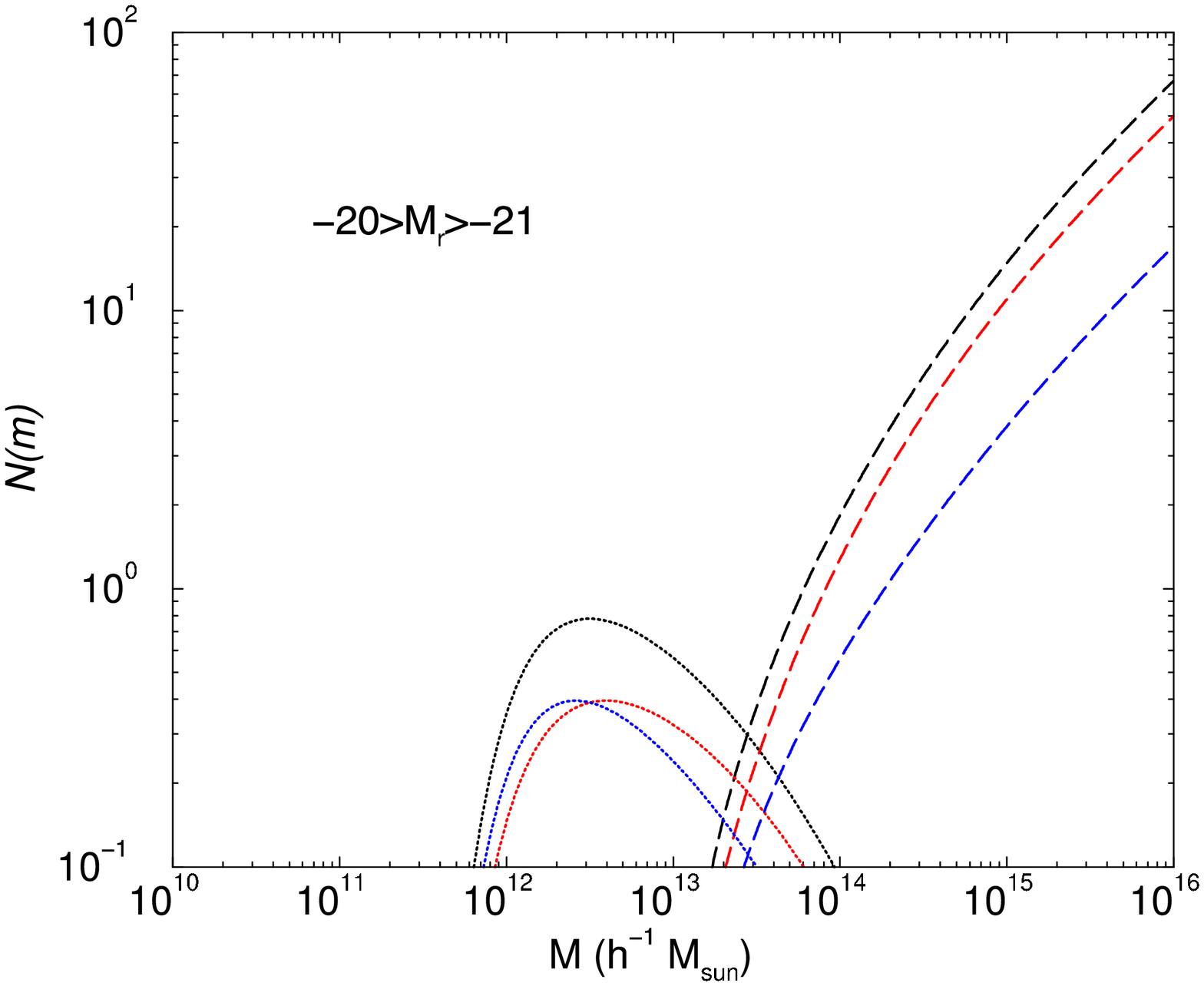,width=\hssize,angle=-90}}
\centerline{\psfig{file=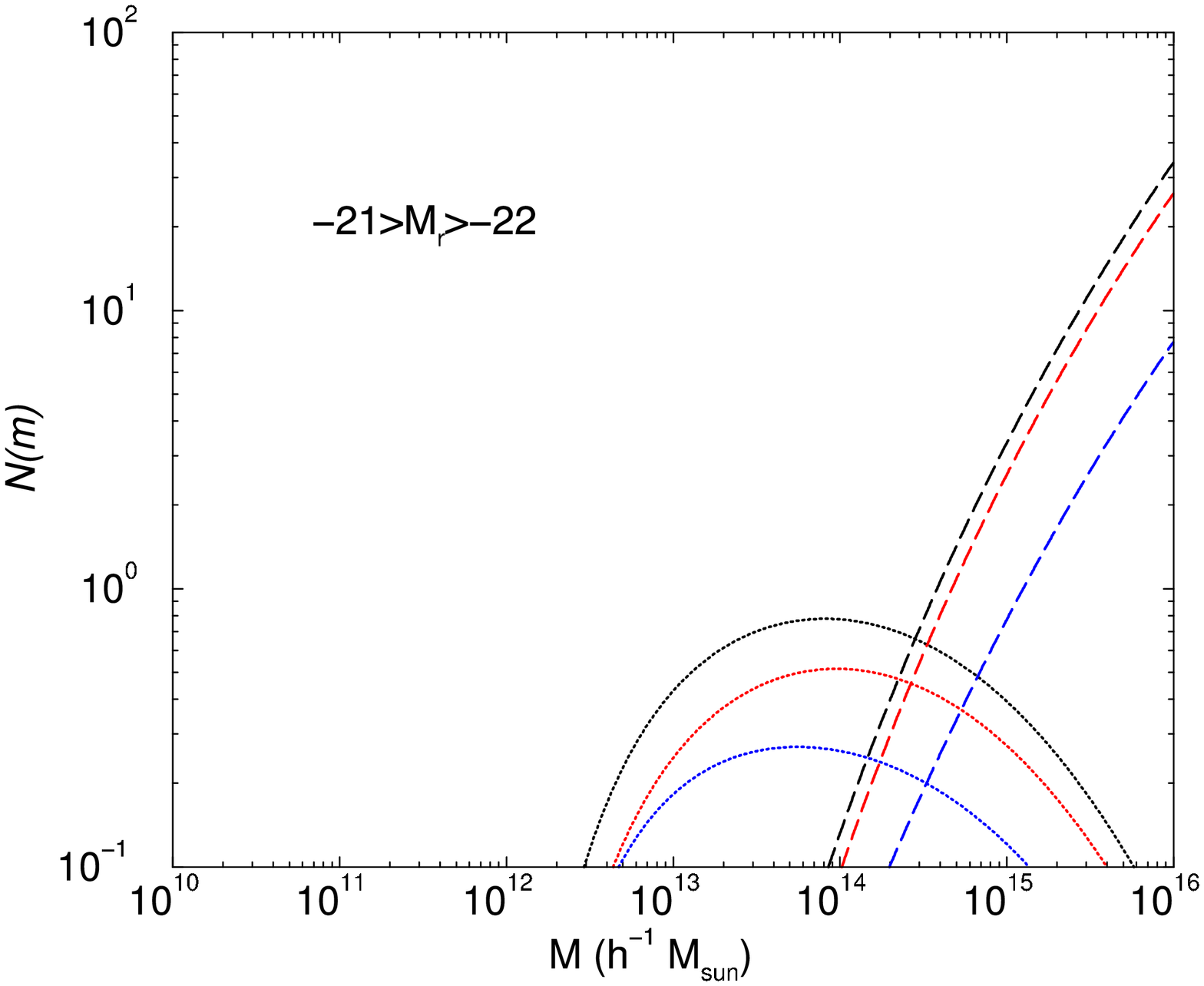,width=\hssize,angle=-90}
\psfig{file=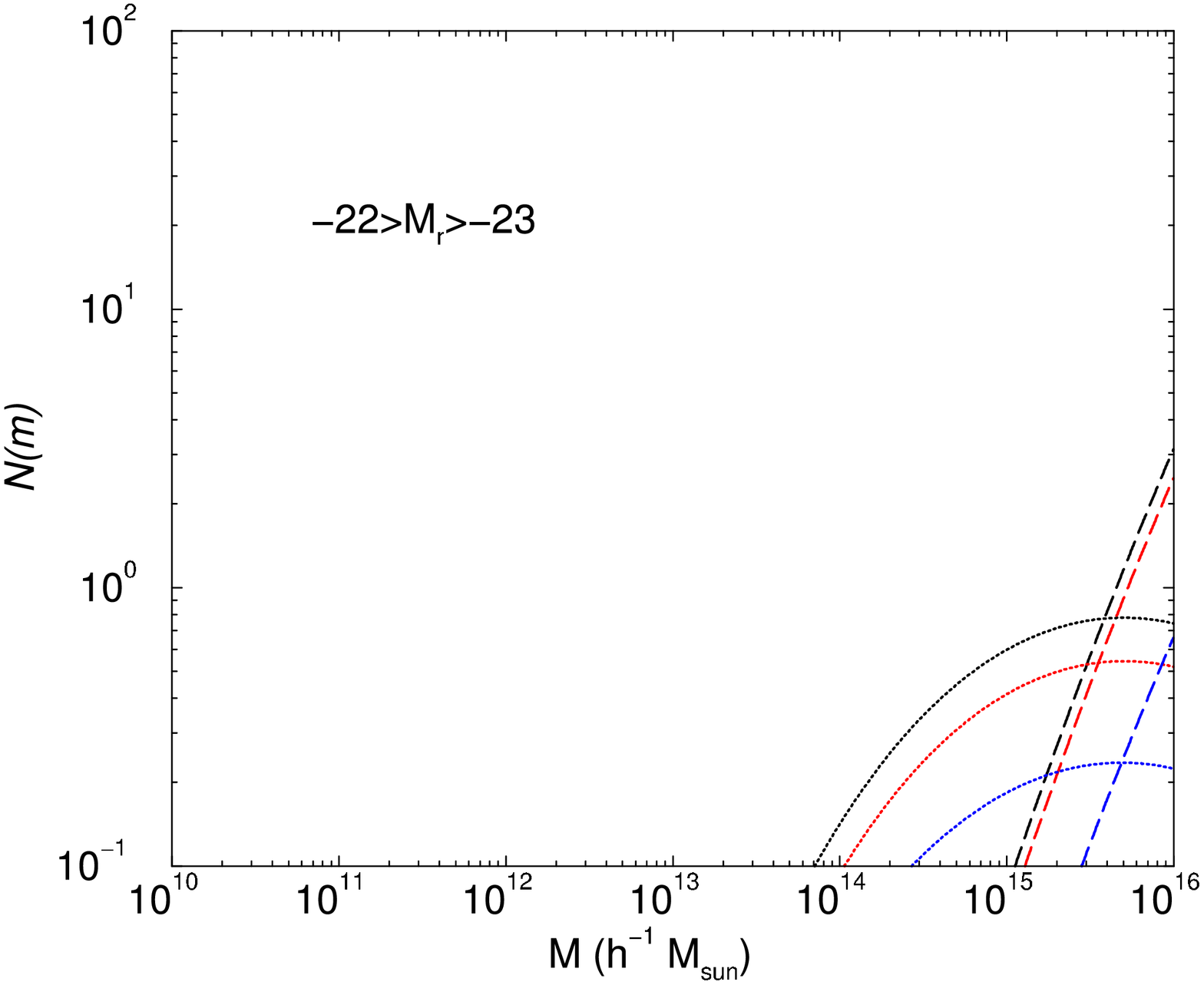,width=\hssize,angle=-90}}
\caption{Halo occupation numbers today as a function of the galaxy luminosity (as labeled on each 
of these plots). For reference, we divide the total occupation number to central (dotted  lines) and satellite (dashed lines) galaxies.
These are again based on our fiducial parameter description and these models are not unique to describe SDSS clustering
data given large degeneracies between parameters. This is also clear from the fact that ``best-fit'' halo occupation models
for same luminosity bins by Zehavi et al. (2004) suggest parameters that are distinctly different and involving
even power-law slopes in mass greater than unity.
}
\end{figure*}

In Equation~1, $L_c(M,z)$ is the relation between central galaxy luminosity of a given dark matter halo and it's halo mass, taken to be a function of the redshift, while $\sigma_{\rm cen}$, with a fiducial value of 0.17, is the log-normal
dispersion in this relation.  For an analytical description of the  $L_{\rm c}(M,z)$ relation, we make use of the  form suggested 
by Vale \& Ostriker (2004)  where this relation as appropriate for $b_J$-band galaxies today was established by inverting 
the 2dFGRS luminosity function given an analytical description for the sub-halo
mass function of the Universe (e.g., De Lucia et al. 2004; Oguri \& Lee 2004).  
The relation is described with a general fitting formula given by
\begin{equation}
\label{eqn:lcm}
L_{\rm c}(M,z) = L_0(1+z)^{\alpha} \frac{(M/M_1)^{a}}{[b+(M/M_1)^{cd(1+z)^{\eta}}]^{1/d}}\, .
\end{equation}
For the rest B-band, the parameters have values of $L_0=5.7\times10^{9} L_{\sun}$, $M_1=10^{11} M_{\sun}$,
$a=4.0$, $b=0.57$, $c=3.72$, and $d=0.23$ (Vale \& Ostriker 2004; Cooray 2005a,b), while
for SDSS $r$-band, we take $M_1=2\times10^{11} M_{\sun}$ and $c=3.78$ with other parameters as above.
The redshift evolution of this relation, based on high-redshift LFs, is discussed in Cooray (2005b).
Following the analysis described there, where we constrained values for redshift-dependent parameters
$\alpha$ and $\eta$, we take fiducial values of -0.5 and -0.1; these were the best-fit parameters
to the LFs of DEEP2 (Willmer et al. 2005), COMBO-17 (Bell et al. 2004), and rest B-band LFs of
Gabasch et al. (2004).

\begin{figure}
\centerline{\psfig{file=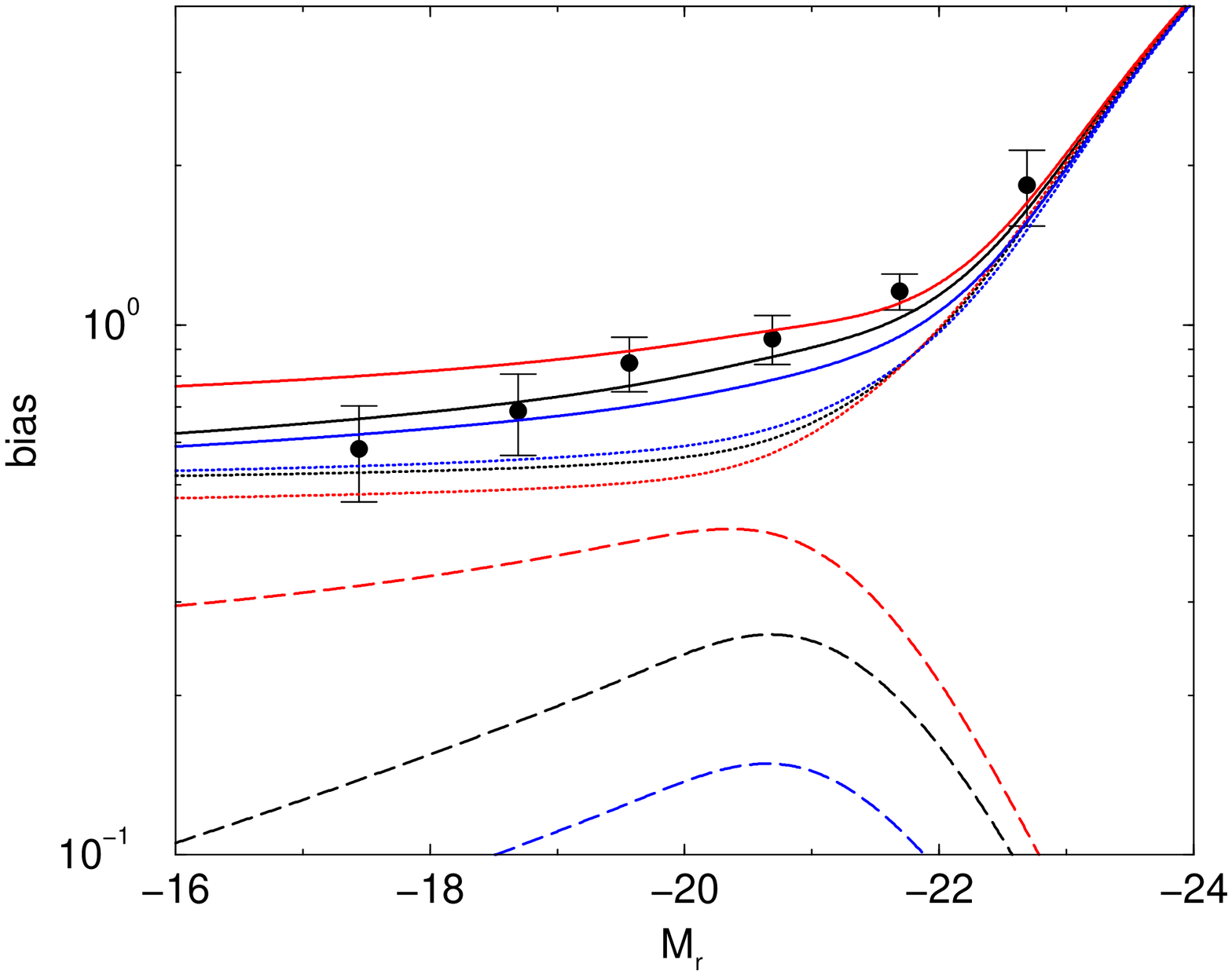,width=\hssize,angle=-90}}
\caption{Galaxy bias as a function of SDSS $r$-band absolute magnitude as calculated from CLFs
({\it solid line}) with SDSS bias measurements shown with data points (from Zehavi et al. 2004). 
We also separate contributions from central galaxies ({\it dotted line})
and satellites ({\it dashed line}) to galaxy bias. 
We also show the bias for galaxy types (early- and late-type galaxies). Late-type galaxies are
expected to be in low-dense regions dominated by low-mass halos and their bias factor, relative to early type galaxies,
would be lower. Satellite galaxies, regardless of the type, are in more massive halos
and, thus, have higher bias factors relative to central galaxies. The average bias factor, shown here
for the whole sample, however, is dominated by central galaxies due to the same reason that the 
LF is also dominated by central galaxies.}
\end{figure}

\begin{figure*}
\centerline{\psfig{file=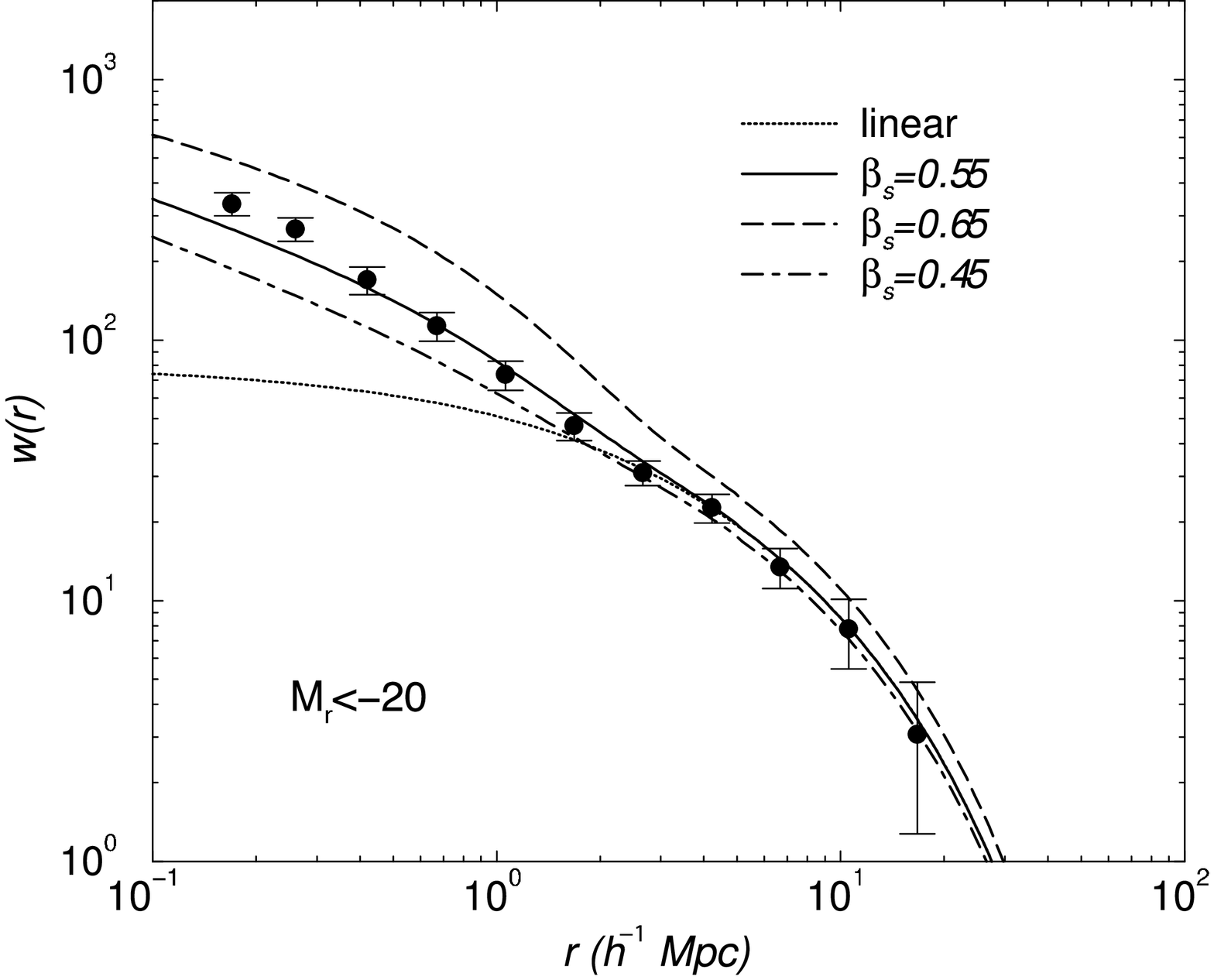,width=\hssize,angle=-90}
\psfig{file=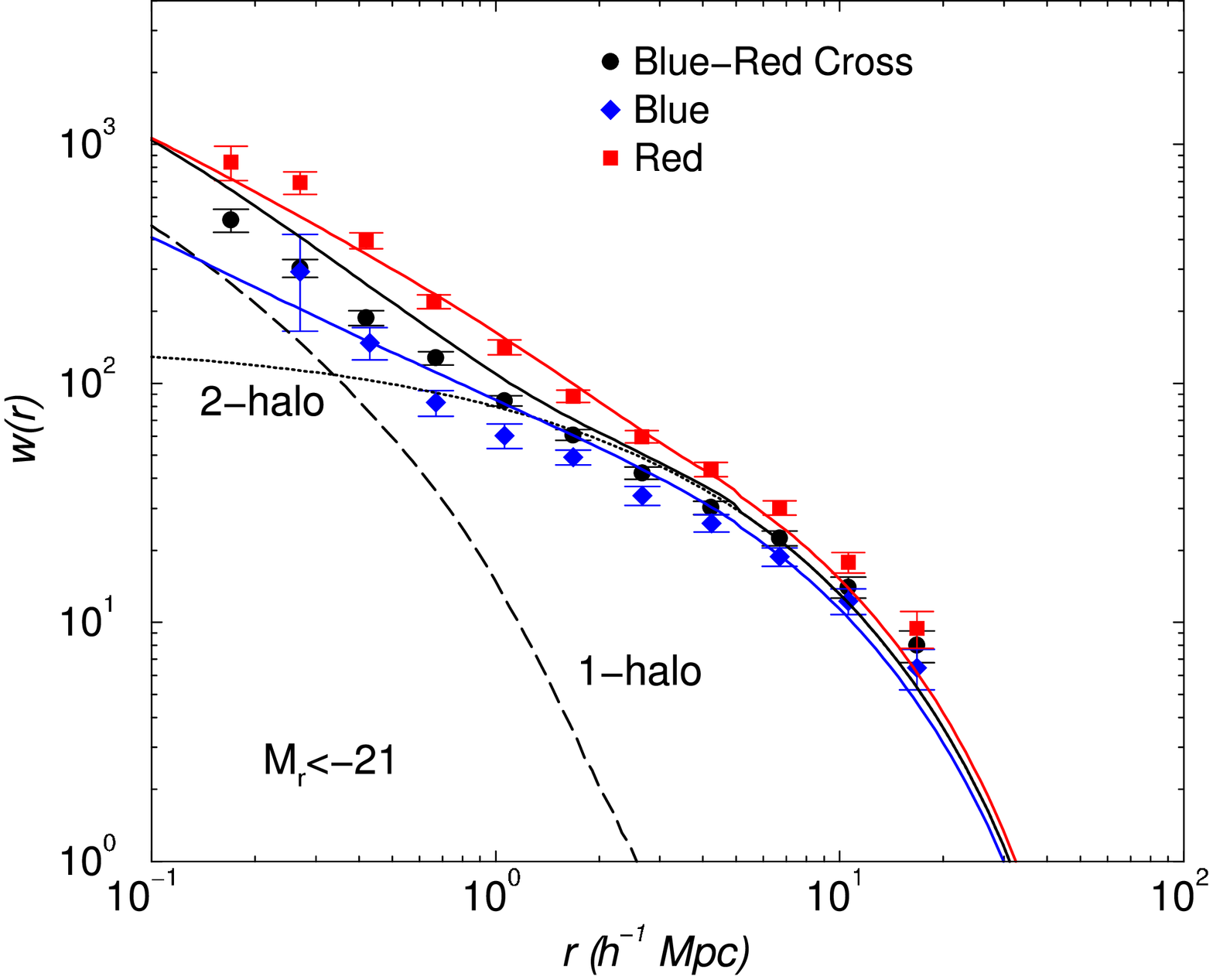,width=\hssize,angle=-90}}
\caption{Projected correlation function of SDSS galaxies (from Zehavi et al. 2004).
{\it Left panel:} For galaxies with $M_r <-20$. Here, we show the prediction based on
CLFs and variations  associated with a change in the power-law slope of the total luminosity--halo mass relation. For reference, we also show the projected clustering power spectrum from the linear
power spectrum alone, but scaled by the large-scale bias factor for galaxies with same luminosities. {\it Right panel:}
Clustering of galaxy types and cross-clustering between galaxy types for the sample with $M_r <-21$.
For reference, we show both 2-halo and 1-halo contributions to the projected
cross-correlation function between the two galaxy types.
}
\end{figure*}

\begin{figure*}
\centerline{\psfig{file=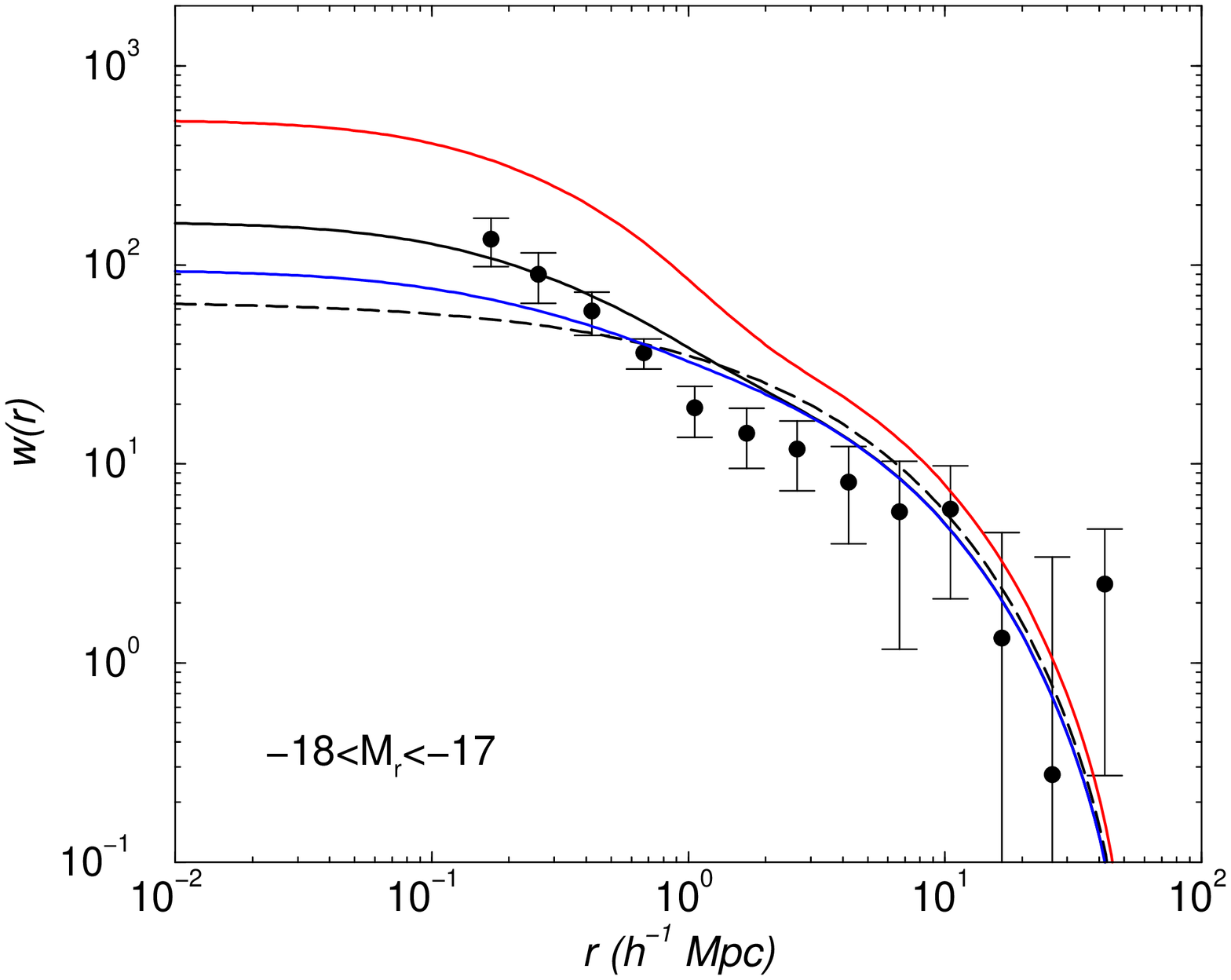,width=\hssize,angle=-90}
\psfig{file=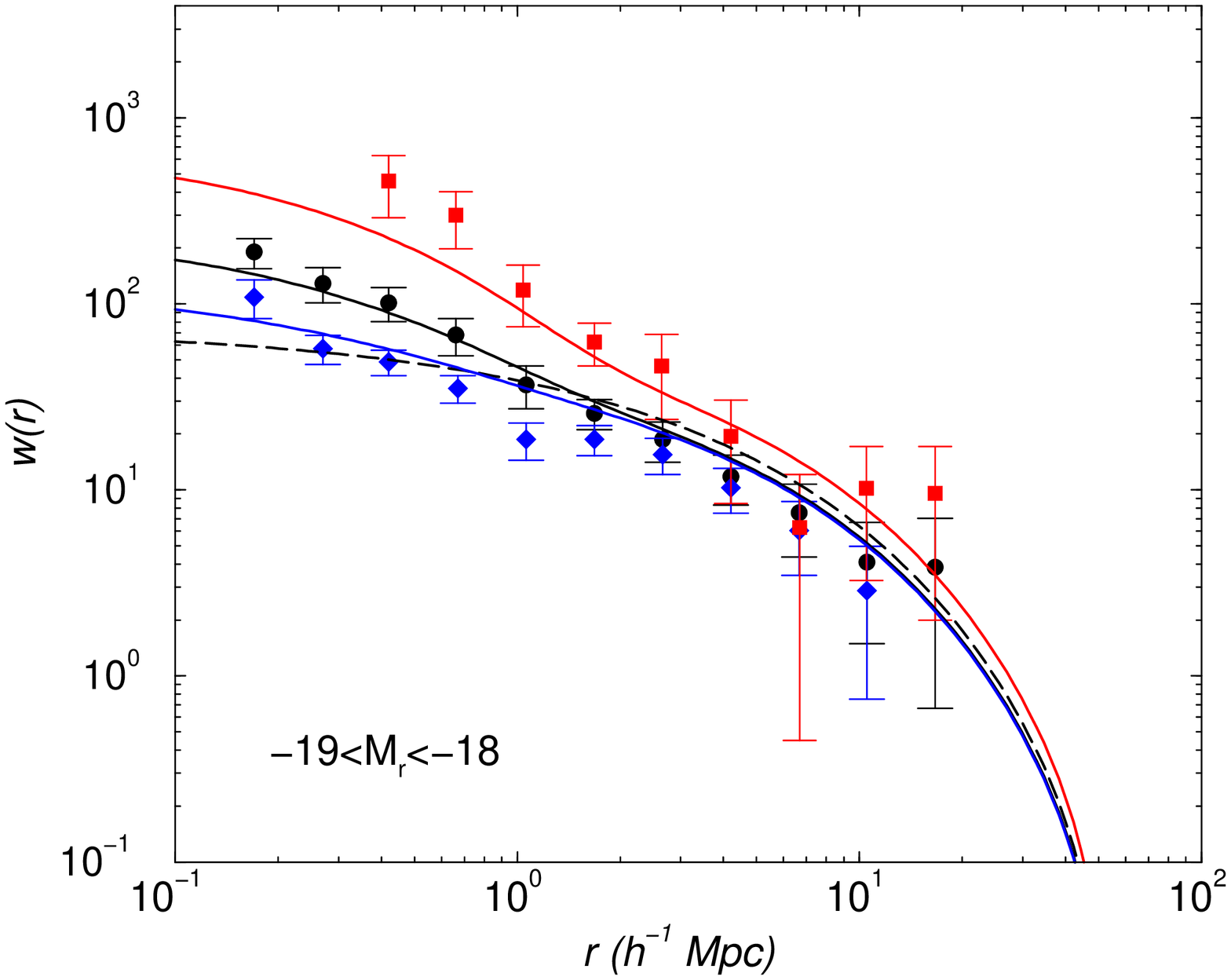,width=\hssize,angle=-90}}
\centerline{\psfig{file=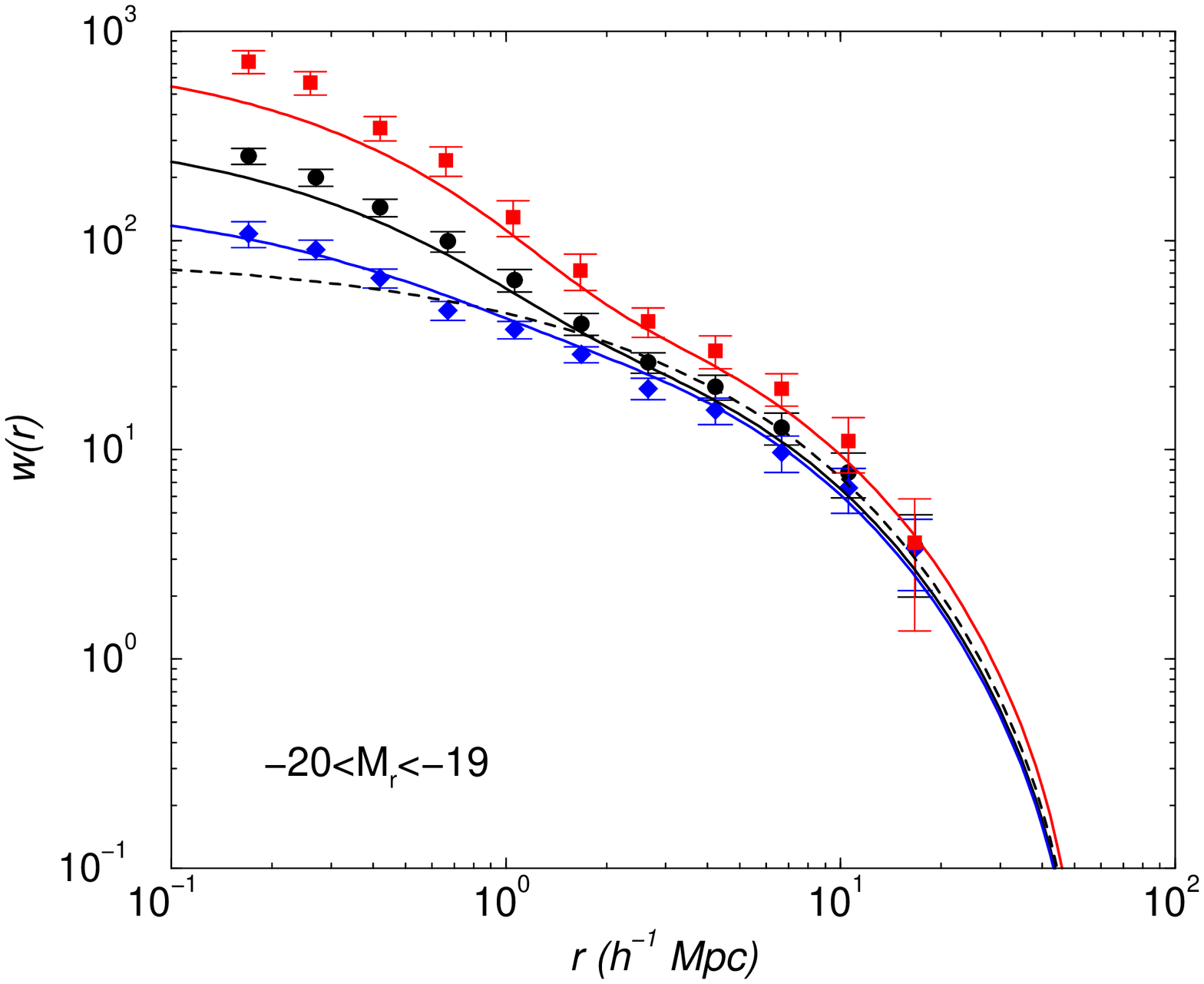,width=\hssize,angle=-90}
\psfig{file=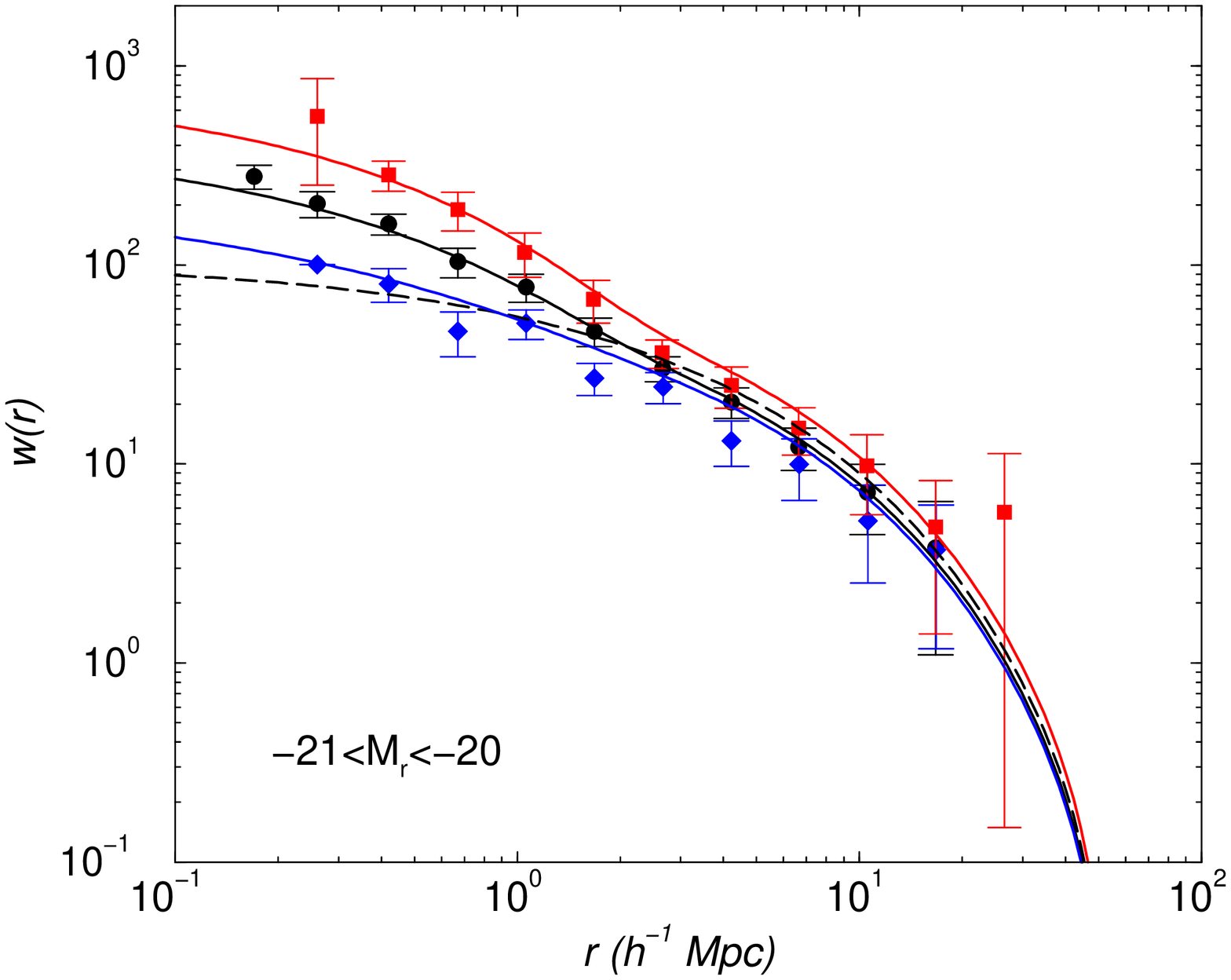,width=\hssize,angle=-90}}
\centerline{\psfig{file=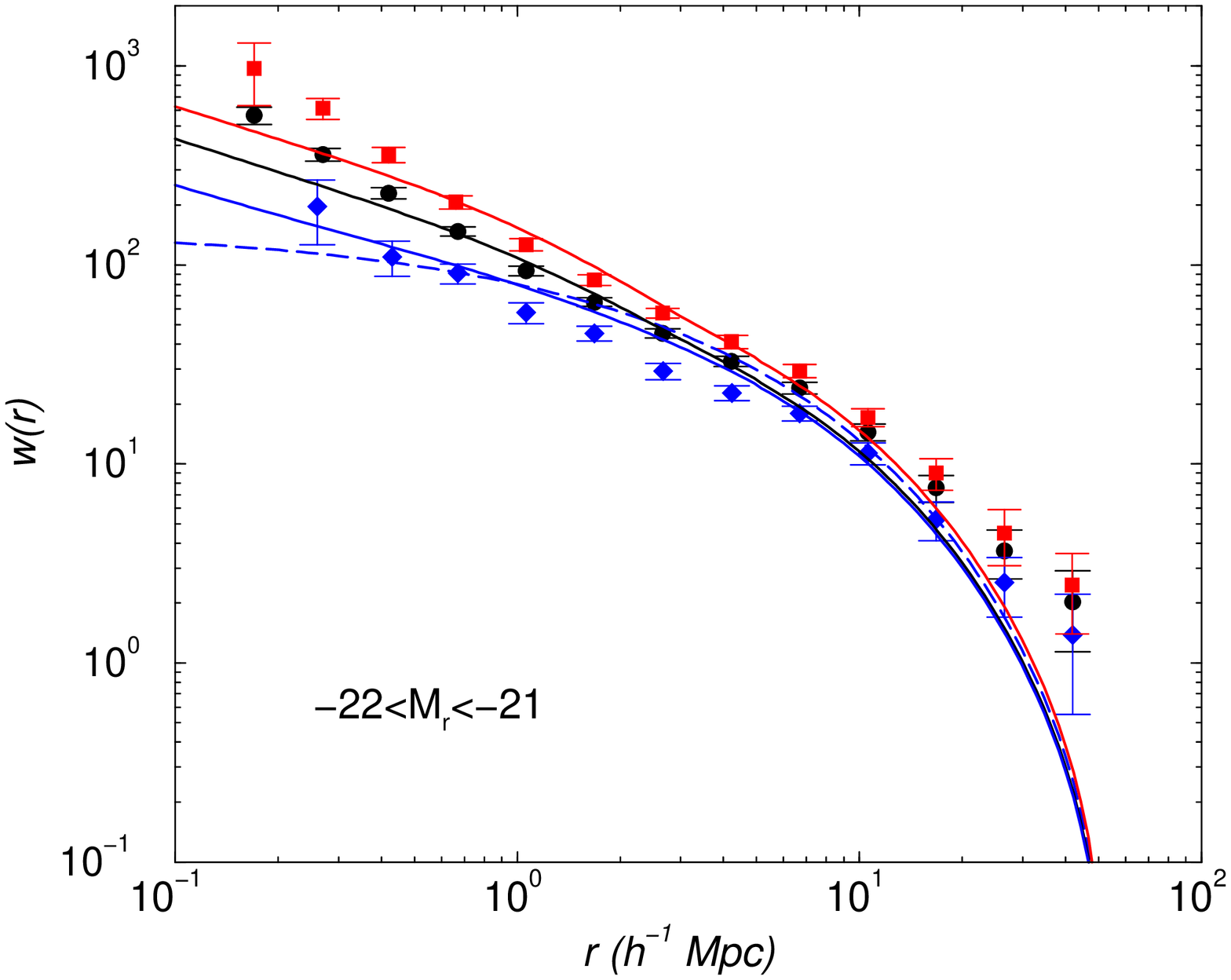,width=\hssize,angle=-90}
\psfig{file=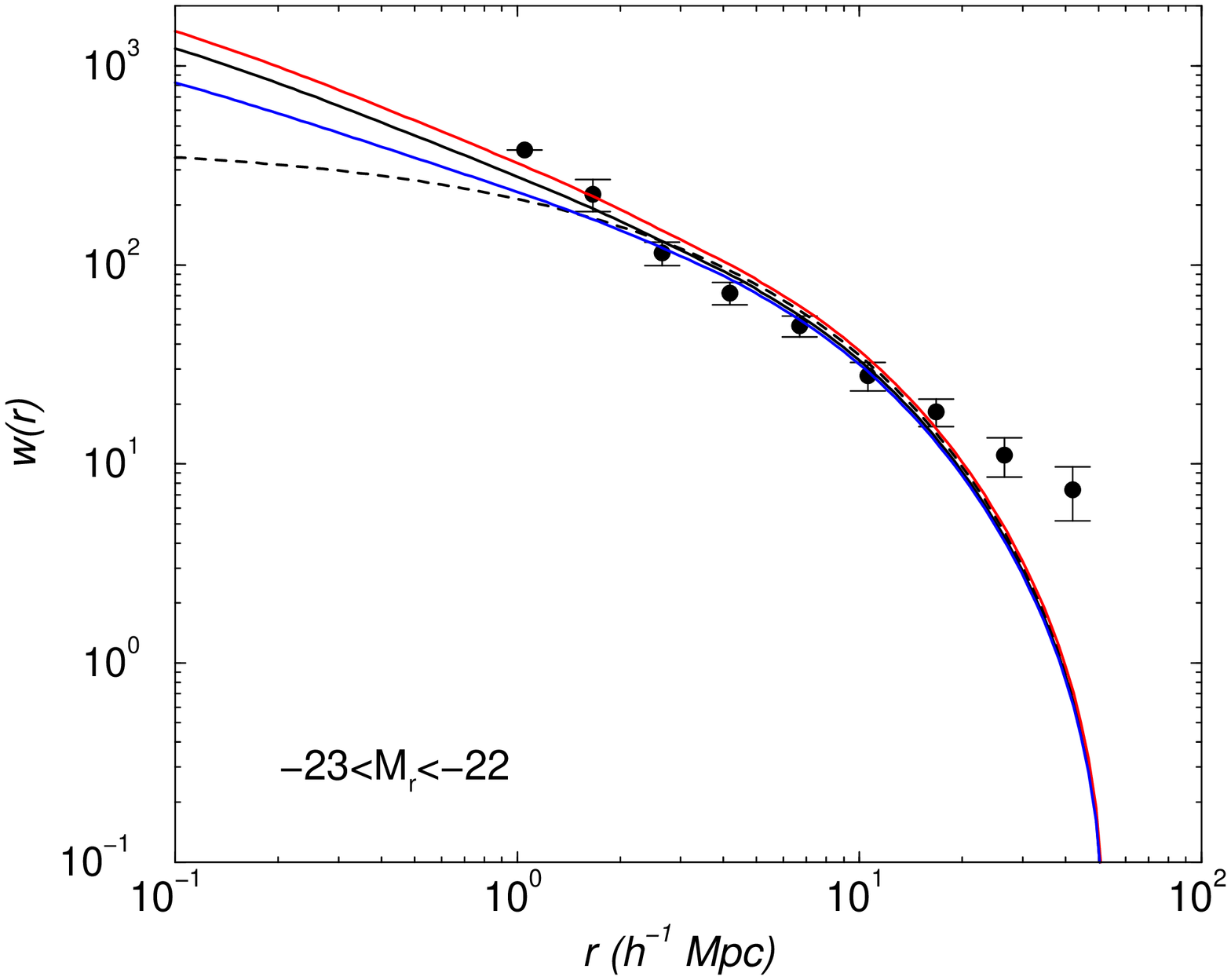,width=\hssize,angle=-90}}
\caption{Projected correlation function as a function galaxy absolute magnitude in the SDSS $r$-band (data from Zehavi et al. 2004). We show the predictions based on our fiducial model parameters.
In addition to the total galaxy sample, when available, we also show the measurements, as well as predictions, for clustering of galaxy types. The CLFs associated with these predictions are shown in Fig.~3 while the halo occupation numbers, 
based on an integration of the CLFs, are shown in Figure~4.}
\end{figure*}

For satellites, the normalization $A(M)$ of the satellite CLF can be obtained by defining 
$L_{\rm s}(M,z)\equiv L_{\rm tot}(M,z)-L_{\rm c}(M,z)$ and requiring 
that $L_{\rm s}(M,z)=\int_{L_{\rm min}}^{L_{\rm max}} \Phi^{\rm sat}(L|M,z)LdL$ with $g_{\rm s}(L)=1$, where
the minimum luminosity  of a satellite is $L_{\rm min}$.
In the luminosity ranges of interest, and due to the numerical value chosen below for the slope $\gamma$, 
our CLFs are mostly independent of the exact value assumed for $L_{\rm min}$ as 
long as it lies in the range $(10^6-10^8)L_{\sun}$.  To describe the total luminosity of a halo, departing from the
model used in Cooray (2005b), we make use of the following phenomenological form:
\begin{eqnarray}
L_{\rm tot}(M,z) = \left\{\begin{array}{ll}
L_{\rm c}(M,z)  & M \leq M_{\rm sat}\\
L_{\rm c}(M,z)\left(\frac{M}{M_{\rm sat}}\right)^{\beta_s(z)} & M>M_{\rm sat}
\end{array}\right. 
\label{eqn:ltot}
\end{eqnarray}
Here, $M_{\rm sat}$ denotes the mass-scale at which satellites begin to appear in dark matter halos
with luminosities as corresponding to those in the given sample of galaxies, while $\beta_s(z)$ 
is the correction to the power-law slope of the
total luminosity--halo mass relation relative to that of the central galaxy--halo mass relation. 
We use this form since other parameterizations we considered resulted in unphysical situations for certain
parameter values in those descriptions, e.g., $L_{\rm tot}(M) < L_c(M)$, while
other parameterizations did not provide useful constraints on parameters used for the description 
due to additional degeneracies. More importantly,
the above form allows us to highlight easily an interesting result, involving the single
parameter best constrained by clustering data, that we will discuss later.
When showing models of CLFs in Figures 3 to 8, motivated by constraints from that data that we will
describe later, we take
$M_{\rm sat}=10^{13}$ M$_{\sun}$ and $\beta_s=0.55$ to describe $r$-band galaxies with absolute magnitudes $M_r <-17$.
In Figures~11 and 12, same numerical values for the parameters  of the satellite 
CLF are also used at high redshifts and in the rest B-band, though we note a redshift-dependent variation in these
parameters, especially when considering $z=4$ clustering data from the Subaru Deep Field.
Though we show figures with $M_{\rm sat}=10^{13}$ M$_{\sun}$ and $\beta_s=0.55$, this does not mean these are the best-fit values or
our preferred values for these parameters. When we model fit the data, we will show constraints on these parameters explicitly and show
that a rather large range of values is allowed by the data. While these two parameters are degenerate with each other,
in addition to SDSS data at $z <0.1$, certain  high-redshift data, such as COMBO-17 at $z \sim 0.6$ and Subaru/XMM-Newton Deep Field with
clustering measurements at $z \sim 4$,   do allow constraints to be placed on these parameters.

While the above form refers to the total luminosity, when $L_{\rm tot}(M,z) > L_{\rm c}(M,z)$, this total luminosity
 must be distributed over a number of satellite galaxies in the halo when describing the satellite CLF. We take a power-law
luminosity distribution and set  $\gamma(M,z)=-1$ in Equation~1 based on previous results derived on the CLF of
galaxy groups and clusters (Cooray \& Milosavljevi\'c 2005b; Cooray 2005a) and direct
measurement in clusters such as Coma (Driver \& De Propis 2002 where $\gamma=-1.01^{+0.04}_{-0.05}$); While the choice of $\gamma\sim -1$
is motivated by the cluster LF, setting $\gamma$ to a different value  between -0.7 and -1.3, over a set of parameter values
we tested, did not change our results significantly.
Furthermore, for the maximum luminosity of satellites in a given halo, following the
result found in Cooray \& Milosavljevi\'c (2005b) based on  a comparison of predictions to the
K-band cluster LF of Lin \& Mohr (2004),  we set $L_{\rm max}=L_{\rm c}/2$.
A comparison to 2dFGRS CLFs  as measured by Yang et al. (2005), however,
suggested that such a sharp cut-off is inconsistent and that to account for scatter in the total galaxy
luminosity, as a function of the halo mass, one must allow for a distribution in
$L_{\rm max}$. Instead of additional numerical integrals, we allow for a luminosity dependence with the introduction of
$g_{\rm s}(L)$ centered around the maximum luminosity of satellites such that
$\Phi^{\rm sat}(L|M)$ does not go to zero rapidly at $L_{\rm max}$. By a comparison to the data,
we again found a log-normal description with
\begin{equation}
g_{\rm s}(L) = \frac{1}{2}\left[1+{\rm erf}\left(\frac{\log(L_{\rm c}/2.0)-\log(L)}{\sigma_{s}}\right)\right] \, ,
\end{equation}
where $\sigma_{s}=0.3$. The description here is such that $f_{\rm L} =1$ when $L < L_{\rm mac}=L_{\rm c}/2$,
but falls to zero at a luminosity beyond $L_{\rm c}/2$ avoiding the sharp drop-off at $L_{\rm c}/2$ with $g_{\rm s}(L)=1$.  
Again, our results are mostly insensitive to parameters of this description since variations here only lead to
small changes to the overall CLF.

The central galaxy CLF takes a log-normal form while the satellite galaxy CLF takes a power-law form in luminosity.
Such a separation describes the LF best with an overall better fit to the data in the K-band as explored by
Cooray \& Milosavljevi\'c (2005b) and 2dFGRS $b_J$-band in Cooray (2005).  Our motivation for log-normal distribution also comes
from measured galaxy cluster LFs that include bright central galaxies where 
a log-normal component, in addition to the Schechter (1976) form, is required to fit the data (e.g., Trentham \& Tully 2002). 
Similarly, the stellar mass function as a function of halos mass in semi-analytical models
is best described with a log-normal component for  central galaxies (Zheng et al. 2004).
As we find later the overall shape of the LF is {\it strongly} sensitive to the shape of the $L_{\rm c}$--$M$ relation,
and it's scatter, and less on details related to the $L_{\rm tot}$--$M$ relation.
The non-linear part of the galaxy correlation function, or any clustering statistic, probes the satellite distribution
and constraints can be put on the $L_{\rm tot}$--$M$ relation. In fact, we find that the average luminosity of satellites, defined
in Section~4, is the single parameter best constrained with current data.

To describe galaxies as a function of color in this analytical description, we must further divide central and satellite
galaxies as a function of their color given the luminosity. Here, motivated by the bimodality of color (e.g., Baldry et al. 2004)
that extends out to high redshifts (e.g., Giallongo et al. 2005), we consider models in terms of galaxy types.
The description in terms of galaxy types is also useful since measurements at high redshifts,
so far, involve the division of galaxy samples to two broad categories involving early-type, or red, and
late-type, or blue, galaxies. Thus, in the case of early type galaxies we write the CLF as
\begin{eqnarray}
\Phi^{\rm cen}_{\rm early}(L|M,z) &=& \Phi^{\rm cen}(L|M,z) f_{\rm early-cen}(M,z) \nonumber \\
\Phi^{\rm sat}_{\rm early}(L|M,z) &=& \Phi^{\rm sat}(L|M,z) f_{\rm early-sat}(M,L,z) \, ,
\end{eqnarray}
where the two functions that divide between early- and late-types are taken to be
functions of mass, in the case of central galaxies, and both mass and luminosity in the case of satellites. 
These functions are described analytically as
\begin{eqnarray}
&&f_{\rm early-cen}(M,z) = \\
&&\frac{f_{\rm cen-E}(z)}{2}\left[1+{\rm erf}\left(\frac{\log(M)-\log(M_{\rm cen}(z))}{\sigma_{\rm early-cen}}\right)\right] \, , \nonumber
\end{eqnarray}
with fiducial parameters of $M_{\rm cen}(z) = 5 \times 10^{11}$ M$_{\sun}$, $\sigma_{\rm early-cen}=2.0$, and $f_{\rm cen-E}(z)=0.6$,
and
\begin{eqnarray}
&&f_{\rm early-sat}(M,L,z) =  \\
&&g_{\rm sat-E}(z)g(M,z) + g_{\rm sat-E}(z)h(L,z)+f_{\rm sat-E}(z) \, , \nonumber
\end{eqnarray}
where,
\begin{eqnarray}
g(M,z) &=& \frac{1}{2}\left[1+{\rm erf}\left(\frac{\log(M)-\log(M_{\rm sat}(z))}{\sigma_{\rm sat}}\right)\right] \nonumber \\
h(L,z) &=& \frac{1}{2}\left[1+{\rm erf}\left(\frac{\log(L)-\log(L_{\rm sat}(z))}{\sigma_{\rm sat}}\right)\right] \, ,
\end{eqnarray}
with $M_{\rm sat}=10^{13}$ M$_{\sun}$, $L_{sat} = 3 \times 10^{9}$ $L_{\rm sun}$, $\sigma_{\rm sat} = 1$,
$f_{\rm sat-E}(z)=0.4$ and $g_{\rm sat-E}(z)=0.2$; Early-type galaxies in the form of satellites varies from
a fraction of $f_{\rm sat-E}(z)$ at low luminosity galaxies in low mass halos to $2g_{\rm sat-E}(z)+f_{\rm sat-E}(z) $ in
halos with masses greater than $ 10^{13}$ M$_{\sun}$. As fractions are defined
with respect to the total galaxy number of a halo,  late-type fractions are simply $[1-f_{\rm early-cen}(M,z)]$
and $[1-f_{\rm early-sat}(M,L,z)]$ for central and satellite galaxies, respectively
and we do not need to specify there parameters separately.

The fractions, following the fiducial values mentioned above --- with some parameters estimated based on model fits
to measurements described later --- are shown in Fig.~1. The late-type fraction  varies from $\sim$ 0.8
at halo masses of $10^{11}$ M$_{\sun}$, in the form of central galaxies, to $\sim$ 0.3 when $M \sim 10^{15}$ M$_{\sun}$
corresponding to galaxy cluster scales, with the fraction essentially dominated by satellite galaxies.

\begin{figure}
\centerline{\psfig{file=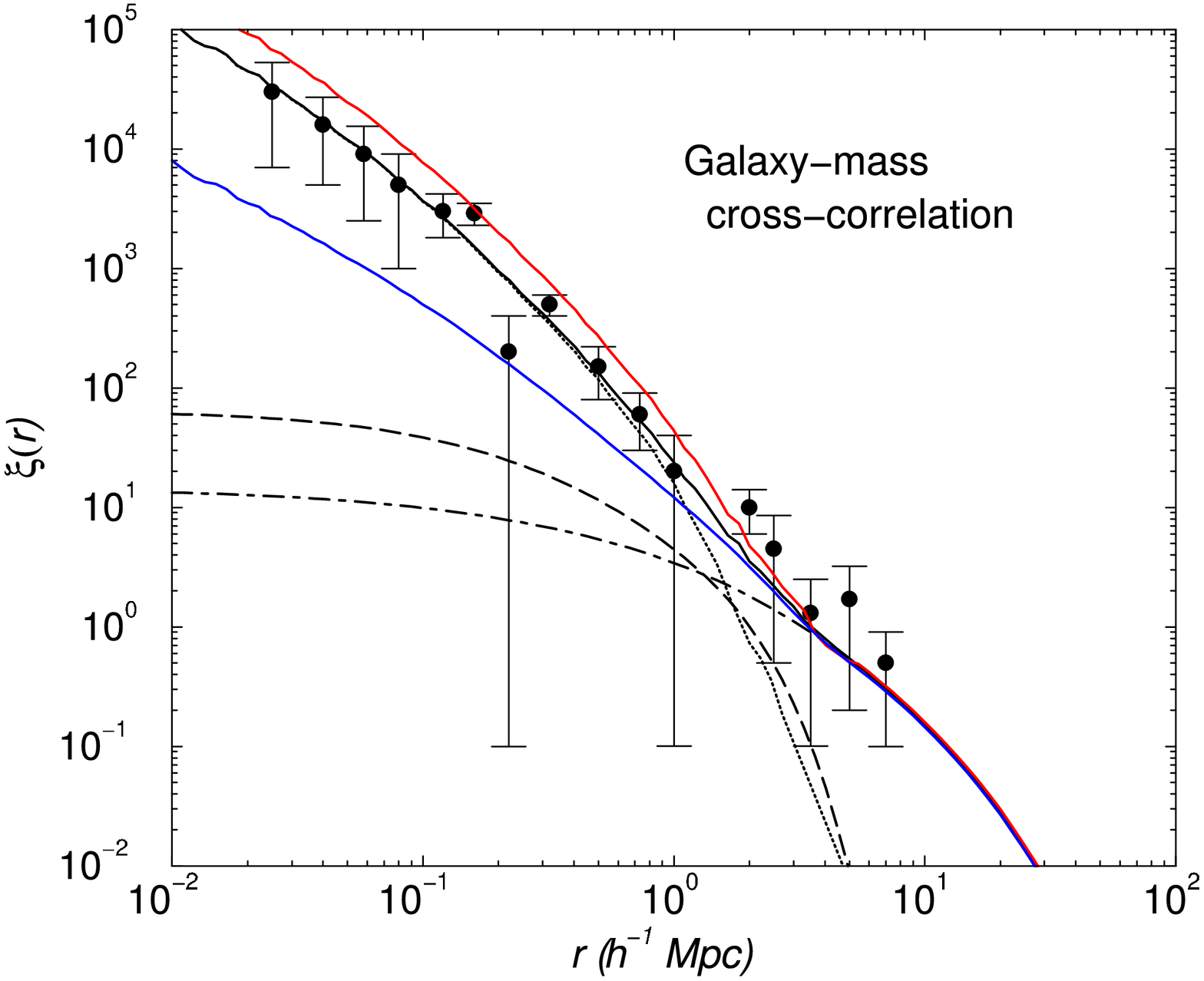,width=\hssize,angle=-90}}
\caption{The three-dimensional cross-correlation function between galaxies  
and dark matter as determined by Sheldon et al. (2004) using SDSS galaxy-galaxy
weak lensing measurements. The galaxy sample associated with this cross-correlation measurement
a volume limited sample in redshifts between 0.1 and 0.174 and in magnitudes between
$-23 < M_r <-21.5$. The volume limited and luminosity-selected sample measurements 
allow an easy prediction based on the same fiducial parameters as those used to
describe projected clustering measurements of galaxies. We also show the
expected cross-correlation between mass and galaxy types. The dotted, dashed and dot-dashed lines show
contribution from the central galaxy 1-halo term, 
the satellite galaxy 1-halo term and from linear theory, scaled
by a bias factor, respectively.
}
\end{figure}

\begin{figure}
\centerline{\psfig{file=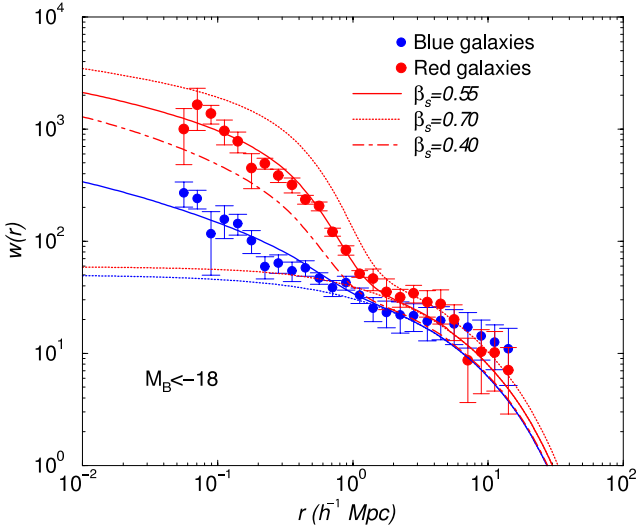,width=\hssize,angle=-90}}
\caption{
Projected correlation function of galaxies at $z \sim 0.6$ as measured by the COMBO-17
survey (Phleps et al. 2005) and divided into clustering of early- and late-type galaxies.
The predictions based on our fiducial model description, with appropriate parameters for
redshift evolution of the $L_{\rm c}(M,z)$ relation, are also shown. In the case of
early-type galaxies, we also show variations in the power-law slope of the total luminosity--halo mass relation. 
While not specified as part of the observations,  we have assumed this sample corresponds to $M_B <-18$ when model
fitting the data.
}
\end{figure}

\begin{figure*}
\centerline{\psfig{file=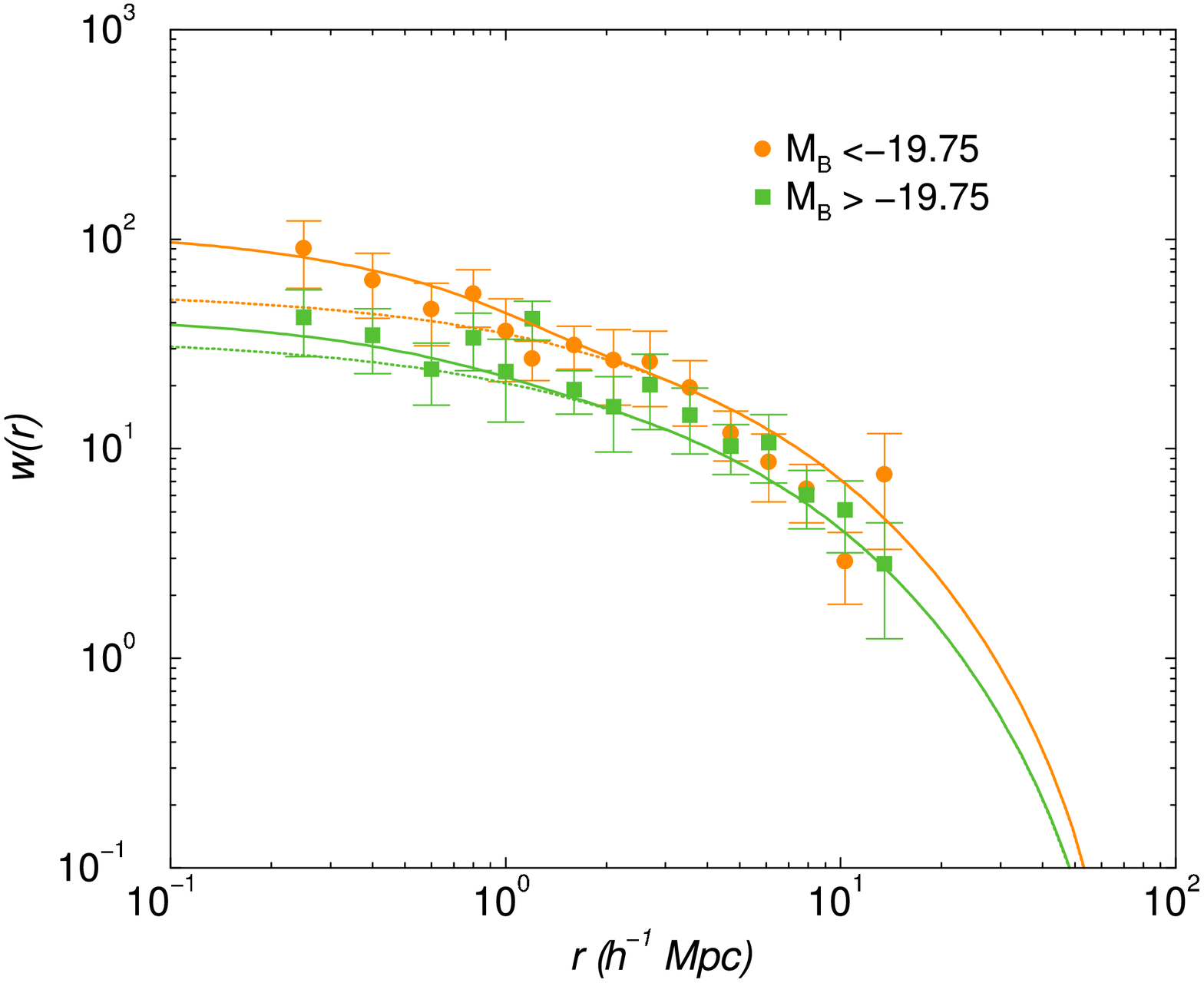,width=\hssize,angle=-90}
\psfig{file=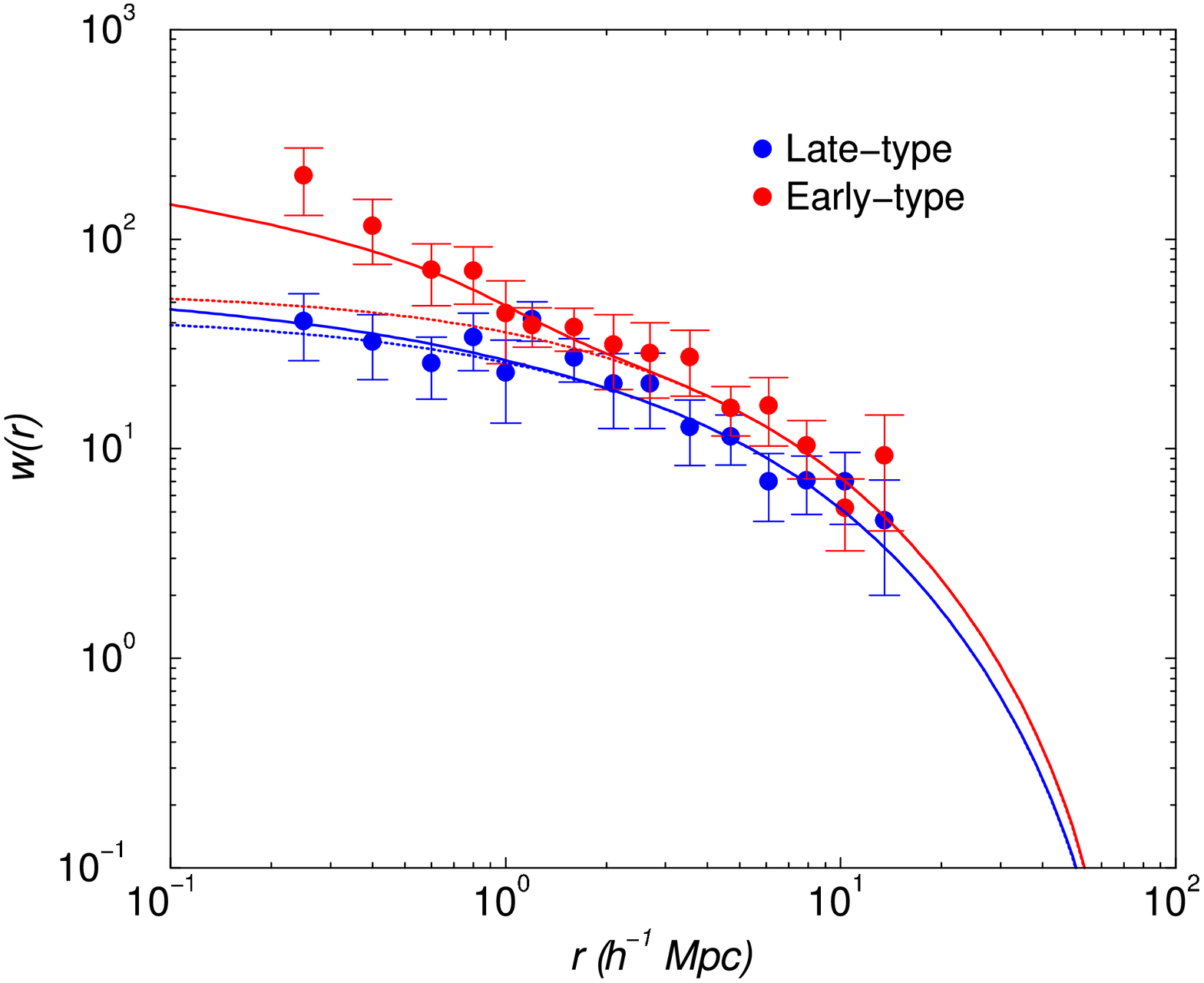,width=\hssize,angle=-90}}
\caption{
Projected correlation function of galaxies at $z \sim 1$ as measured by the DEEP2
survey (Coil et al. 2004). {\it Left panel:}  Clustering of galaxies divided into
two luminosity samples  with $M_B < -19.75$ (circles) and $M_B > -19.75$ (squares).
The predictions based on CLFs are also shown; We assume a low-end magnitude of -18 for the faint sample,
while no such assumption is needed at the bright-end due to the cut-off associated with the LF.
{\it Right panel:} Galaxy clustering in the total sample divided to galaxy types.
In both panels, dotted lines are predictions based on the linear theory power spectrum.
}
\end{figure*}

\begin{figure}
\centerline{\psfig{file=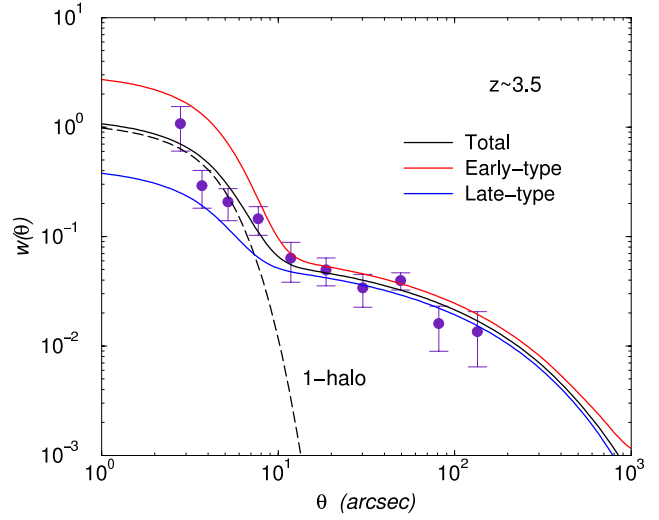,width=\hssize,angle=-90}}
\caption{
Projected angular correlation function of galaxies at $z \sim 3$ 
as measured by the GOODS
survey (Lee et al. 2005). The measurements are for the total sample, but for comparison,
we also show the expected clustering of red- and blue-galaxies if the sample had been divided to galaxy types.
For comparison, we also show the 1-halo contribution.
}
\end{figure}

\begin{figure*}
\centerline{\psfig{file=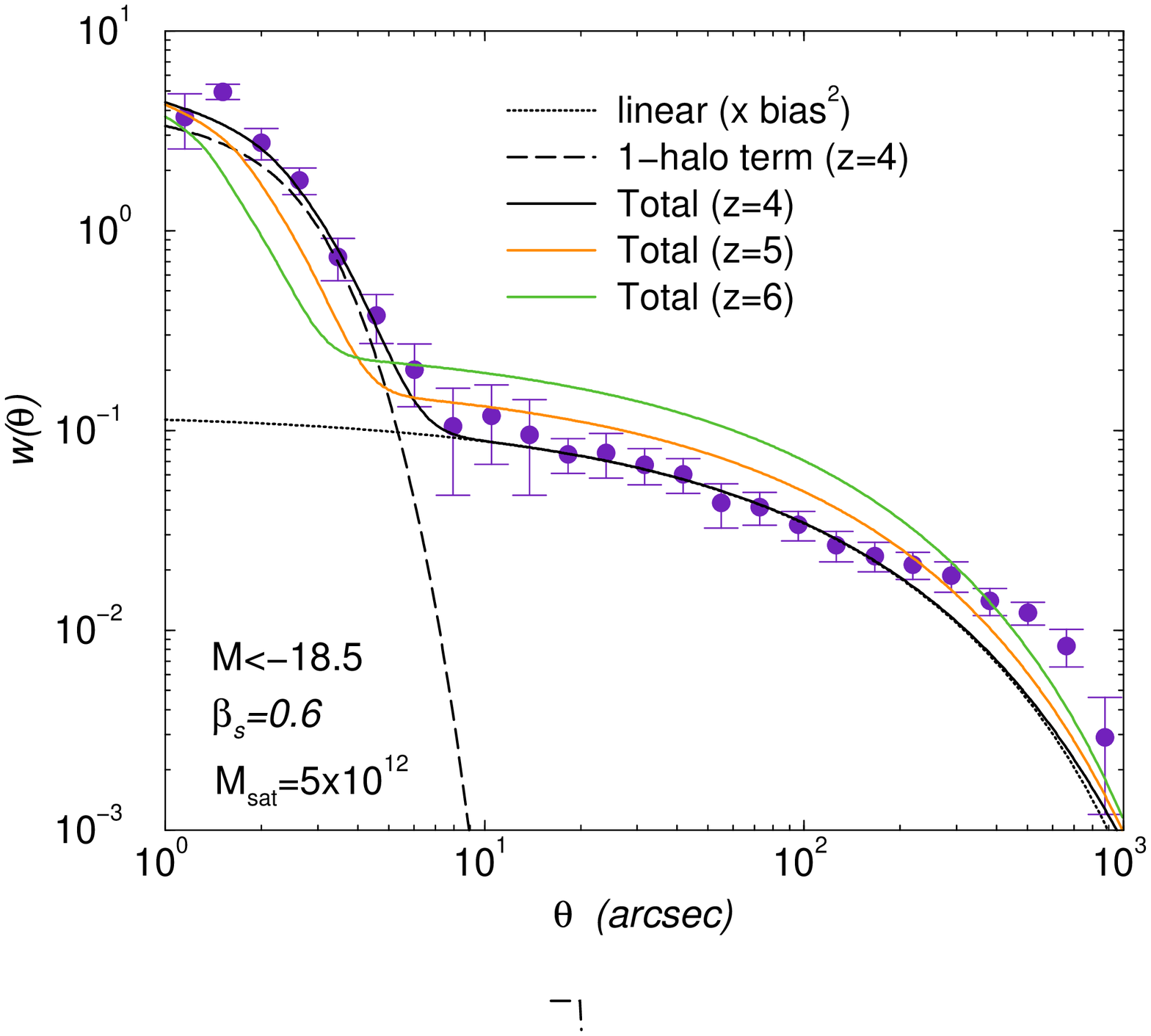,width=\hssize,angle=-90}
\psfig{file=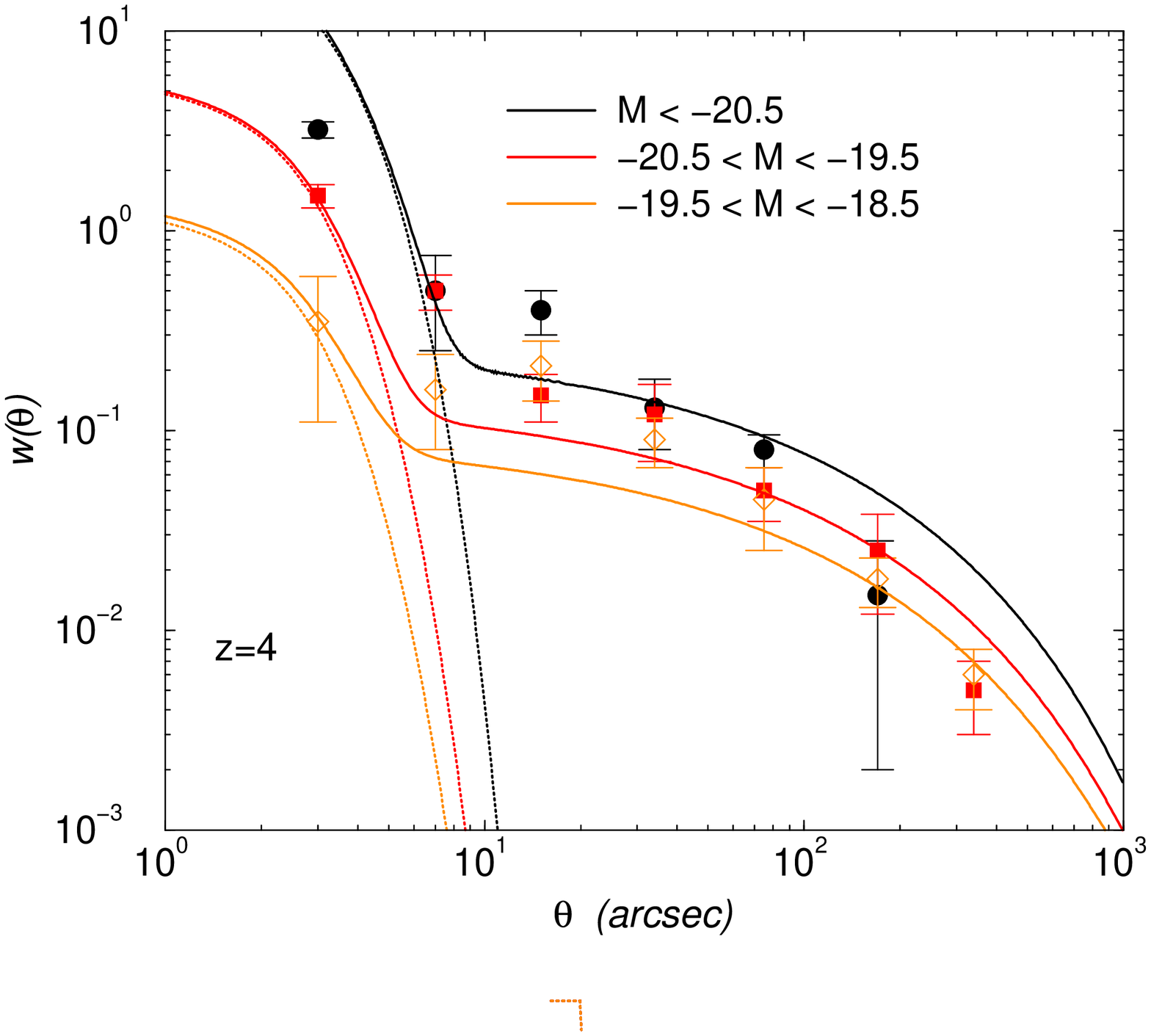,width=\hssize,angle=-90}}
\caption{
Projected angular correlation function of LBGs at $z \sim 4$ as measured in the Subaru/XMM-Newton Deep Field
 (Ouchi et al. 2005). {\it Left panel:}  Clustering of galaxies with $i$-band magnitudes brighter than 27.5,
corresponding to rest-frame $M_B <-18.5$. For comparison, we also show expected clustering with
the same luminosity cut at $z \sim 5$ and 6; High signal-to-noise ratio clustering measurements at such high redshifts are soon
expected from Subaru and other deep drop-out surveys. At high redshifts, large-scale clustering increases
due to the evolution in the halo clustering bias factor, but at the same time, non-linear clustering
decreases as the number of galaxies that appear as satellites at a given luminosity begins to decrease at high redshifts.
The dotted line shows the prediction based on linear theory at $z \sim 4$, scaled by the large-scale
bias factor for galaxies with $M_B <-18.5$ (see, Cooray 2005b).
{\it Right panel:} Galaxy clustering at $z \sim 4$, divided to luminosity bins (measurements from
Kashikawa et al. 2005 using the Subaru Deep Field data) as indicated on the figure.
At each of these luminosity bins, 
we assume same parameters related to satellites as the ones used to describe clustering
for the sample in the left panel. The differences in the one-halo non-linear term prediction and small-scale
observed clustering are due most likely due to variations  in the parameters related to the satellite CLF
as a function of satellite luminosity. For example, with $M_{\rm sat}=5 \times 10^{12}$ M$_{\sun}$ and $\beta_s=0.6$,
we over predict the small-scale clustering in the highest luminosity bin with $M_B < -20.5$, while
small-scale clustering is well reproduced in the low luminosity bins. The excess in the high luminosity bin 
is due to the large number of satellites allowed, while a higher value for $M_{\rm sat}$ can be used to make
the agreement with data better. This suggests that satellites with high luminosities appear in higher mass halos than
compared to halos in which low luminosity galaxies appear; This is clearly consistent with the general expectation. 
A detailed analysis of luminosity dependent clustering at $z \sim  4$, 
using a new set of measurements from the imaging data in Ouchi et al. (2005), will be described
in detail elsewhere (Cooray \& Ouchi, in preparation).}
\end{figure*}

\section{Galaxy Luminosity Function}

Given the CLF, the galaxy LF is obtained through
\begin{equation}
\Phi_i(L,z) = \int dM\, \frac{dn}{dM}(z)\, \Phi_i(L|M,z) \, ,
\end{equation}
where $i$ represents the division to galaxy types. Here $dn/dM(z)$ denotes the mass function of dark matter halos
and we use the formalism of Sheth \& Tormen (1999) in our numerical calculations.
This mass function is in better agreement with numerical simulations (Jenkins et al. 2001)
when compared to the Press-Schechter mass function (Press \& Schechter 1974).
Using our fiducial values for CLF parameters, in Fig.~2, we show the
SDSS galaxy LF (from Blanton et al. 2004) and the separation to early- and late-type
galaxies. We only concentrate on galaxies with $M_r <-17$ since this sample overlaps with galaxies
used by Zehavi et al. (2004) for clustering measurements that are also used in the present analysis. The CLFs related to
this description are shown in Fig.~3. At the faint-end, these CLFs flatten due to our assumption that the power-law
slope of the luminosity distribution within halos is $\gamma=-1$, which is consistent with 
the LF of galaxies in clusters over the magnitude range of interest to this paper.  At fainter magnitudes 
the CLF becomes complicated due to effects associated with the luminosity
distribution of dwarf galaxies (e.g., Cooray \& Cen 2005). In the present analysis,
we do not consider such low-luminosity galaxies with $M_r > -16$ and issues related to the subhalo mass function and associated
substructure can therefore be easily ignored.

In previous discussions of galaxy clustering under the halo model
the occupation number has been widely used as a way to relate statistics of dark matter to galaxies 
(e.g., Kauffmann et al. 1999; Benson et al. 2000; Berlind et al. 2003;
see, review in Cooray \& Sheth 2002). To compare with models of the halo occupation number, CLFs can be easily integrated
such that
\begin{eqnarray}
N_{\rm cen}(M,z) &=& \int dL \, \Phi^{\rm cen}(L|M,z) \nonumber \\
N_{\rm sat}(M,z) &=& \int dL \, \Phi^{\rm sat}(L|M,z) \, .\nonumber \\
\end{eqnarray}
Since the halo occupation number captures no information on the luminosity distribution of galaxies,
models involving the halo occupation number
cannot be used to model the galaxy LF easily.  We show the halo occupation number in Fig.~4 for $-23 <M_r <-18$
galaxies in the SDSS sample (left panel) and redshift dependence of the halo occupation number (right panel).
At high redshifts, the occupation numbers are at the B-band since galaxy samples in COMBO-17, DEEP2, and
GOODS are defined at this wavelength. While Subaru/XMM-Newton Deep Field sample is defined in the observed
$i$-band, we assume rest-frame B-band luminosities when model fitting the data.

In the case of satellites,  since $N_{\rm sat}(M,z) = A(M,z) \int dL\, L^{\gamma(M)} f_{\rm s}(L)$
at the high mass limit of the halo mass with $\gamma(M)$ a constant and when
$L_{\rm tot}(M,z) \gg L_{\rm c}(M,z)$ we expect $N_{\rm sat}(M) \propto L_{\rm tot}(M,z) \sim M^{\beta_s+\alpha}$,
where $\beta_s$ is the slope introduced in equation~4 and $\alpha$ is the slope of the $L_c(M)$ relation at the same
halo masses. With $\alpha_s \sim 0.2$, and $\beta_s =0.55$, the fiducial slope of occupation number
is around 0.75, though this slope is mass dependent given the rapid variation of central galaxy luminosity with halo mass.

In Fig.~5, we present the halo occupation number as a function of luminosity considered
by Zehavi et al. (2004) for clustering measurements. These occupation numbers, based on CLFs,
can be compared with {\it best-fit} halo occupation models suggested in Zehavi et al. (2004).
In our fiducial model,
satellites with $M_r <-17$ only appear in halos with masses greater than $M_{\rm sat}=10^{13}$ M$_{\sun}$.
We see a cut-off in the halo occupation number of central galaxies at masses around $10^{11}$ M$_{\sun}$.
At $M_r <-20$, this cut-off is around $\sim 7 \times 10^{11}$ M$_{\sun}$. This value can
be compared to the suggested minimum value  of $\sim 10^{12}$ M$_{\sun}$ for the halo occupation number down to
the same magnitude in Zehavi et al. (2004).
The difference can be understood based on the fact that Zehavi et al. (2004)
description of the satellite halo occupation number is $(M/M_1)^\alpha$ with a no cut-off
at a lower mass, while the minimum halo mass cut-off only applies to the central galaxy occupation number.
It could be that the degeneracy between the central and satellite galaxy occupation numbers leads to an overestimate in
the minimum mass for central galaxies to appear, while that overestimate is 
partly accounted with an increase in the slope of  the halo occupation number for satellite galaxies. 

As stated in Zehavi et al. (2004), the halo occupation model parameters suggested there are not unique.
The mass cut-off detected based on the halo occupation number model fits to galaxy statistics
should not be treated as a general lower limit on halo mass to host galaxies.
The cut-off usually one detects with occupation numbers is the minimum halo mass to host a galaxy given the
minimum luminosity of galaxies in the sample under consideration (for example, the minimum mass of
the central galaxy halo occupation number as a function of luminosity in Fig.~5).
It could be that halos with a mass lower than the cut-off continue to host galaxies, but with a lower
luminosity, and due to sample selection criteria such halos would not be included in the sample used for
clustering studies.  We will return to this below in the context of model
fits to clustering data where we find no conclusive evidence for a general minimum halo mass
to host galaxies, for galaxies with $M_r <-17$.

In addition to the low mass cut-off  of central galaxies, we also have the freedom to
select a low mass for the appearance of satellites with the parameter $M_{\rm sat}$.
In Figures~4 and 5, we have set $M_{\rm sat}=10^{13}$ M$_{\sun}$, though best-fit
halo occupation numbers from Zehavi et al. (2004) suggest the presence of satellites,
as appropriate for the same sample of galaxies, in halos with a lower mass than this.
While this could be due to differences between the model, as stated before,
the occupation models as well as our CLFs may not be unique. Later, we will use
data to constrain parameters such as $M_{\rm sat}$ and find large degeneracies between
$\beta_s$, the power-law slope, and $M_{\rm sat}$ such that as $M_{\rm sat}$ is lowered,
$\beta_s$ is increased. The same degeneracy should also appear in model fits based on the
halo occupation number. For example, with a larger value for M$_{\rm min}$,  the minimum mass for
the central galaxy halo occupation number as in model descriptions of Zehavi et al. (2004), one should find
a larger slope $\alpha$ for the satellite halo occupation number such that the total number of satellite galaxies remains the same;
This behavior could partly explain the unusually large values for the slope suggested in
Zehavi et al. (2004). The degeneracy between $M_{\rm sat}$  and $\beta_s$ suggests that a single parameter involving
the combination of these two parameters can be best determined with the data.
As we find later, this parameter is the total luminosity of satellite galaxies
averaged over the halo mass distribution that hosts galaxies between $-17 > M_r >-23$
in the SDSS sample.

\begin{figure*}
\centerline{\psfig{file=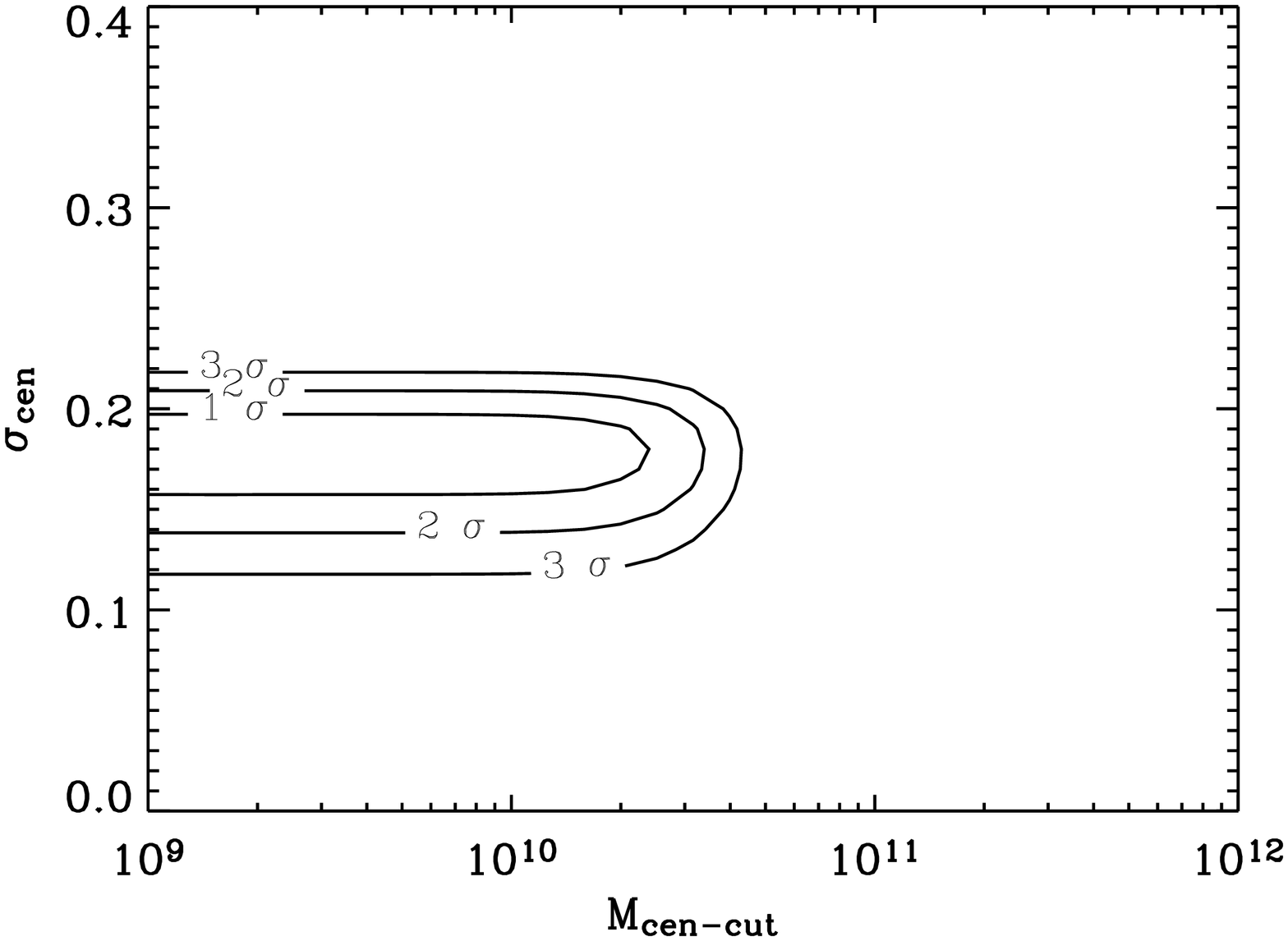,width=\hssize,angle=90}
\psfig{file=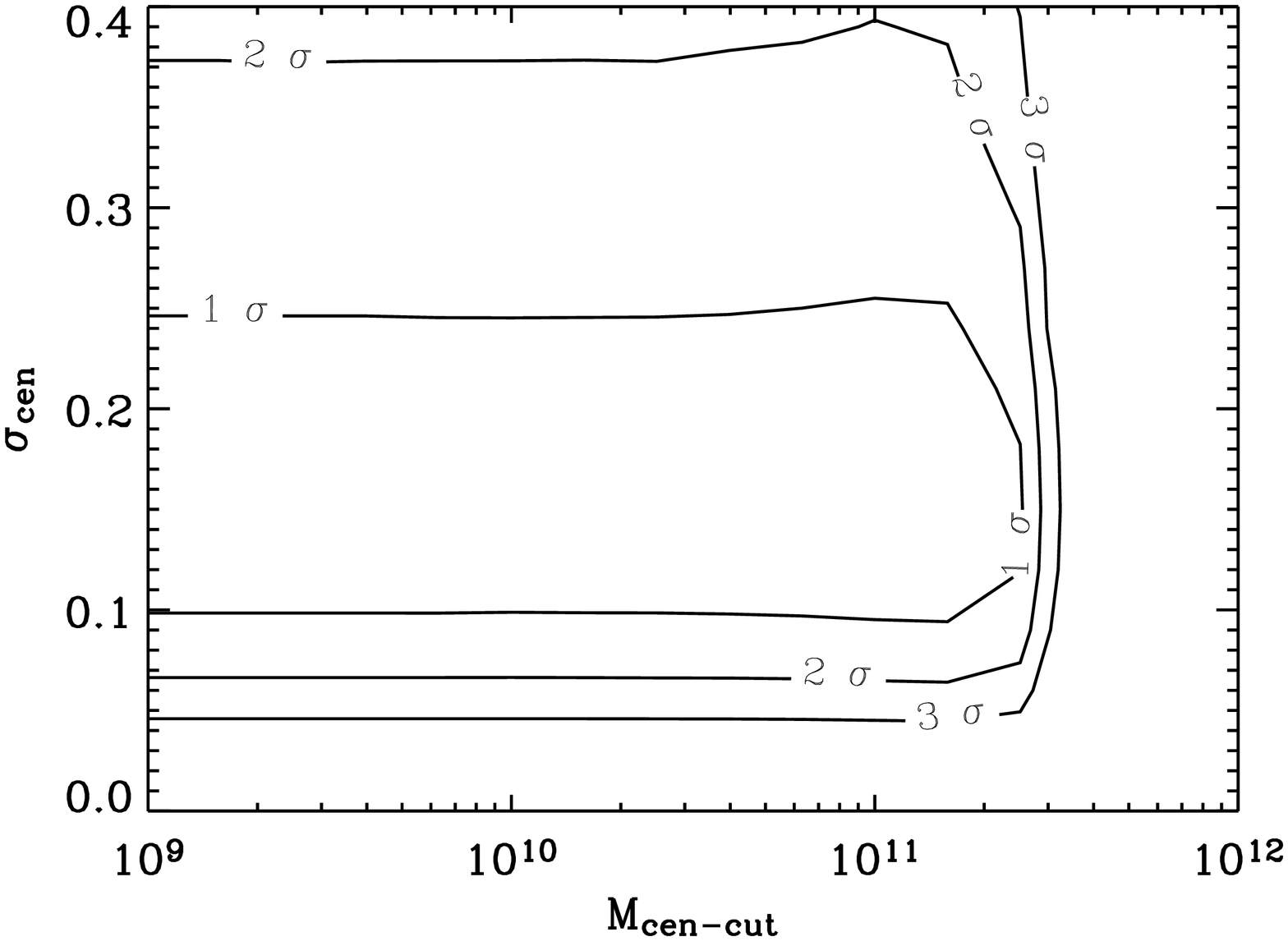,width=\hssize,angle=90}}
\caption{Constraints on parameters $\sigma_{\rm cen}$, the log-normal dispersion of
the central galaxy---halo mass relation, and $M-{\rm cen-cut}$, the lower
halo mass to host a central galaxy, independent of luminosity, related to the
central galaxy CLF description.
The left panel shows the constraint based on the SDSS LF (from Blanton et al. 2004), down to
$M_r$ of -17, and the right panel shows the constraints from SDSS galaxy clustering measurements
(from Zehavi et al. 2004). While the LF strongly constrains these parameters, clustering
measurements do not. The difference comes from the fact that clustering measurements are more sensitive to
satellite galaxies while, as shown in Figure~2, LF measurements are sensitive to statistics of central galaxies.}
\end{figure*} 

\begin{figure*}
\centerline{\psfig{file=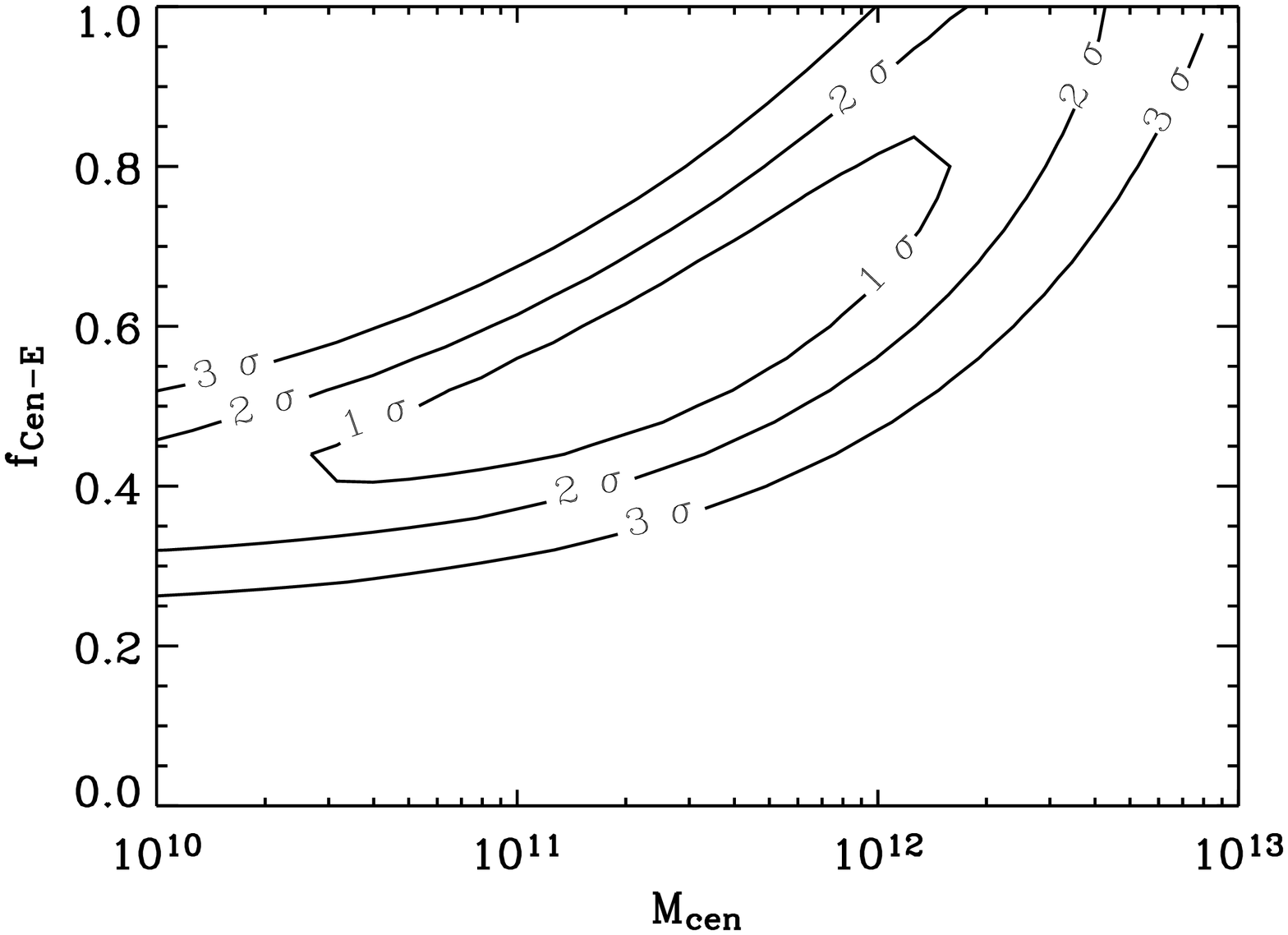,width=9cm,angle=90}
\psfig{file=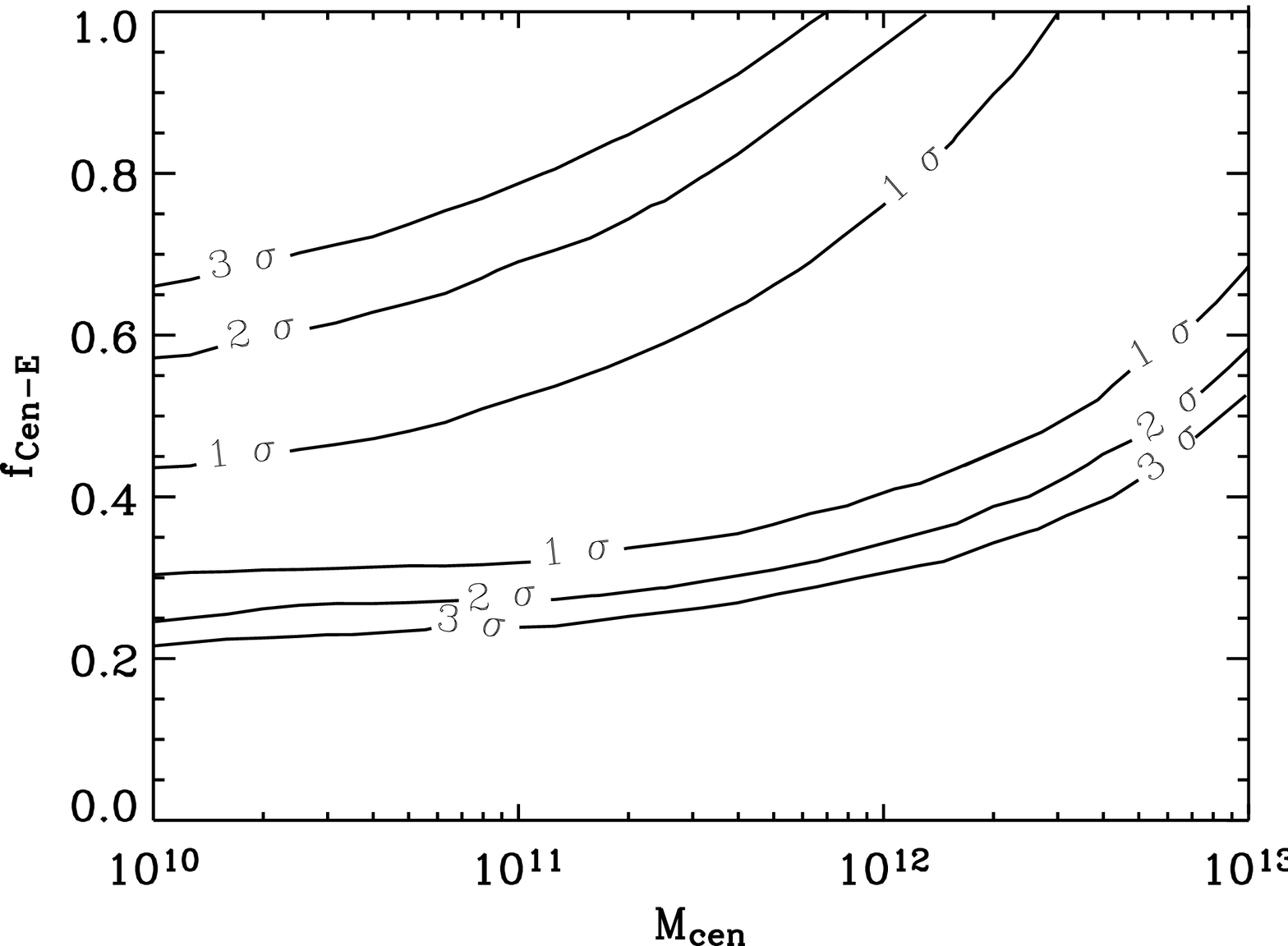,width=\hssize,angle=0}}
\vspace{0.3in}
\caption{
Constraints on parameters $f_{\rm cen-E}$, the fraction of early-type central galaxies 
at the high mass-end and $M_{\rm cen}$ related to the analytical description of the early-type
galaxy fraction of central galaxies.
The left panel shows the constraint based on SDSS galaxy type LFs (from Blanton et al. 2004), down to
$M_r$ of -17, and the right panel shows the constraints from SDSS galaxy clustering measurements
divided to galaxy types between magnitudes bins from -18 to -21 (from Zehavi et al. 2004). As above, while LF
 strongly constrain parameters related to central galaxies clustering
measurements do not. 
}
\end{figure*}

\begin{figure*}
\centerline{\psfig{file=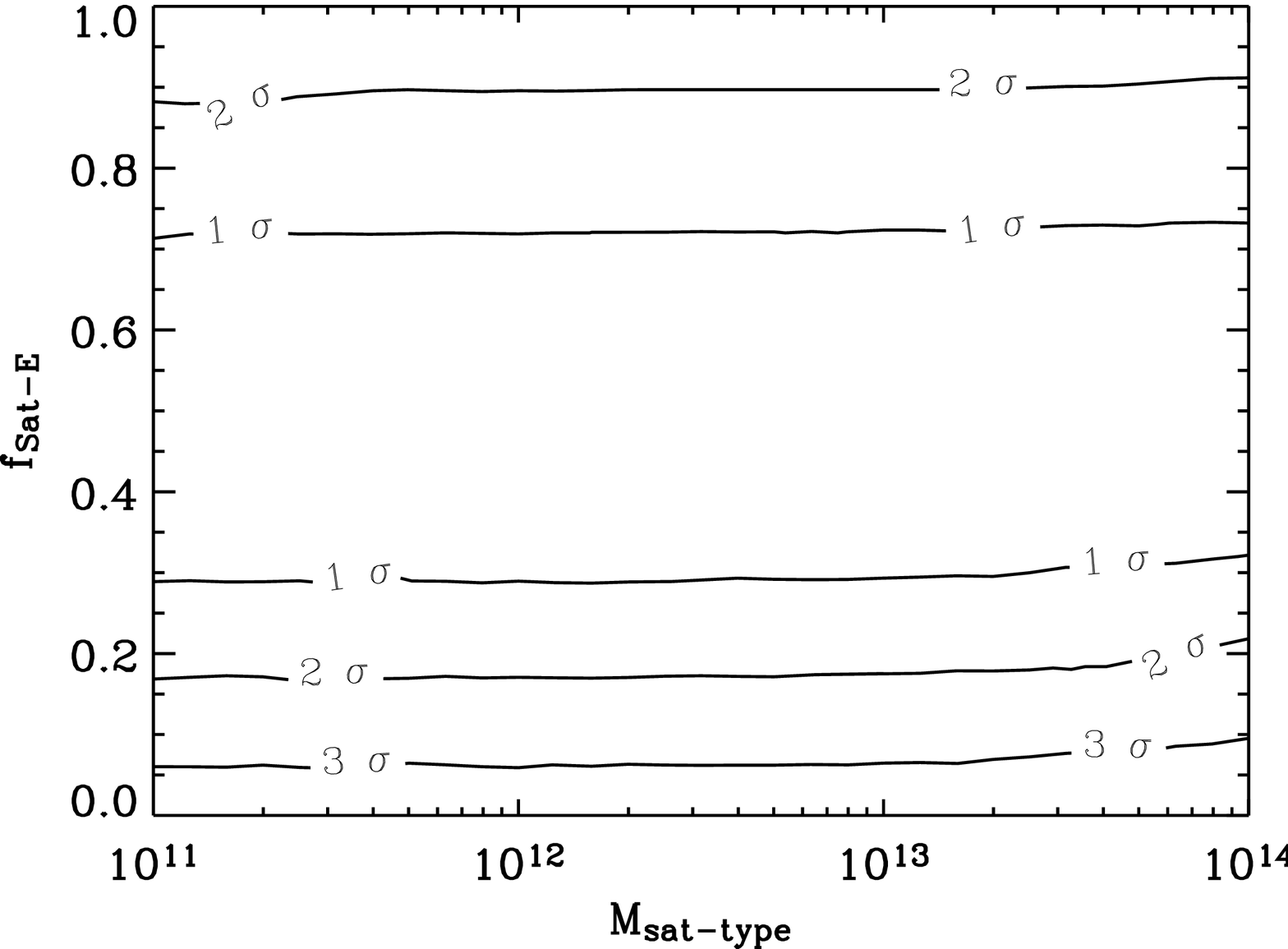,width=7cm,angle=-0}
\hspace{0.7in}
\psfig{file=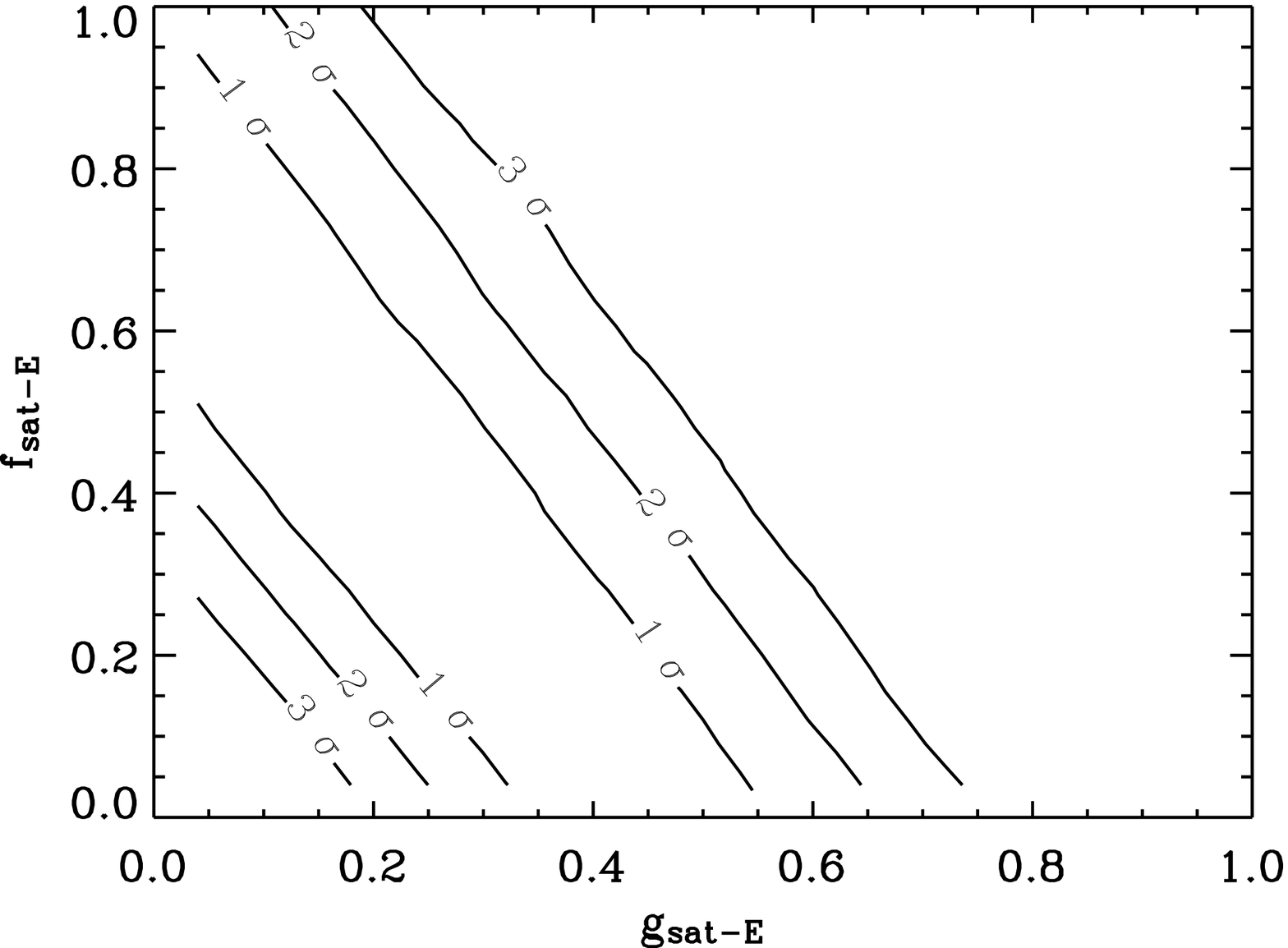,width=7cm,angle=0}}
\vspace{0.5in}
\caption{
Constraints on parameters that describe galaxy types related to the satellite CLF (see, equation~\ref{eqn:ltot})
based on galaxy clustering data divided to galaxy types between $-18 > M_r >-22$ in SDSS. 
While M$_{\rm sat}$ is not strongly constrained with clustering data, $f_{\rm sat-E}$ describing the fraction of
early type galaxies as satellites in low mass halos is constrained to be between 0.5 $\pm 0.15$ at the 68\% confidence level.
The right-panel shows constraints on parameters $f_{\rm sat-E}$ and $g_{\rm sat-E}$.}
\end{figure*} 

\begin{figure*}
\centerline{\psfig{file=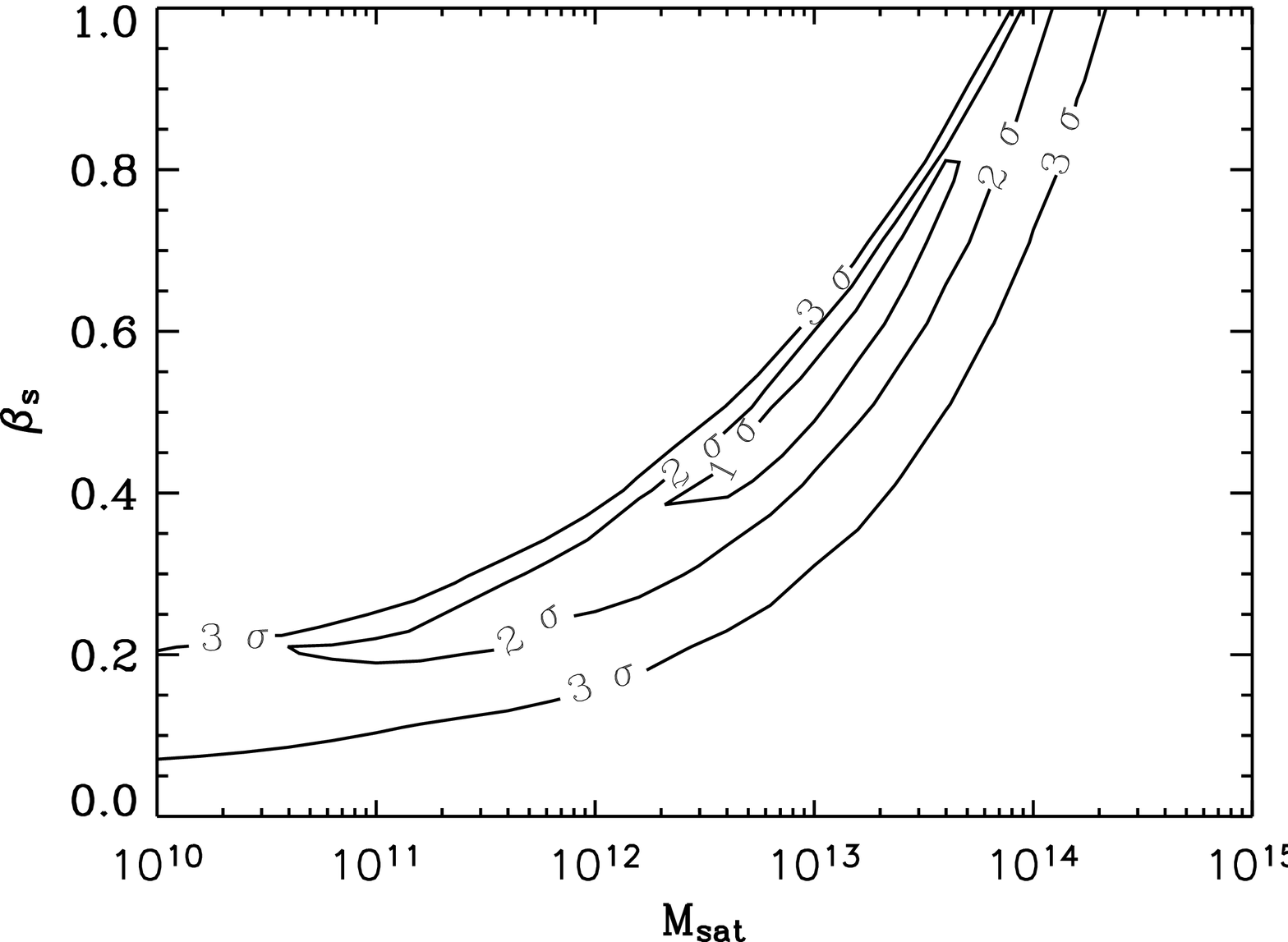,width=\hssize,angle=0}
\hspace{0.4in}
\psfig{file=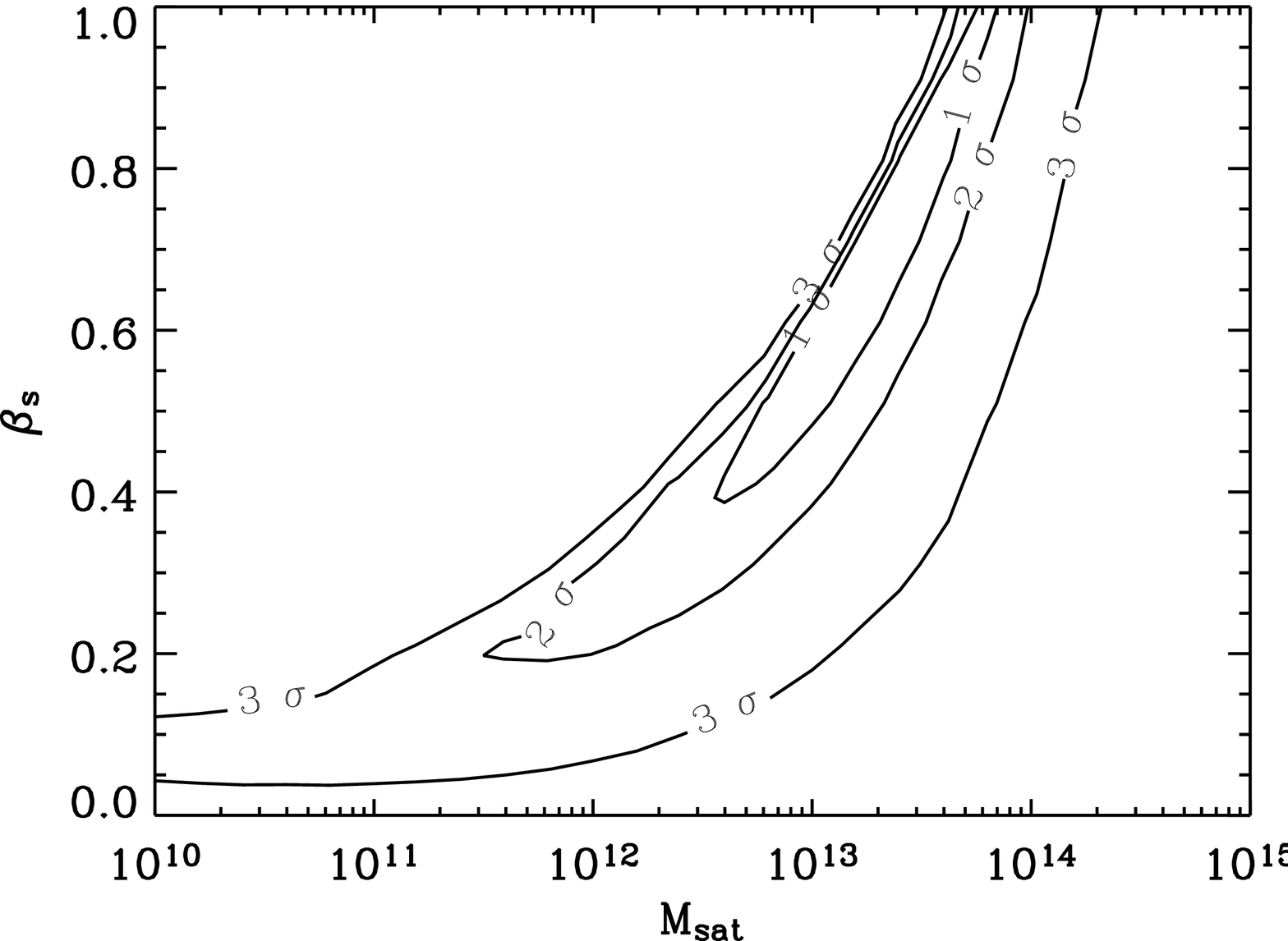,width=\hssize,angle=0}}
\vspace{0.5in}
\centerline{\psfig{file=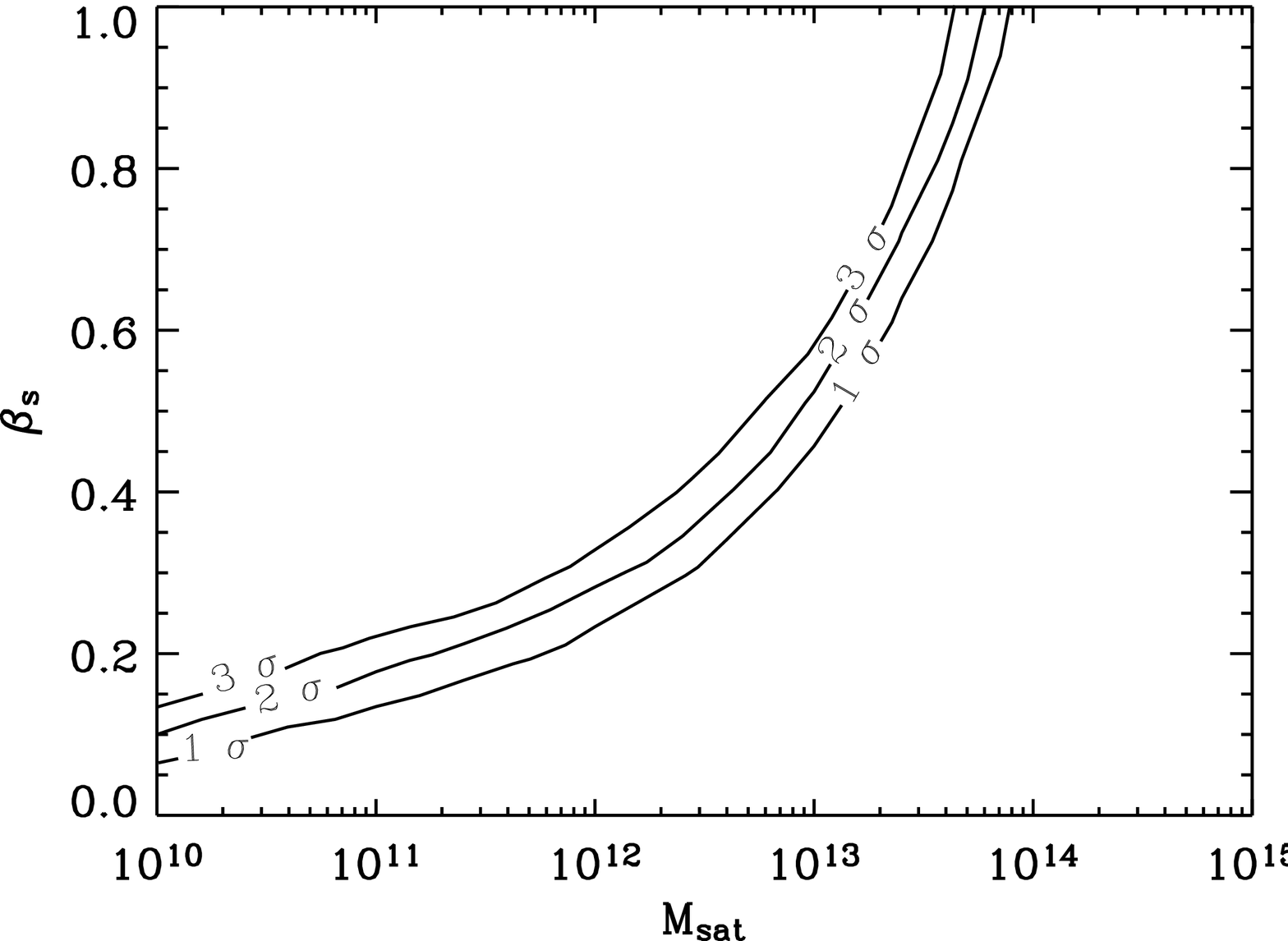,width=\hssize,angle=0}
\hspace{0.4in}
\psfig{file=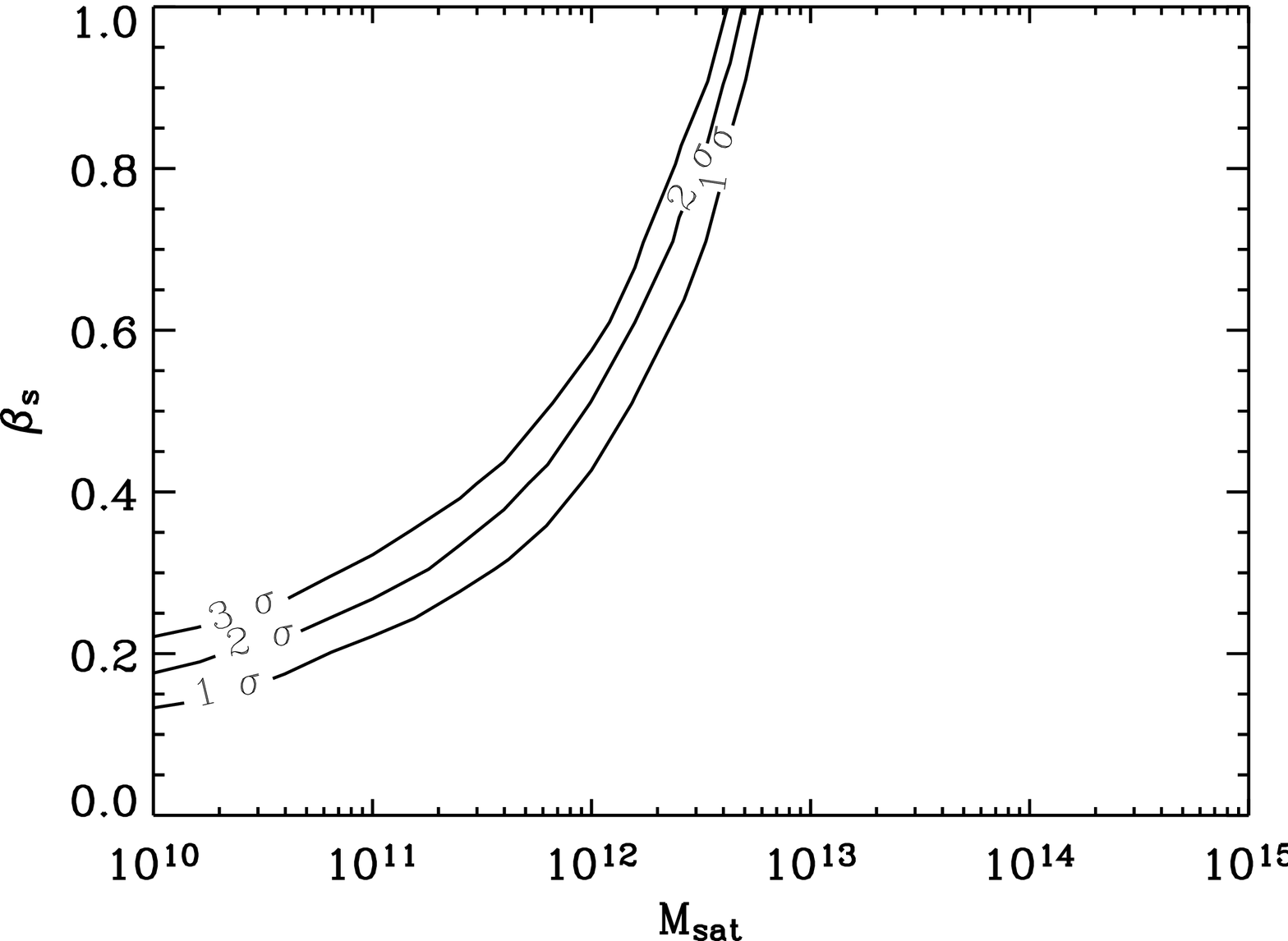,width=\hssize,angle=0}}
\vspace{0.5in}
\centerline{\psfig{file=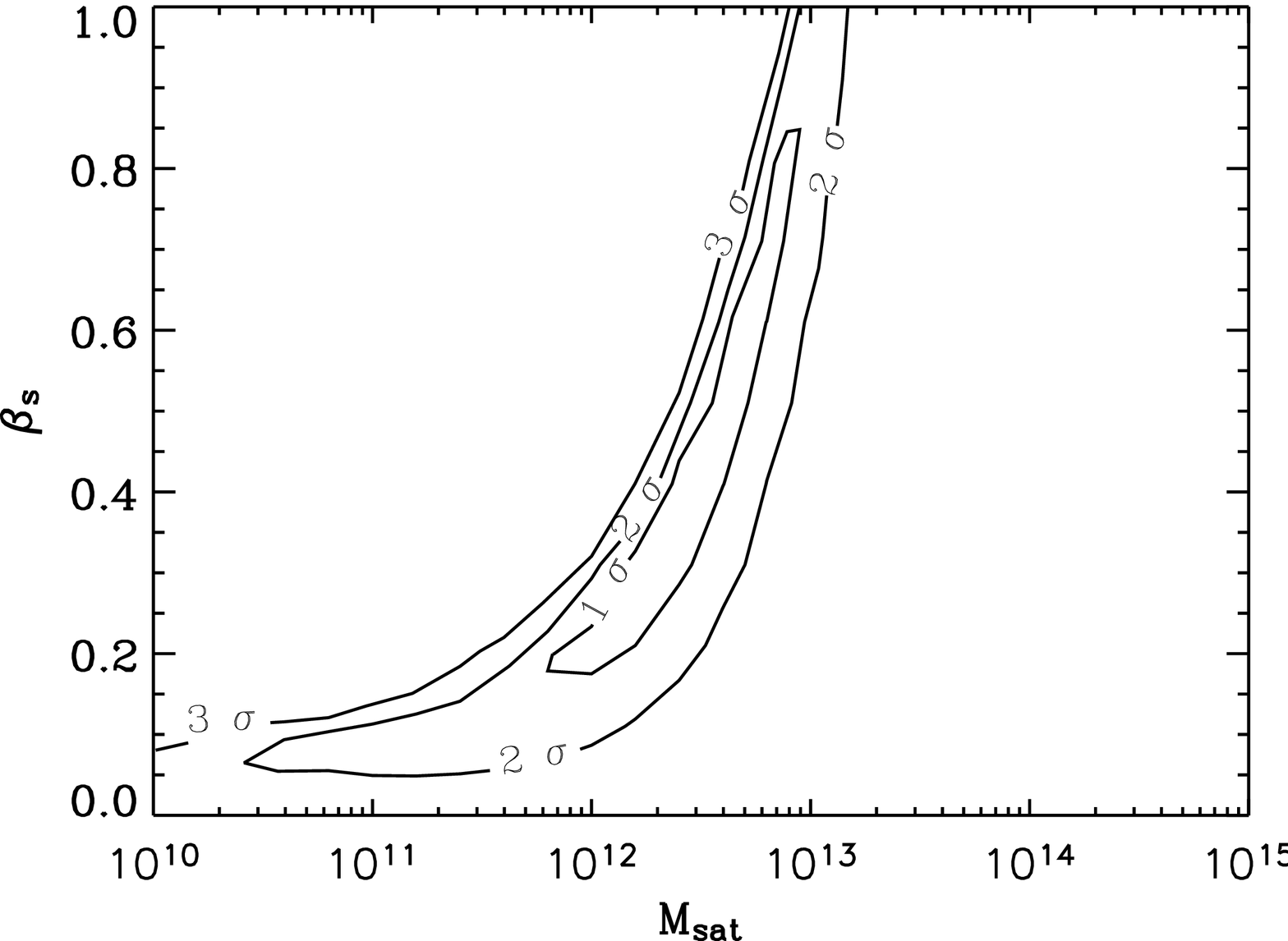,width=\hssize,angle=0}}
\vspace{0.5in}
\caption{
Constraints on parameters $\beta_{\rm s}$, the additional power-law slope of total luminosity--halo mass relation (in addition to
the slope of central galaxy--halo mass relation), and $M_{\rm sat}$, the halo mass scale at which satellites begin to appear, 
related to the satellite CLF. These constraints come from clustering measurements
from SDSS (top left), COMBO-17 (top right), DEEP2 (middle left), GOODS (middle right), and
Subaru/XMM-Newton Deep Field (bottom panel)
at redshifts less than 0.1, around 0.6, around unity, between 2.5 and 3.5, and at 4, respectively.
We only make use of total clustering data divided to galaxy luminosity bins here, but the constraints
shown above are for the combine data set at each of the redshifts. In later figures, we will
highlight  differences between certain luminosity bins instead of the overall constraint shown here.
}
\end{figure*}

\begin{figure*}
\centerline{\psfig{file=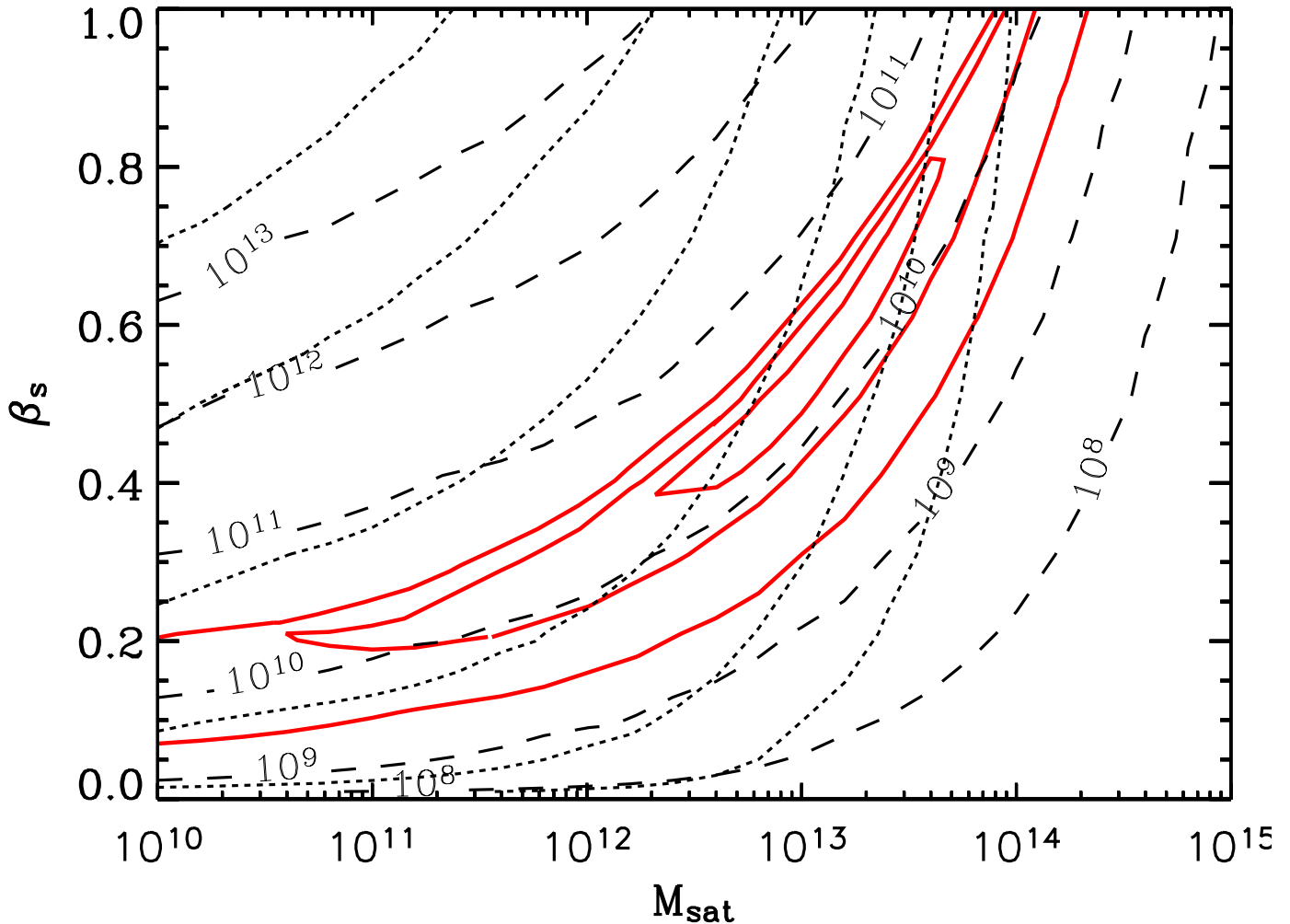,width=\hdsize,angle=0}}
\vspace{1in}
\caption{
$\langle L_{\rm sat} \rangle$, the sample-averaged 
luminosity of satellites for the given sample of galaxies (equation~\ref{eqn:lsatave}), 
as a function of $\beta_{\rm s}$, the power-law slope of total luminosity--halo mass relation,
and $M_{\rm sat}$, the halo mass scale at which satellites appears.
The dashed lines show the average satellite luminosity for SDSS sample while dotted lines show
the same at $z \sim 3$ as appropriate for the GOODS survey. For reference, we overlap constraints on
this parameter space from SDSS (same as Figure~17, top left panel). The degeneracy in
$\beta_s$--$M_{\rm sat}$ plane traces contours of equal average satellite luminosity and
this single parameter is best constrained by current galaxy clustering measurements.
}
\end{figure*}

\begin{figure*}
\centerline{
\psfig{file=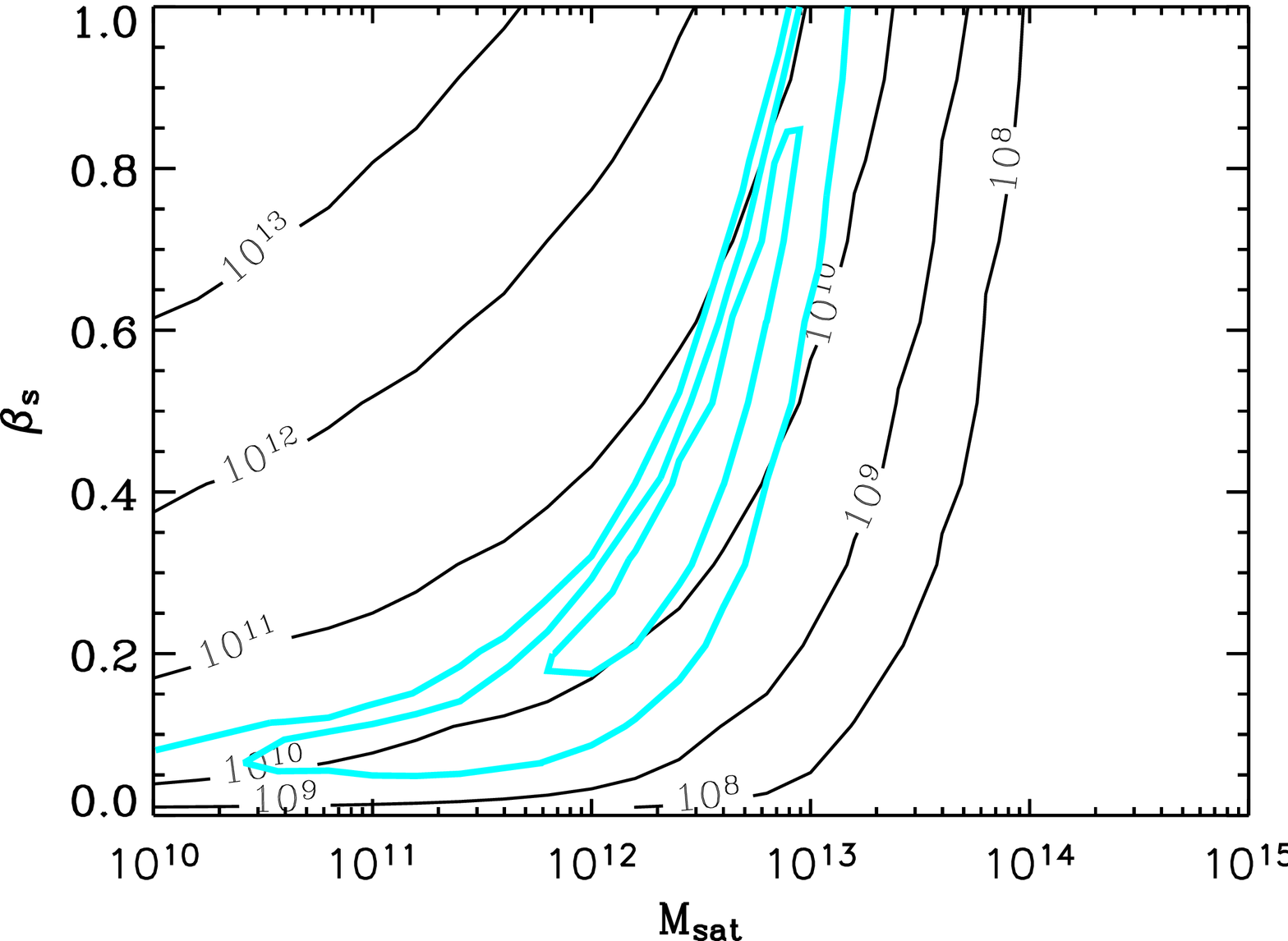,width=\hssize,angle=0}
\hspace{0.4in}
\psfig{file=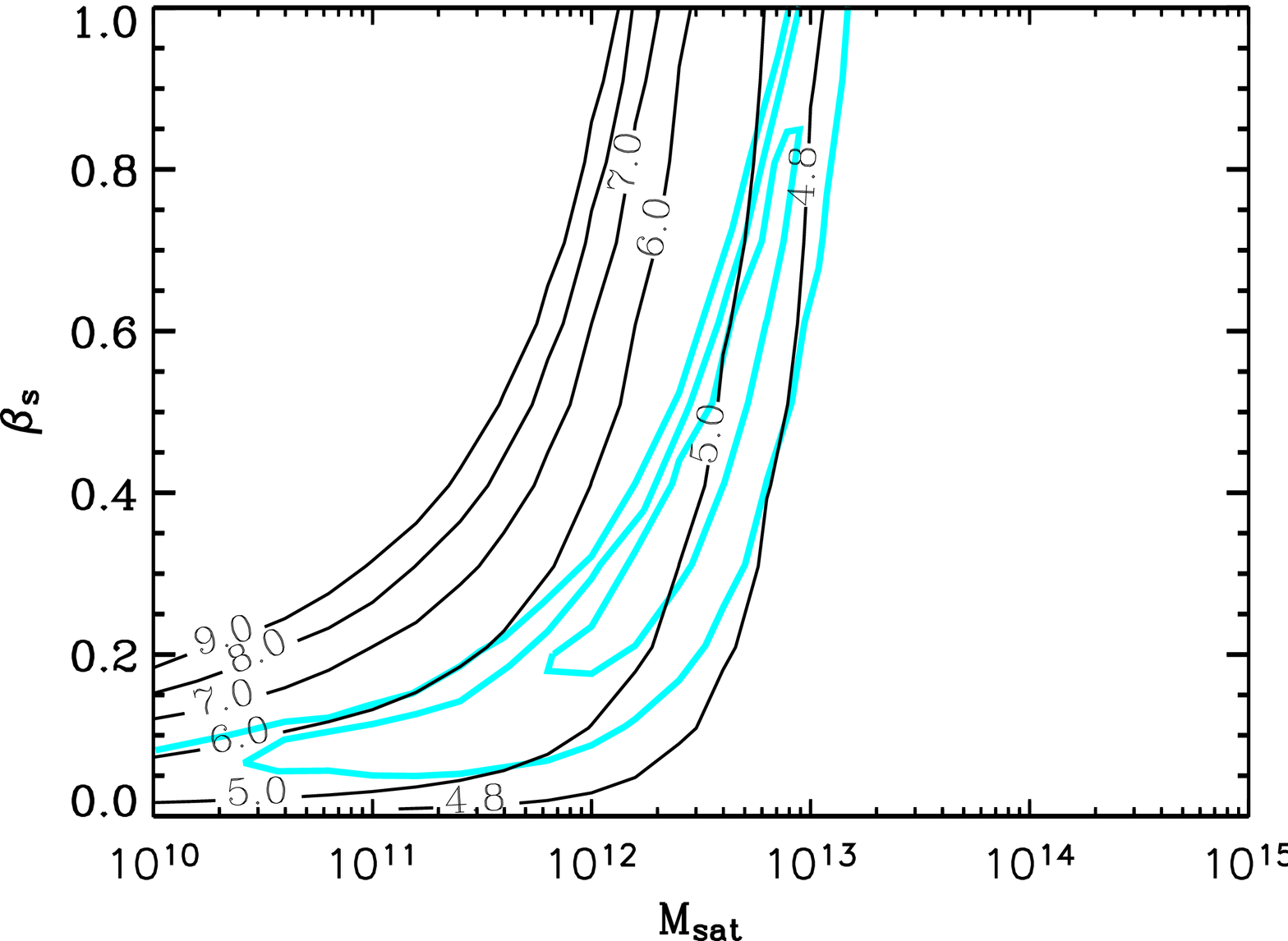,width=\hssize,angle=0}}
\vspace{0.5in}
\caption{ {\it Left panel:}
$\langle L_{\rm sat} \rangle$, the sample-averaged 
luminosity of satellites 
as a function of $\beta_{\rm s}$
and $M_{\rm sat}$ at $z=4$ for galaxies with $M_B <-18.5$.
For reference, we overlap constraints on
this parameter space from Subaru (same as Figure~19, bottom panel). 
{\it Right panel:} $\bar{n}(z)$, the number density of galaxies at $z \sim 4$ with $M_B <-18.5$, as a function of 
as a function of $\beta_{\rm s}$, the power-law slope of total luminosity--halo mass relation,
and $M_{\rm sat}$, the halo mass scale at which satellites appears (in units of $10^{-3}$ h$_{70}^{3}$ Mpc$^{-3}$).
For reference, we overlap constraints on
this parameter space from Subaru/XMM-Newton Deep Field (same as Figure~17, bottom panel). The degeneracy in
$\beta_s$--$M_{\rm sat}$ plane also traces essentially contours of equal number density of galaxies
as well as equal values for $\langle L_{\rm sat} \rangle$.}
\end{figure*}

\begin{figure*}
\centerline{
\psfig{file=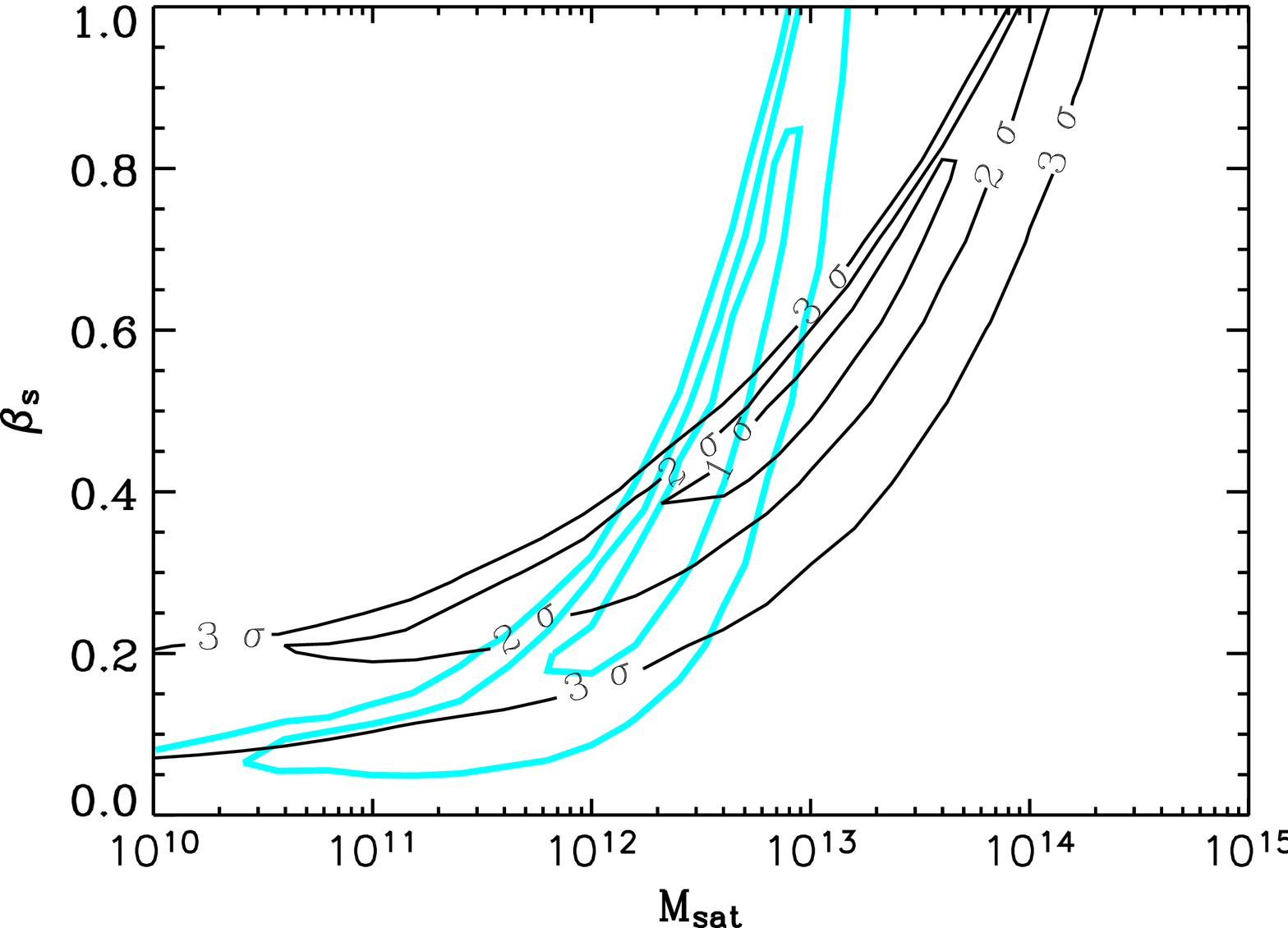,width=\hssize,angle=0}
\hspace{0.4in}
\psfig{file=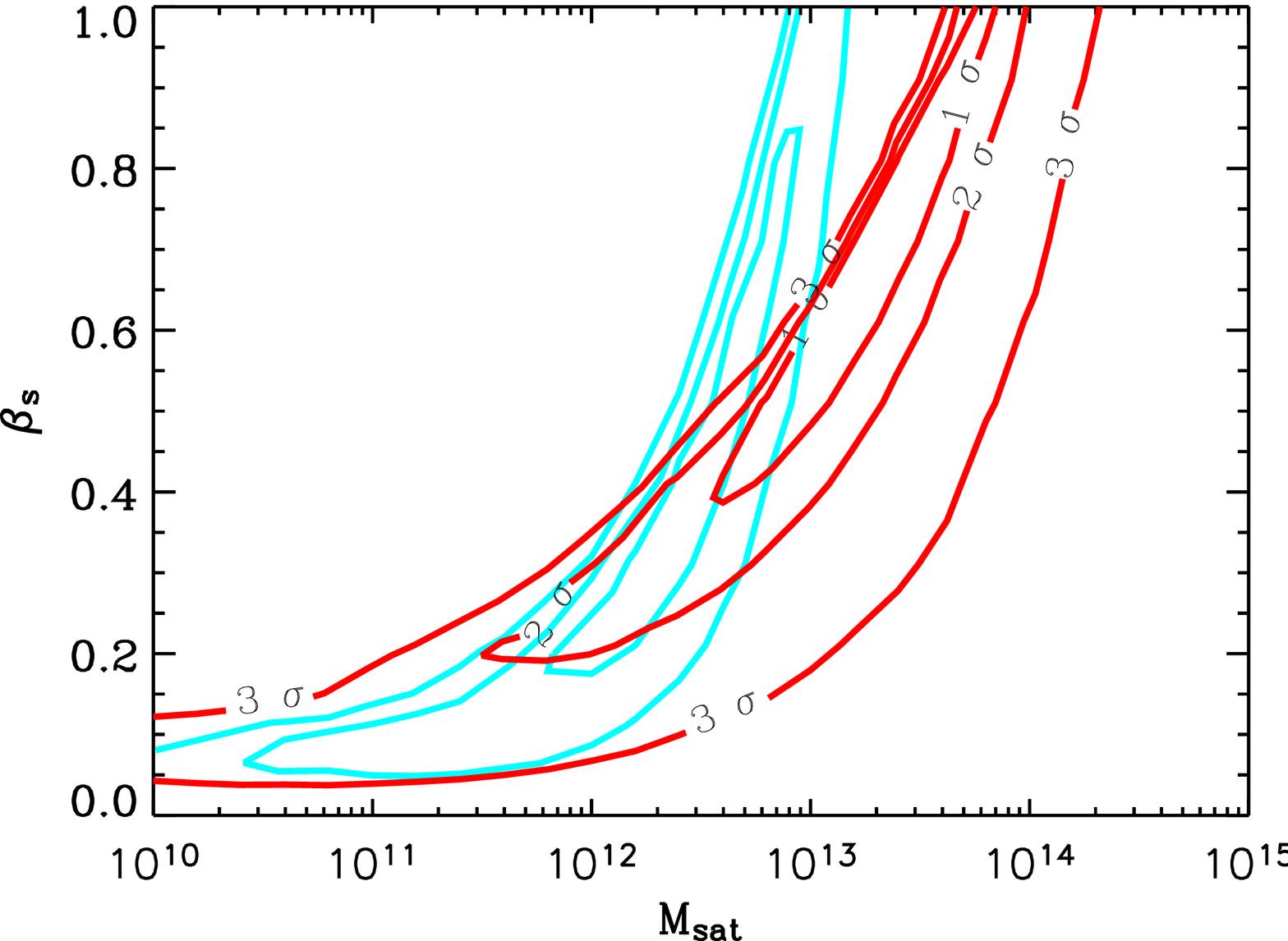,width=\hssize,angle=0}}
\vspace{0.4in}
\caption{
Constraints on parameters $\beta_{\rm s}$, the additional power-law slope of total luminosity--halo mass relation\
 (in addition to
the slope of central galaxy--halo mass relation), and $M_{\rm sat}$, the halo mass scale at which satellites begi\
n to appear,
related to the satellite CLF from Subaru/XMM-Newton Deep Field at $z=4$ compared with constraints from
SDSS (left panel) and COMBO-17 (right panel). At the 1$\sigma$ confidence, a clear difference between
 constraints at $z\sim4$ and at $z \sim 0.6$ is clear. In the text, we discuss this difference in the context tha\
t
we find no difference in the redshift dependence of  $\langle L_{\rm sat} \rangle$.
}
\end{figure*}

\begin{figure*}
\centerline{\psfig{file=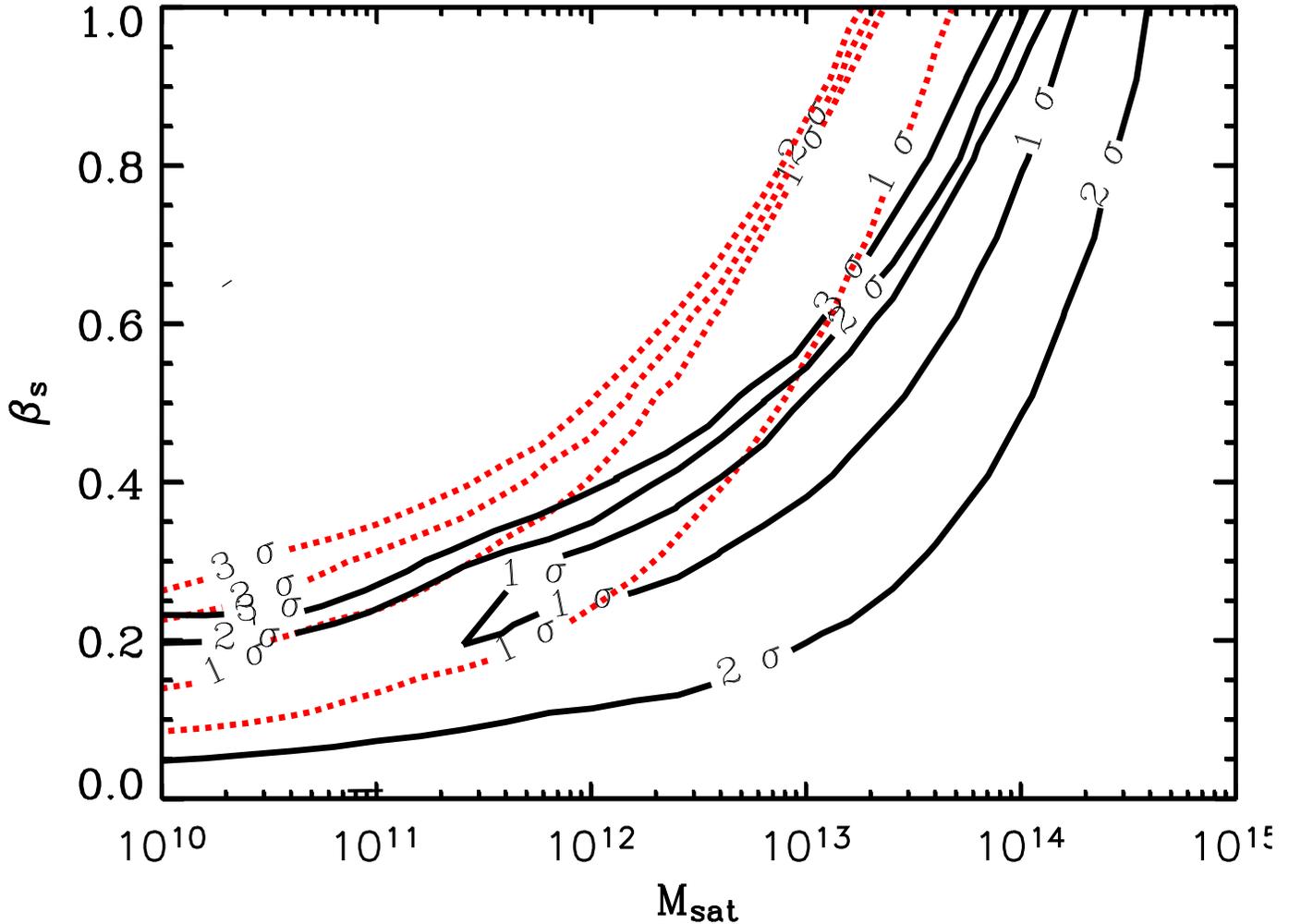,width=\hdsize,angle=0}}
\vspace{1in}
\caption{The constraints on $M_{\rm sat}$--$\beta_s$ plane at two different
luminosity bins: dotted lines show the constraint when $-18 > M_r > -19$ and
solid lines show the case for $-21 > M_r >-22$. We find some evidence
for an increase in $M_{\rm sat}$ as the galaxy luminosity is increased, but the
exact dependence between $M_{\rm sat}$ and galaxy luminosity is not well
established with current data.}
\end{figure*}

\begin{figure*}
\centerline{\psfig{file=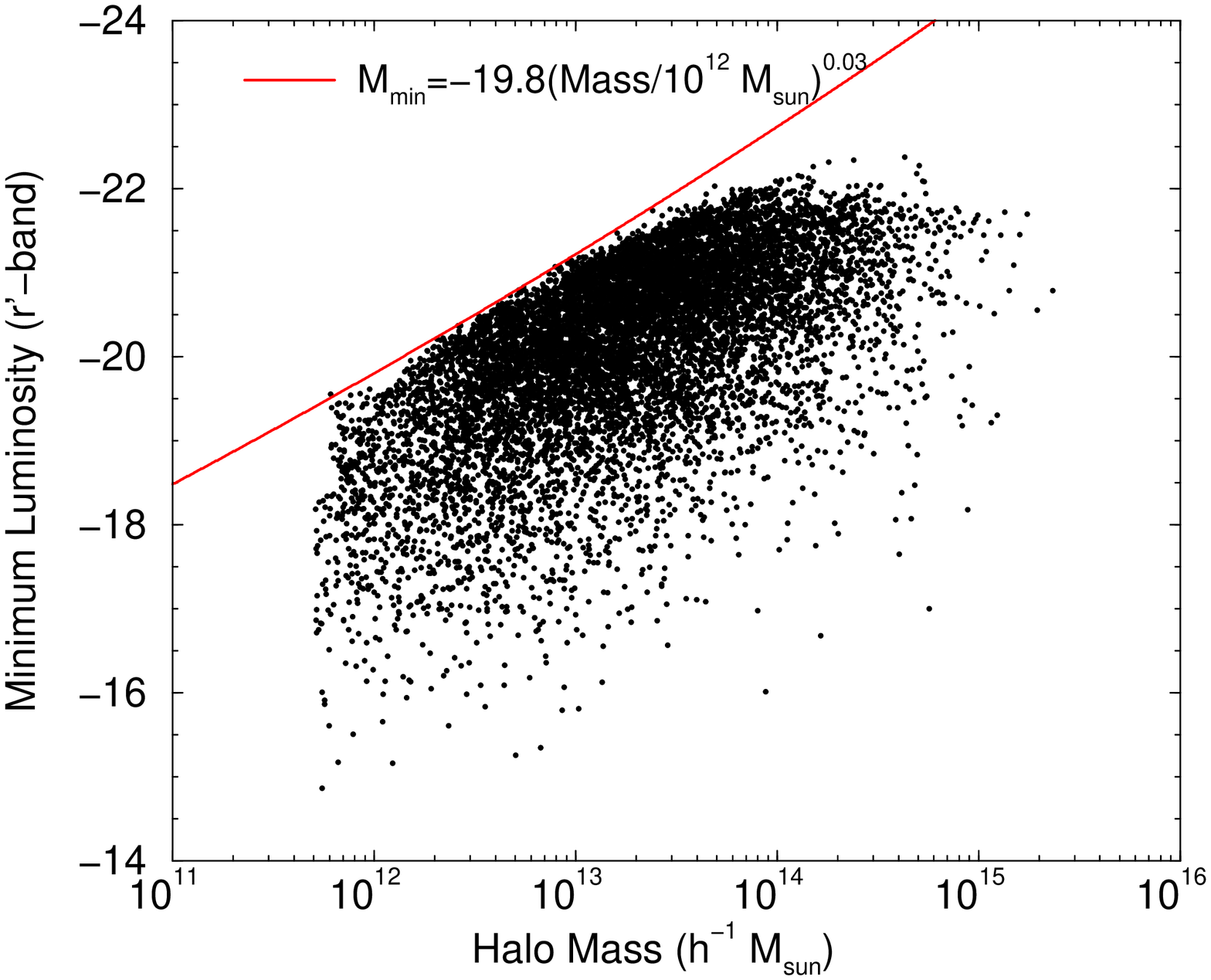,width=\hssize,angle=-90}
\psfig{file=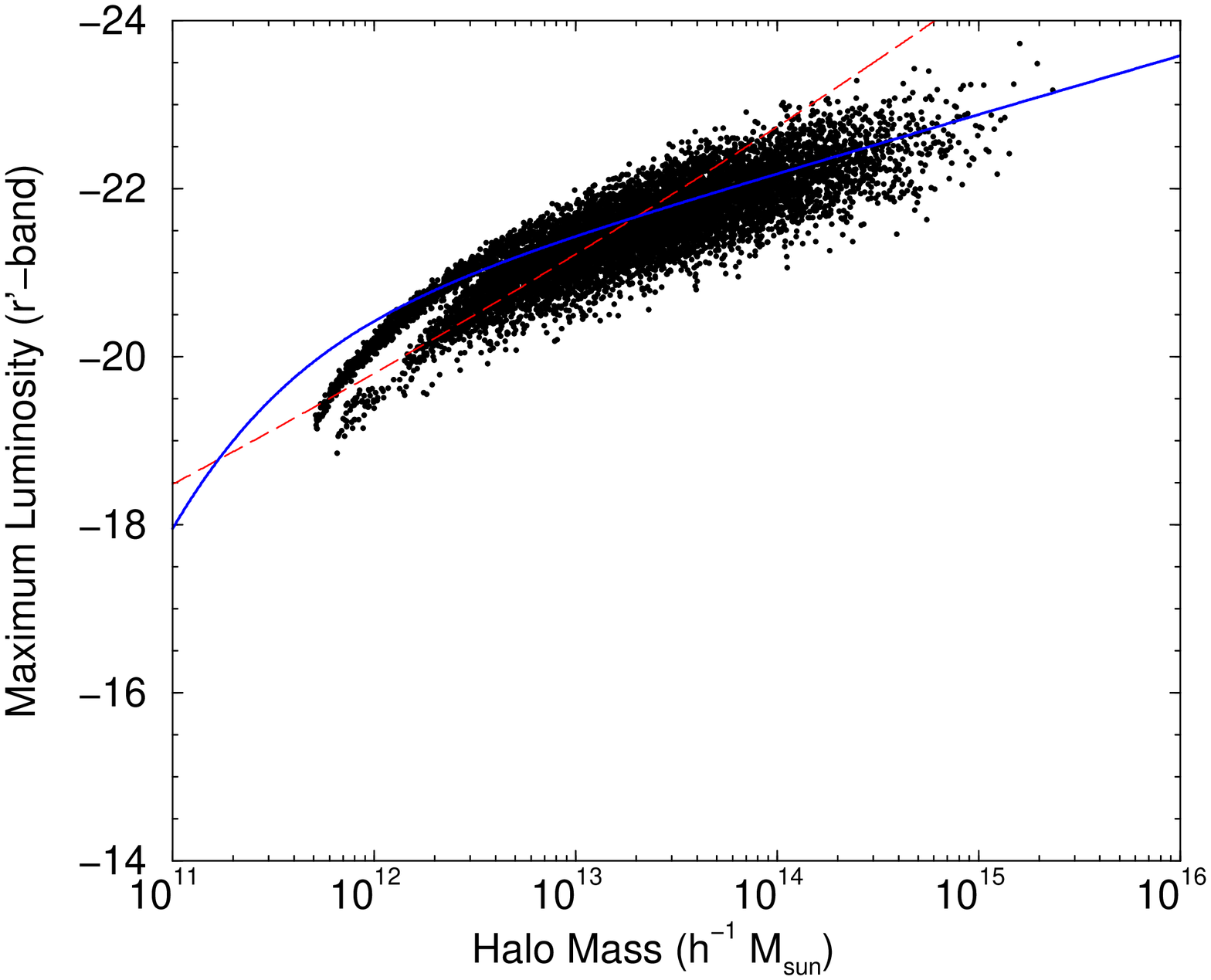,width=\hssize,angle=-90}}
\caption{{\it Left:} The minimum mass for the appearance of satellite galaxies at a given
luminosity as a function of the halo mass
based on a catalog of galaxy groups and clusters in the SDSS (described in
Weinmann et al. 2005). Here, each data point represents a group or a cluster 
where the the halo mass was determined based on the total luminosity of the halo.
Each data point represents the luminosity of the faintest galaxy assigned to each halo.
Here, we focus on the $r'$-band luminosity as the sample used for galaxy clustering in
Zehavi et al. (2004) is defined for that sample. The solid line shows the relation
established from this catalog between the minimum luminosity  in the $r'$-band and the halo mass:
$M_r(min) \approx -19.8(M/10^{12}\; M_{\sun})^{0.03}$.
{\it Right:} The maximum luminosity of
a halo as a function of the halo mass. Here, we plot the luminosity of the brightest
galaxy assigned to that halo (which may or may not the central galaxy in terms of
cluster/group dynamics). The long-dashed line is the same relation as that in the left panel.
The solid line is the relation between central galaxy luminosity and the halo mass as described in
equation~(3) at $z=0$ as required to explain the SDSS luminosity function from Blanton et al. (2004).
 The mass estimate is highly uncertain when halo masses are below few times 10$^{12}$ M$_{\sun}$
due to the small number of galaxies and the scatter in the luminosity of the dominant 
galaxy. Due to this, and the uncertain assignment of galaxies that are satellites of a bigger halo to
less massive halos, we do not consider the difference at the low mass end between the brightest
galaxy luminosity and the expected luminosity from the $L_c(M)$ relation to be any concern.}
\end{figure*}

\begin{figure*}
\centerline{\psfig{file=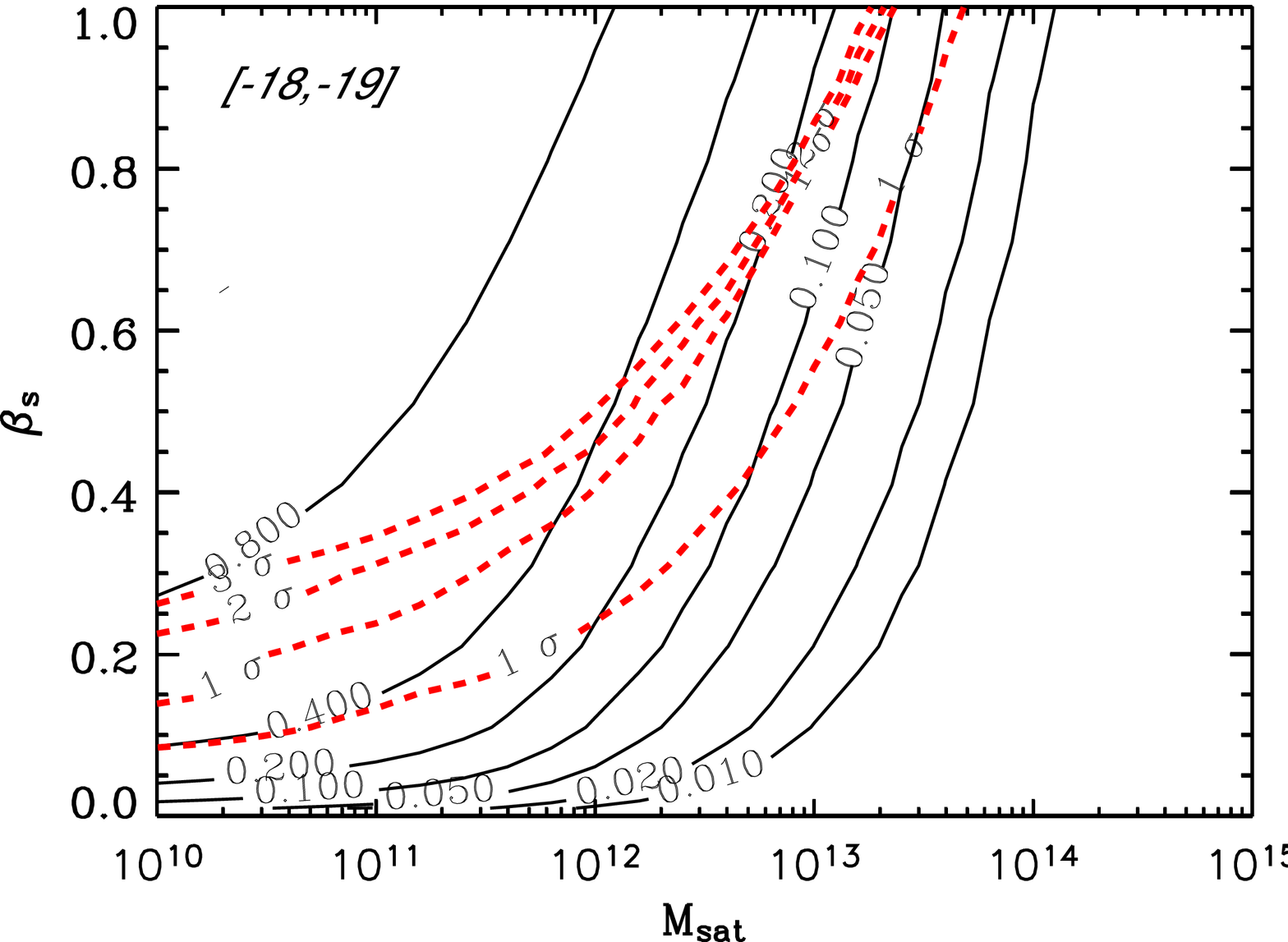,width=\hssize,angle=0}
\hspace{0.3in}
\psfig{file=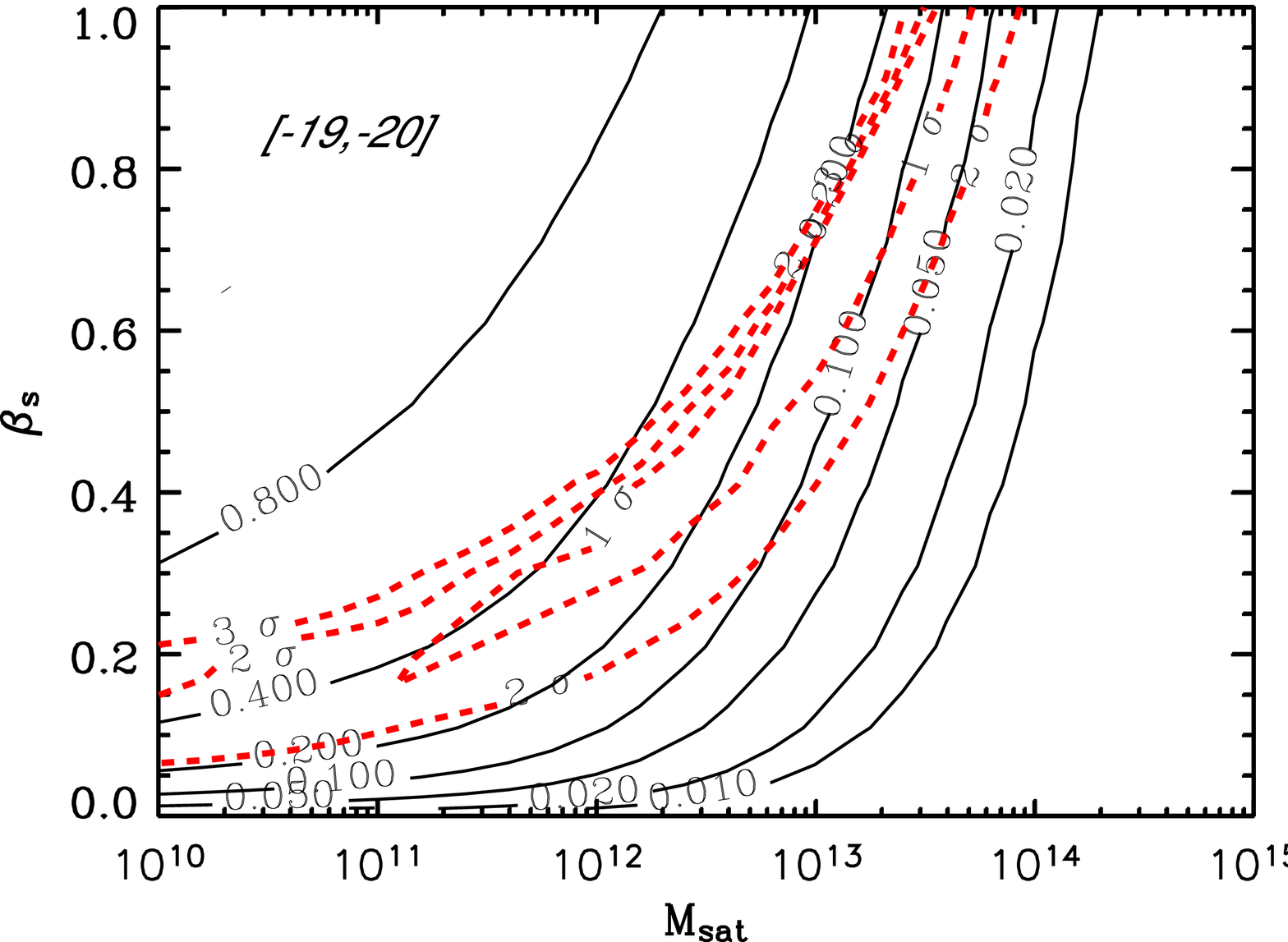,width=\hssize,angle=0}}
\vspace{0.5in}
\centerline{\psfig{file=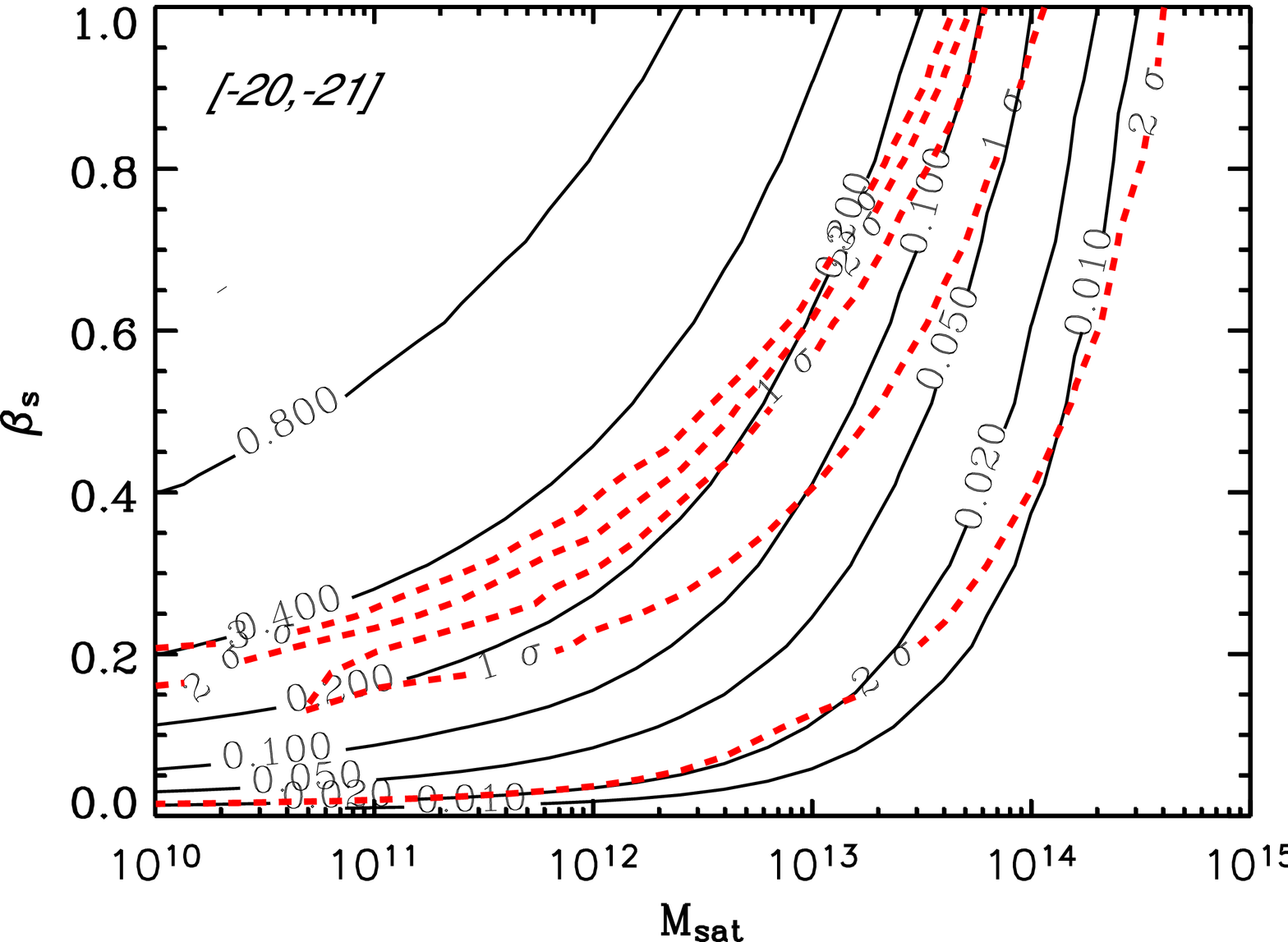,width=\hssize,angle=0}
\hspace{0.3in}
\psfig{file=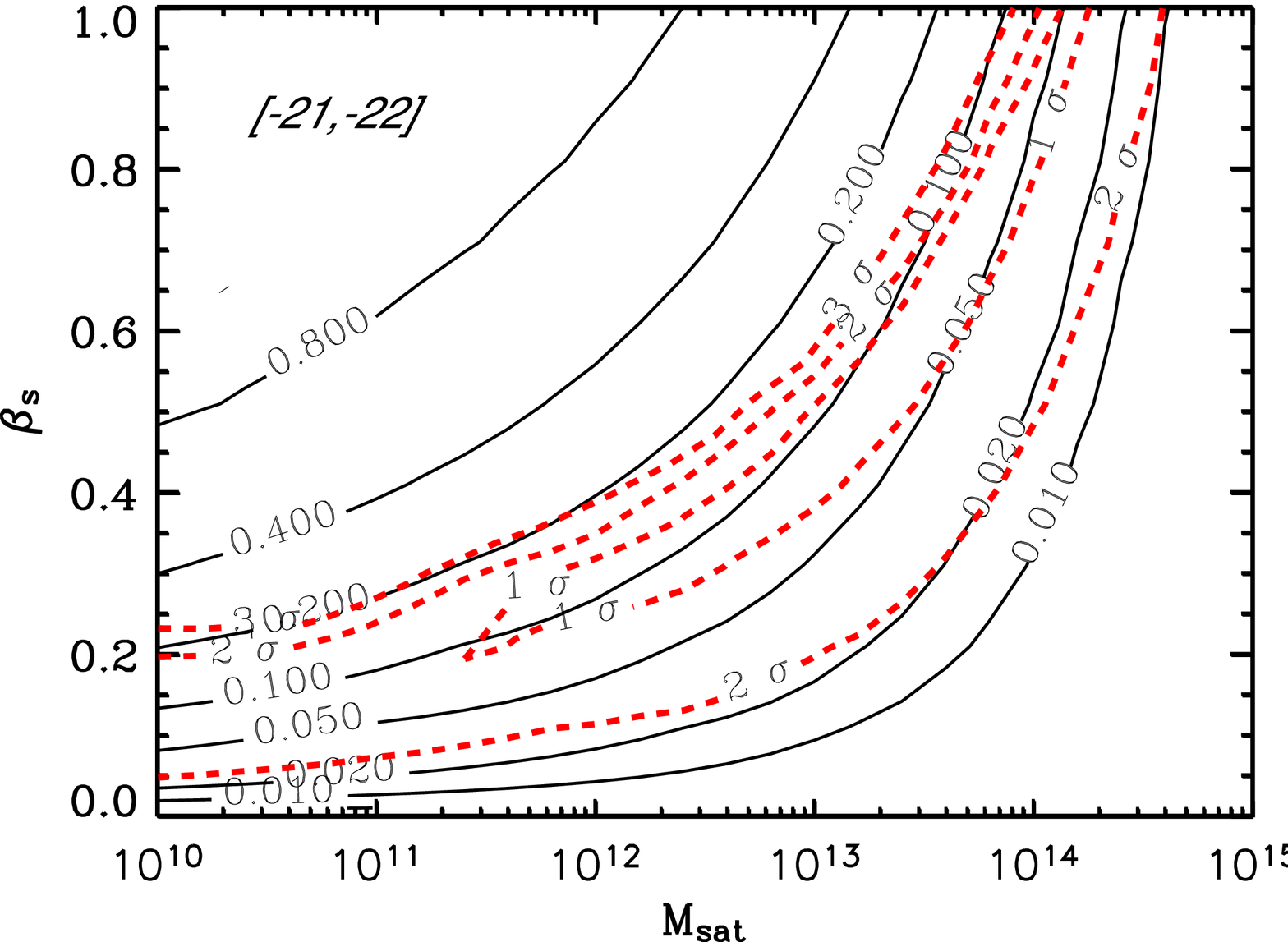,width=\hssize,angle=0}}
\vspace{0.5in}
\caption{
Fraction of satellites as a function of the
luminosity bin in $r$-band, as labeled on each of the four panels,
as a function of
$\beta_{\rm s}$, the power-law slope of total luminosity--halo mass relation,
and $M_{\rm sat}$, the halo mass scale at which satellites appears.
For reference, we overlap the constraints on this parameter space from SDSS 
as relevant for each of the luminosity bins; Note that these constraints are worse than
the overall constraint on this plane when the galaxy sample is combined.
Satellite fractions range from 0.05 to 0.15, when $-22 < M_r <-21$ to $\sim 0.1$ to 0.5
when $-19 < M_r <-18$ at the 68\% confidence level. These fractions are consistent with
values suggested in Mandelbaum et al. (2004) in the three low-luminosity bins
based on an analysis of galaxy-galaxy lensing data with numerical simulations.
}
\end{figure*}

\begin{figure*}
\centerline{\psfig{file=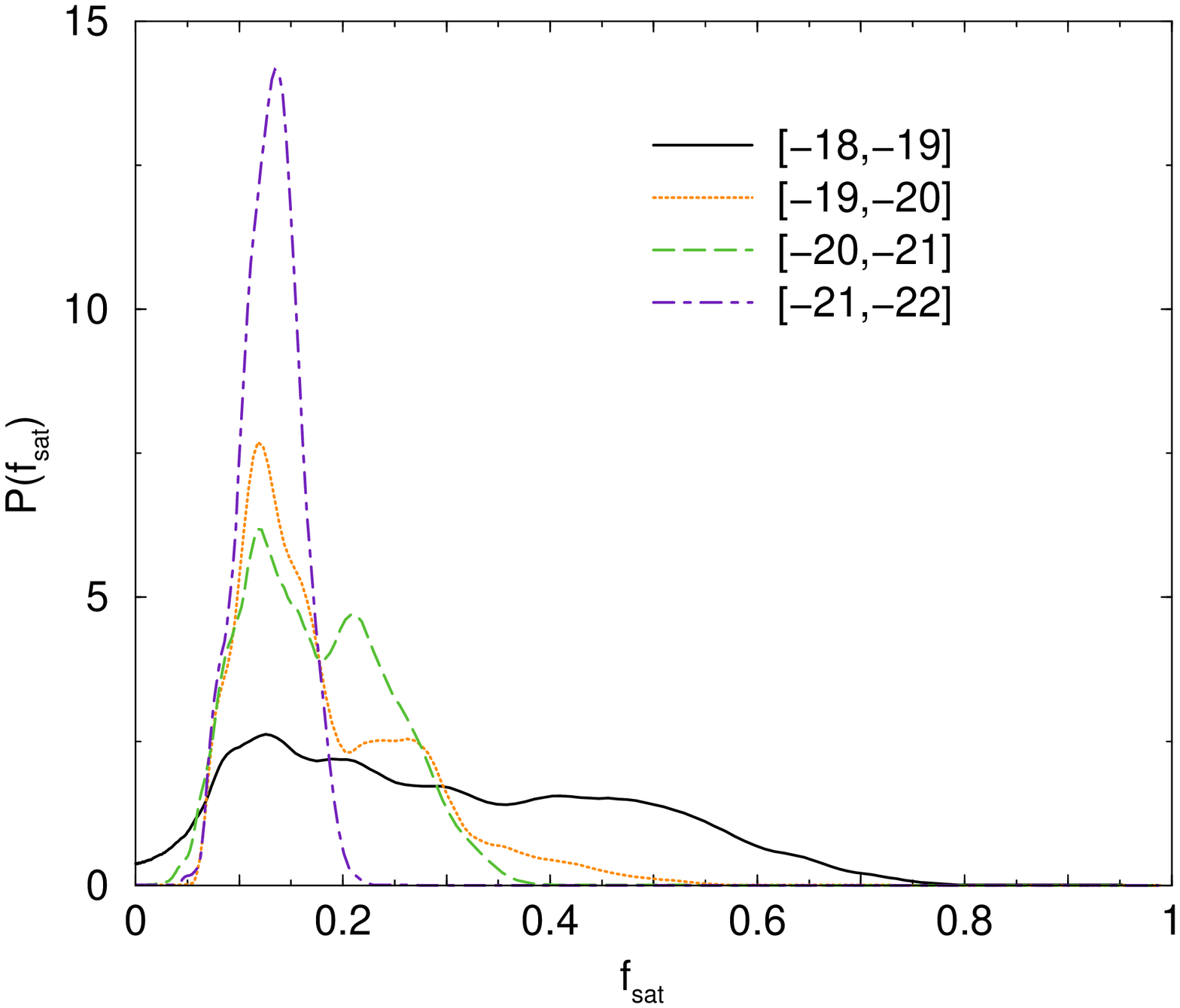,width=\hssize,angle=-90}
\psfig{file=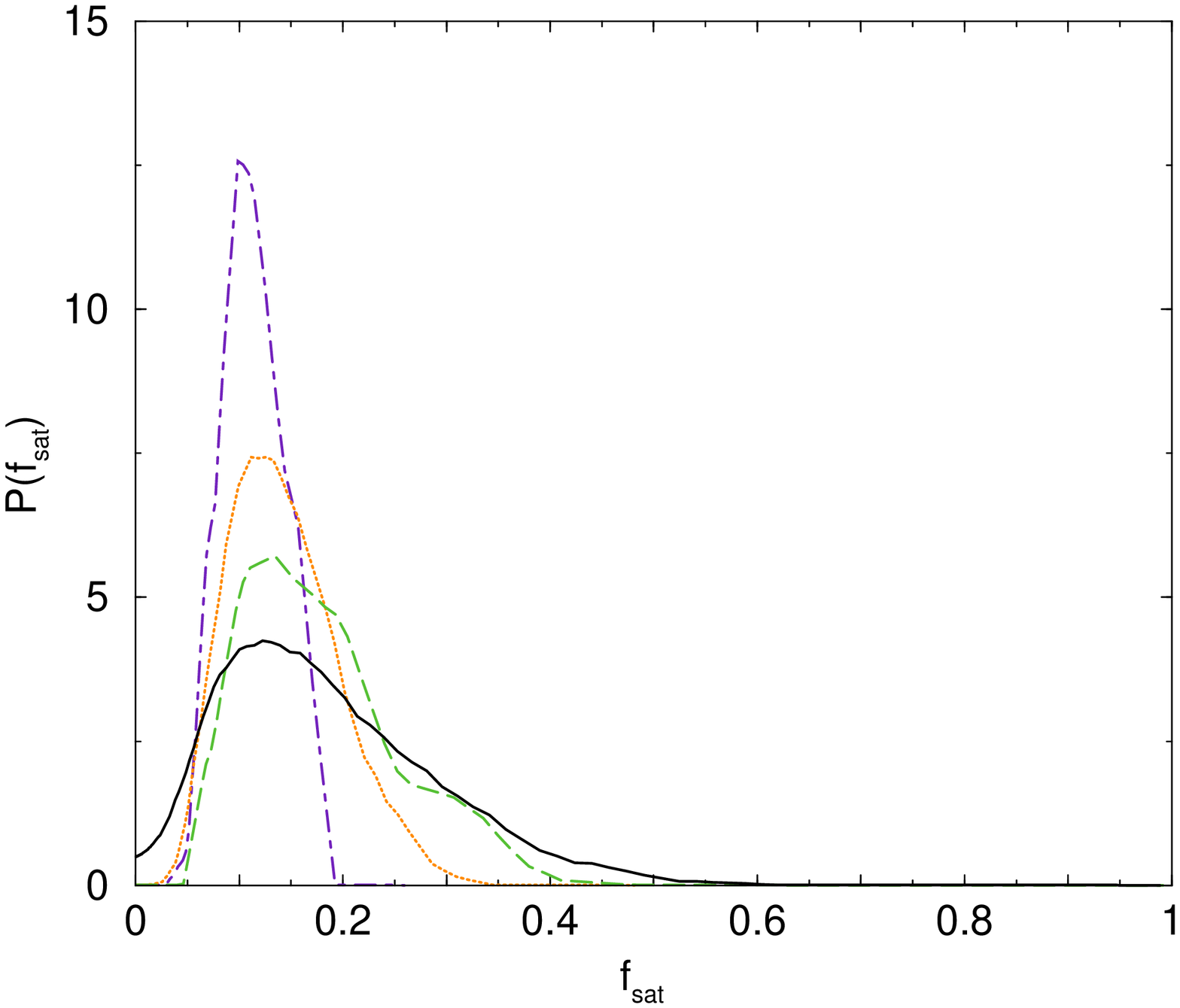,width=\hssize,angle=-90}}
\caption{
Probability distribution for the fraction of satellites as a function of the
luminosity bin in $r$-band, as labeled on each on the left panel.
{\it Left panel:} Satellite fraction with $\beta_s$ taken as a uniform prior between 0 and 1.
{\it Right panel:} Satellite fraction with the constraint that $0.4 < \beta_s < 1$. The lower
estimate was taken to be roughly consistent with the minimum luminosity---halo mass relation suggested by
the SDSS galaxy group catalog data (Fig.~22), combined with clustering constraints shown in Fig.~21 and 23.}
\end{figure*}

\begin{figure*}
\centerline{\psfig{file=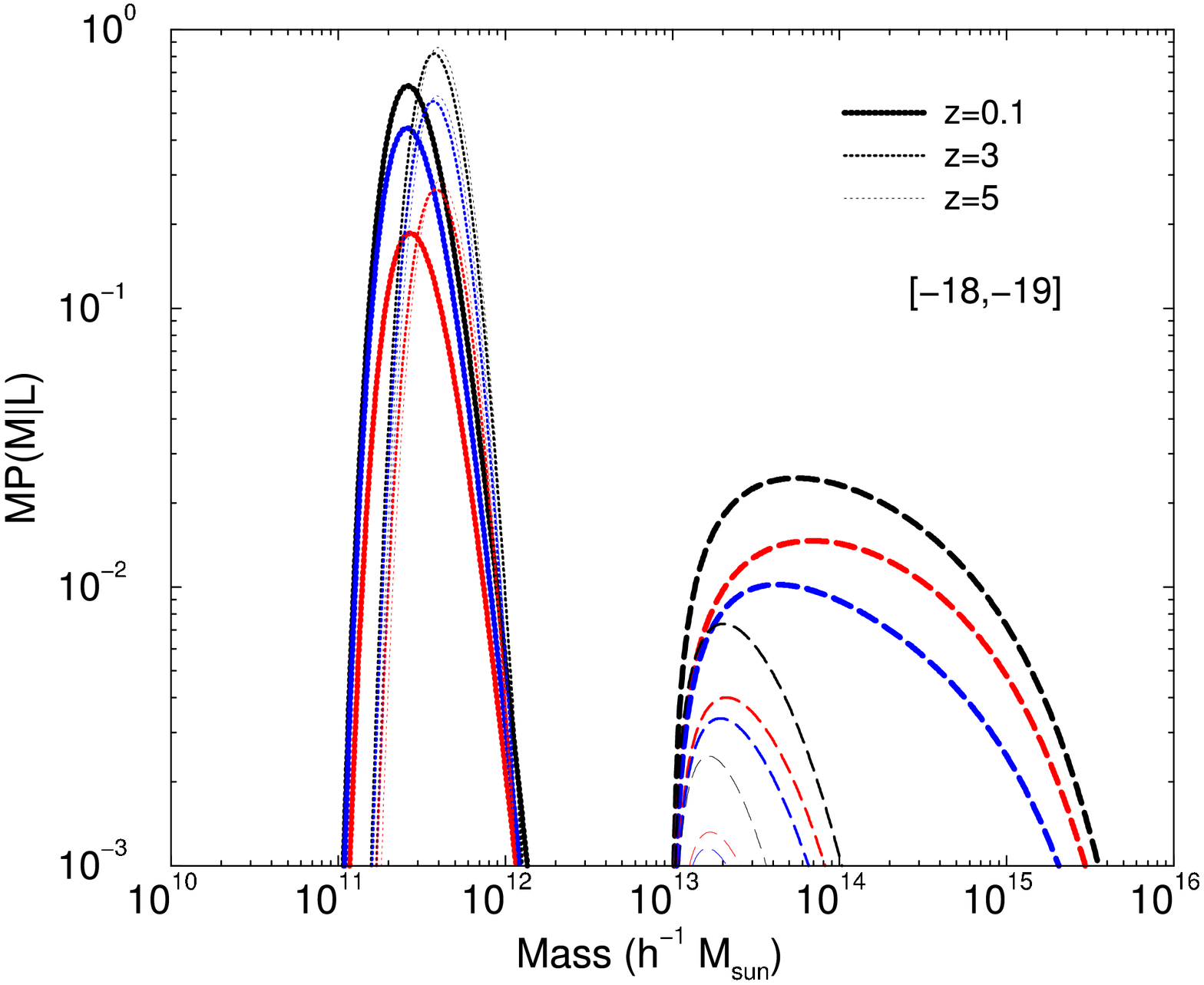,width=\hssize,angle=-90}
\psfig{file=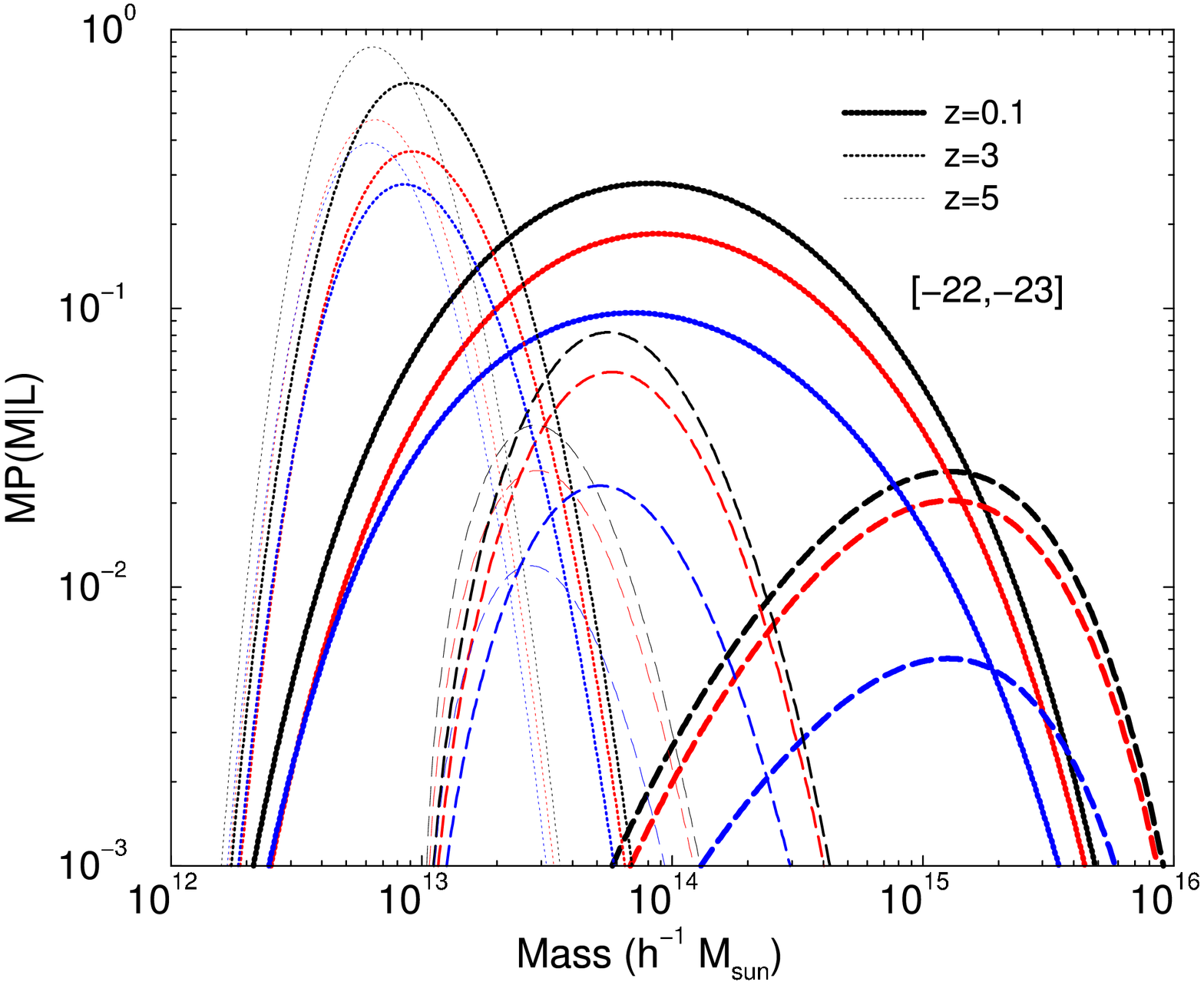,width=\hssize,angle=-90}}
\caption{The conditional probability distribution function of halo mass $P(M|L,z)$ to host a galaxy
of the given luminosity and at the given redshift as a function of the halo mass. The black lines
are the total galaxy sample while red and blue lines show the sample divided to early- and
late-type galaxies.  Left- and  right-panels show these
probabilities for $M_r$ or $M_B$ magnitudes between -18 and -19, and between -22 and -23, respectively
at redshifts of 0.1 (in r-band), 3 and 5 (in B-band), in decreasing thickness of lines.
}
\end{figure*}

\begin{figure*}
\centerline{\psfig{file=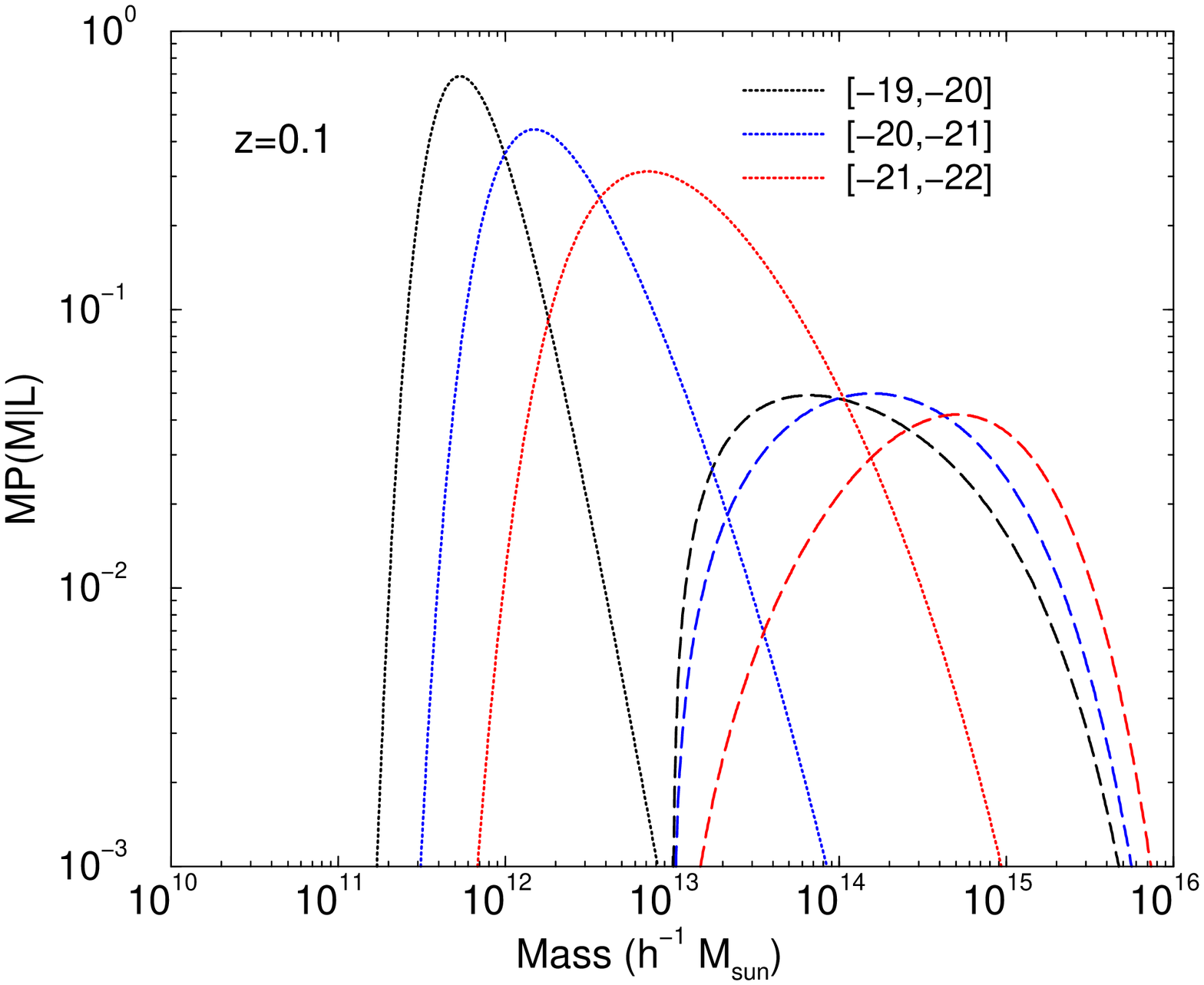,width=\hssize,angle=-90}
\psfig{file=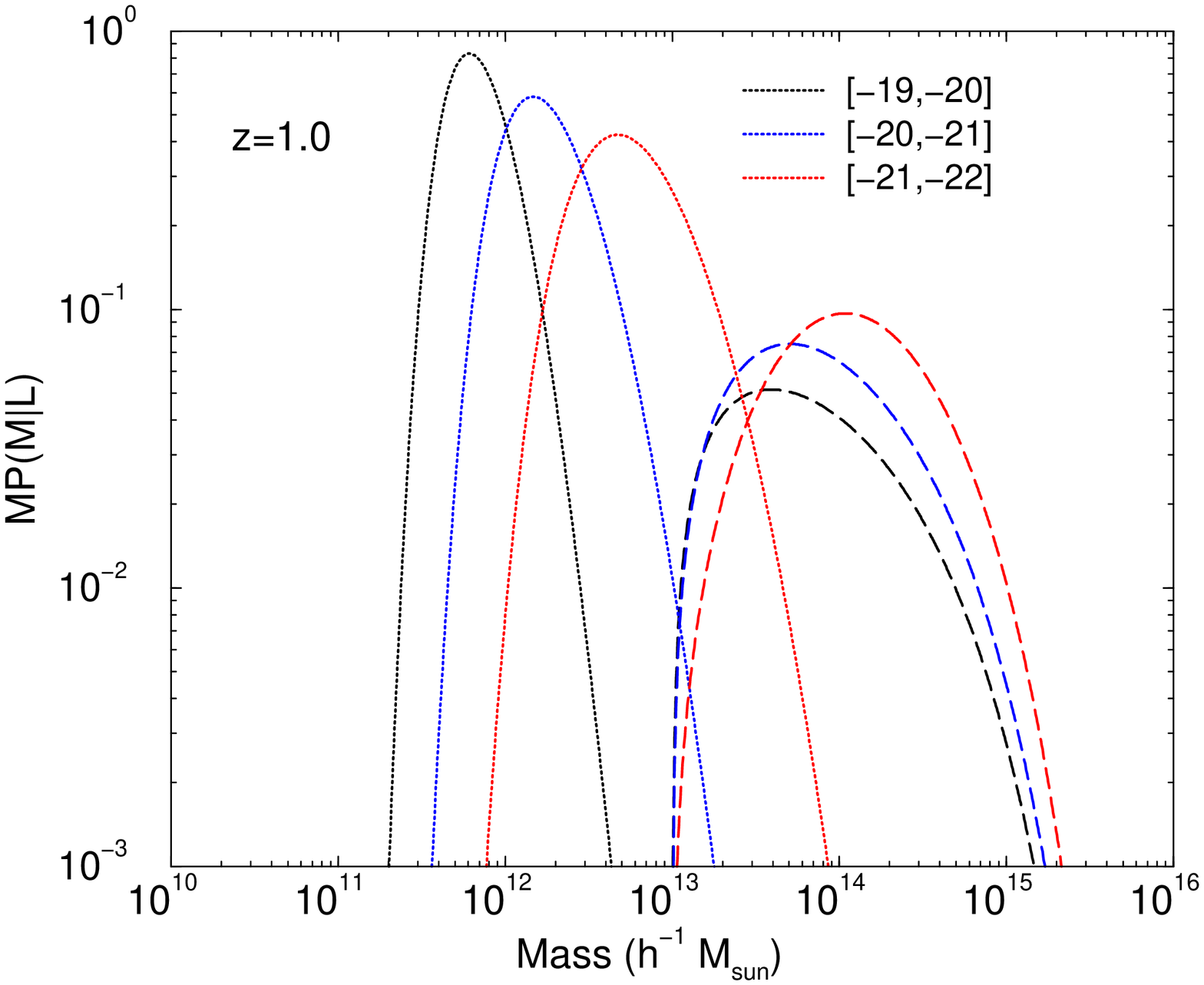,width=\hssize,angle=-90}}
\centerline{\psfig{file=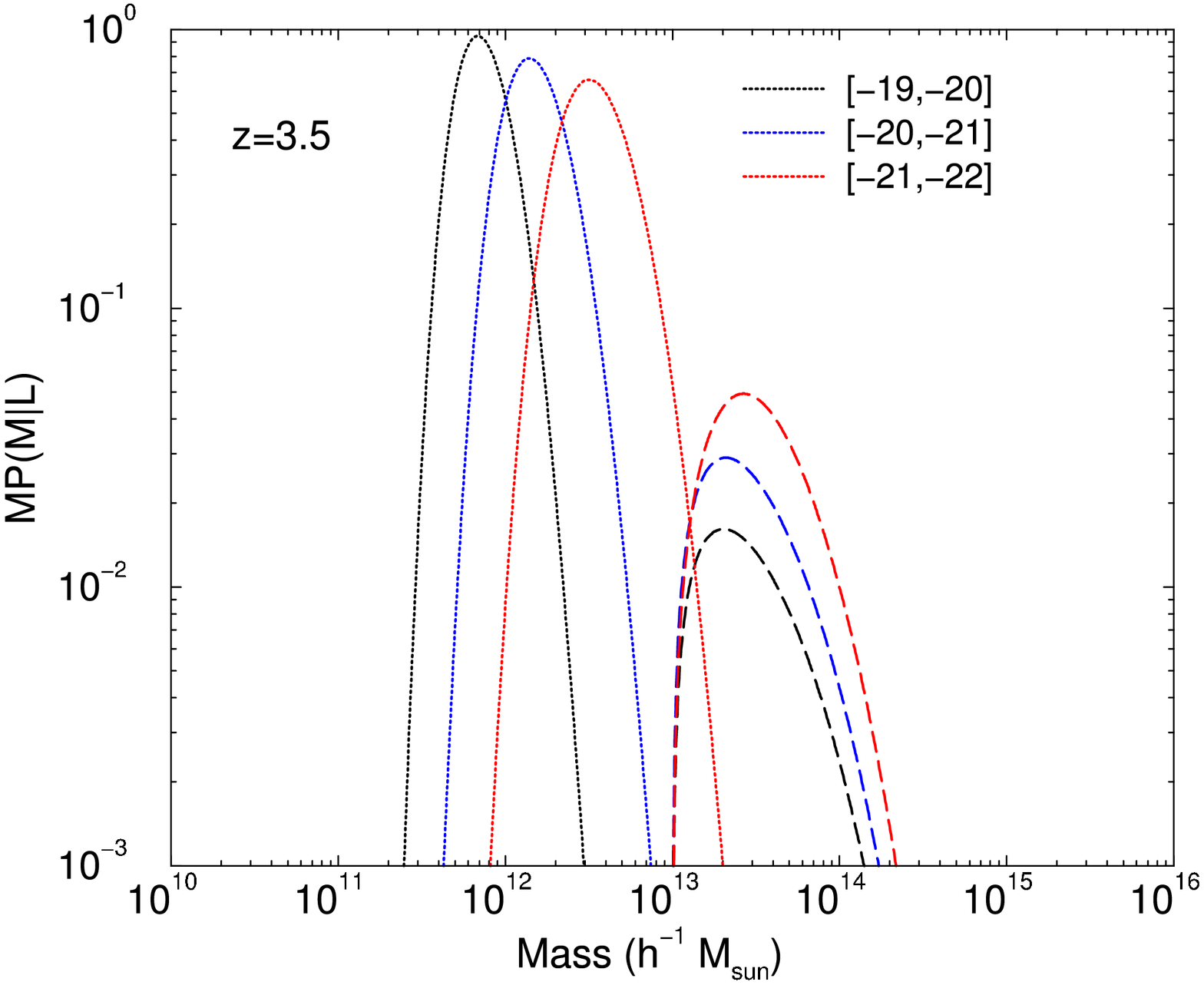,width=\hssize,angle=-90}
\psfig{file=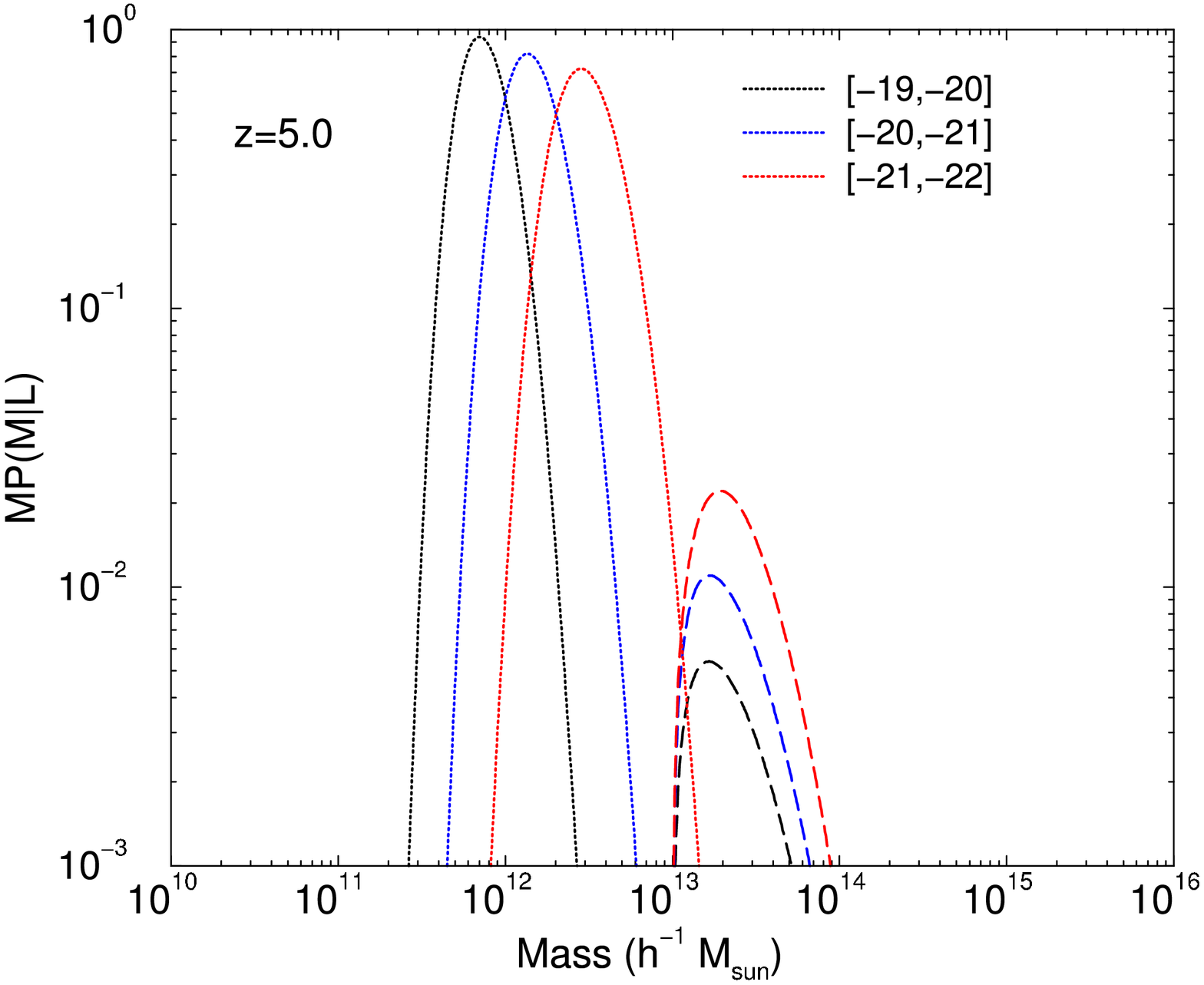,width=\hssize,angle=-90}}
\caption{The conditional probability distribution function of halo mass $P(M|L,z)$ to host a galaxy
of the given luminosity at a given redshift as a function of the halo mass. 
The four panels show these probabilities at different redshifts as labeled on each of the panels,
while the plotted curves are for magnitudes between [-19,20], [-20,-21] and [-21,-22], in r-band at a redshift of
0.1 and B-band for other redshifts, with 
probabilities shown separately for central (dotted lines) and satellite (dashed lines) galaxies.
These probabilities based on CLFs can be compared with the same probabilities
extracted from an analysis of SDSS galaxy-galaxy  lensing data in Mandelbaum et al. (2004; see,
their figures~3 and 4) using numerical simulations.}
\end{figure*}

\begin{figure}
\centerline{\psfig{file=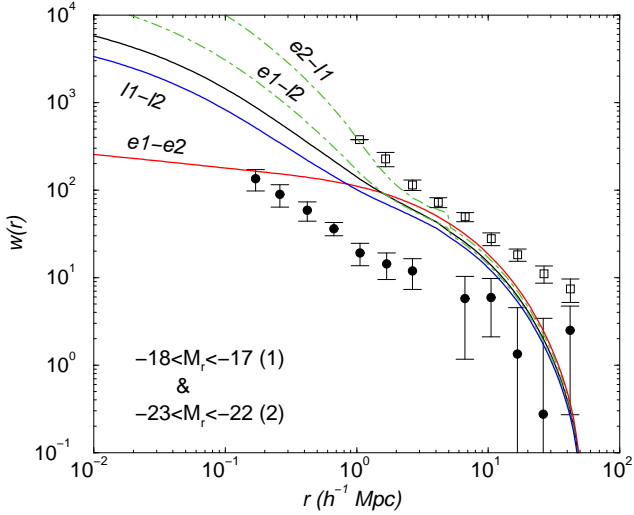,width=\hssize,angle=-90}}
\caption{The predicted cross-correlation between SDSS faint and bright samples of galaxies.
For comparison we show the clustering of galaxies in each of the luminosity bins (from Zehavi et al. 2004), but
cross-clustering between luminosity bins is yet to be measured. We propose such a measurement
as a way to improve constraints on parameters related to the satellite CLF of
the fainter sample. In addition to cross-clustering between galaxies in separate luminosity
bins one can also consider cross-clustering between galaxy types (shown as the upper dot-dashed line
for clustering between early type galaxies in the brighter sample and late-type galaxies in the fainter sample).
A complete set of such measurements across several luminosity bins, in the form of
a covariance matrix of cross-correlations $C(L_a^i,L_b^j,r)$, 
will provide all the information related to galaxy
clustering at the two-point level and will provide additional information to constrain parameters
related to the galaxy type CLFs.
}
\end{figure}

\section{Galaxy Clustering with CLFs}

Using the CLF, instead of the halo occupation number, we can write the power spectrum of galaxies
between type $i$ and type $j$ in terms of the 1- and 2-halo terms (see, review in Cooray \& Sheth 2002)
at a redshift $z$ as
\begin{eqnarray}
&& P_\gal^{ij}(k|L,z) = P^{ij}_{1h}(k|L,z) + P^{ij}_{2h}(k|L,z)\, ,
               \qquad{\rm where}\nonumber\\
&& P^{ij}_{1h}(k|L,z) = \frac{1}{\bar{n}_i(L,z)\bar{n}_j(L,z)}\int dM \, \frac{dn(z)}{dM} \nonumber \\
&\times& \Big[\Phi_i^{\rm sat}(L|M,z)\Phi_j^{\rm sat}(L|M,z) u^2_{\rm gal}(k|M,z) \nonumber \\
&+& \Phi_i^{\rm cen}(L|M,z)\Phi_j^{\rm sat}(L|M,z)  u_{\rm gal}(k|M,z) \nonumber \\
&+&\Phi_j^{\rm cen}(L|M,z)\Phi_i^{\rm sat}(L|M,z) u_{\rm gal}(k|M,z) \Big]\nonumber \\
&& \quad \quad {\rm and} \nonumber \\
&& P^{ij}_{2h}(k|L,z) = P^\lin(k,z) \Big[I^{\rm cen}_i(k|L,z)I^{\rm cen}_j(k|L,z) \nonumber \\
&+&		 I^{\rm cen}_i(k|L,z)I^{\rm sat}_j(k|L,z) +
		 I^{\rm sat}_i(k|L,z)I^{\rm cen}_j(k|L,z) \nonumber\\
&+&		 I^{\rm sat}_i(k|L,z)I^{\rm sat}_j(k|L,z) \Big]
 \label{eqn:gal}
\end{eqnarray}
with the integrals $I^{\rm cen}(k|L,z)$ and $I^{\rm sat}(k|L,z)$ given  by
\begin{eqnarray}
&&I^{\rm cen}_i(k|L,z) = \int dM\, \frac{dn(z)}{dM} b_1(M,z) \frac{\Phi_i^{\rm cen}(L|M,z)}{\bar{n}_i(L,z)} \quad {\rm and} \nonumber \\
&&I^{\rm sat}_i(k|L,z) = \int dM\, \frac{dn(z)}{dM} b_1(M,z) \frac{\Phi_i^{\rm sat}(L|M,z)}{\bar{n}_i(L,z)} u_{\rm gal}(k|M,z) \, , \nonumber \\
\end{eqnarray}
respectively.
Here, and above,
\begin{equation}
 \bar{n}_i(L,z) = \int dM \, \frac{dn(z)}{dM}\, \left[\Phi_i^{\rm cen}(L|M,z)+\Phi_i^{\rm sat}(L|M,z)\right]
 \label{eqn:barngal}
\end{equation}
denotes the mean number density of galaxies of type $i$ while
\begin{equation}
 u_{\rm gal}(k|M,z) = \int_0^{r_{vir}} dr\ 4\pi r^2\,{\sin kr\over kr}\
{\rho_{\rm gal}(r|M,z)\over M} \, ,
\label{eqn:yint}
\end{equation}
denotes the normalized Fourier transform of the galaxy density distribution within
a halo of mass $M$ when $b_1(M,z)$ is the first-order bias factor of dark matter halos.
Here for dark matter halo bias we use the bias factor derived by
Sheth, Mo \& Tormen (2001) which corrects earlier
calculations by Mo et al. (1997; Efstathiou et al. 1988; Cole \& Kaiser 1989)
based on spherical collapse arguments.

The standard assumption in above equations is that galaxies trace dark matter within halos such that
one can utilize the dark matter distribution given by analytic forms such as the NFW (Navarro et al. 1996) profile.
An improved approximation will be to use the density distribution defined by sub-halos
to describe galaxies and, instead of the halo mass function,
use a combination of halo mass function and the subhalo mass function
to describe the satellite contribution to galaxy clustering
that also accounts for effects associated with substructure (e.g., Sheth \& Jain 2002).
Even if corrections exist for the power spectrum from the subhalo mass function, these only modify the strongly
non-linear regime and leave the transitional regime from linear to non-linear clustering probed
by current galaxy clustering measurements unaffected. 
Since relevant profiles related to substructure is still not well studied numerically we make use of
the NFW dark matter density profile (Navarro et al. 1997) to describe the galaxy distribution within halos.
The concentration parameter is defined following the scaling relation of Bullock et al. (2001).
The relevant expressions in our calculation are summarized in Cooray \& Sheth (2002).
In a future paper we plan to combine galaxy-galaxy lensing measurements with galaxy clustering measurements
to test the extent to which galaxies trace dark matter. For now, we will ignore any differences in
the galaxy profile relative to dark matter and concentrate only on basic parameters related to the CLF rather than
statistics such as profiles.

In Equation~\ref{eqn:gal}, when $i=j$, the expression reduces to the power spectrum of galaxies of the same galaxy type.
Similarly one can ignore the index $i$ and $j$ and replace the CLF with the total CLF
to calculate the power spectrum of the total galaxy sample at a given luminosity. Furthermore,
one can also consider the cross power spectrum of samples between $(L_1,i)$ with
$(L_2,j)$, where $i$ and $j$ denote the type, but instead of at a fixed luminosity, cross-correlations are
considered between different luminosities. In this case, the above expressions must be
generalized for the case with two different luminosity bins. Since this is
straightforward, we do not reproduce the appropriate expressions here.
These cross-correlation measurements between two different luminosity bins and different
galaxy types across those bins are yet to be measured. These measurements provide the full
set of clustering measurements related to galaxies and can be thought of as a
covariance matrix of the form $C(L_a^i,L_b^j,r)$ where $a$ and $b$ are indices over the luminosity bins
and $i$ and $j$ are indices over the galaxy types, while $r$ is the projected length at which
clustering is measured. Towards the end of our discussion, we will motivate such a full set of
measurements as a way to establish the satellite CLF more accurately.

For reference, to compare with lensing-lensing galaxy measurements,
the cross-power spectrum between galaxies of type $i$ and  the dark matter distribution is
\begin{eqnarray}
&& P_{\delta-i}(k|L,z) = P^{\delta-i}_{1h}(k|L,z) + P^{\delta-i}_{2h}(k|L,z)\, ,
               \qquad{\rm where}\nonumber\\
&& P^{\delta-i}_{1h}(k|L,z) = \frac{1}{\bar{n}_i(L,z)}\int dM \, \frac{M}{\bar{\rho}}\frac{dn(z)}{dM} \nonumber \\
&\times& \Big[\Phi_i^{\rm sat}(L|M,z) u^2_{\rm gal}(k|M,z) \nonumber \\
&+& \Phi_i^{\rm cen}(L|M,z)  u_{\rm gal}(k|M,z) \Big]\nonumber \\
&& \quad \quad {\rm and} \nonumber \\
&& P^{\delta-i}_{2h}(k|L,z) = P^\lin(k,z) \Big[I^{\rm cen}_i(k|L,z)I^{\delta}(k,z) \nonumber \\
&+&		 I^{\rm sat}_i(k|L,z)I^{\delta}(k,z) \Big]
\label{eqn:dmgal}
\end{eqnarray}
with the integral $I^\delta(k,z)$ given  by
\begin{eqnarray}
I^{\delta}(k,z) &=& \int dM\, \frac{M}{\bar{\rho}}\frac{dn(z)}{dM} b_1(M,z)  u_{\rm gal}(k|M,z) \, ,
\end{eqnarray}
and $\bar{\rho}$ is mean comoving density of dark matter.

At large scales, the galaxy power spectrum or the cross-power spectrum,
 reduces to that of the linear power spectrum scaled by the constant
galaxy bias factor(s). One can understand this by noting that
at large scales, $u_\gal(k|M,z)\to 1$ and the galaxy power spectrum simplifies to
\begin{equation}
 P_{\gal}(k|L,z) \approx b_i(L,z) b_j(L,z)\, P^\lin(k,z),
\end{equation}
where
\begin{eqnarray}
&& b_i(L,z) = \\
&&\int dM\, \frac{dn(z)}{dM}\, b_1(M,z)\,
          \frac{\left[\Phi_i^{\rm cen}(L|M,z)+\Phi_i^{\rm sat}(L|M,z)\right]}{\bar{n}_i(L,z)} \, , \nonumber
\end{eqnarray}
is  the mean large-scale bias factor of the $i$-type galaxy population.
This large-scale bias factor has already been discussed  using CLFs previously (see, Cooray \& Milosavljevi\'c 2005b; Cooray 2005a,b)
and we summarize results based on the current analysis in Fig.~6.

Given the power spectrum, the three-dimensional correlation function of galaxies of type $i$ and $j$ with luminosity $L$ 
at a redshift of $z$ is
\begin{equation}
\xi_{ij}(r|L,z) =\int \frac{k^2 dk}{2\pi^2} P_{ij}(k|L,z) \frac{\sin (kr)}{kr} \, .
\end{equation}
Given limited statistics, most measurements are averaged over samples of galaxies distributed over a certain redshift range.
In this case the projected correlation function follows as
\begin{equation}
w_p^{ij}(r|L,z) =\int \frac{k dk}{2\pi} P_{ij}(k|L,z) J_0(k r) \, .
\end{equation}
In the case of a broad redshift distribution of galaxies over which clustering is projected
the same correlation function is generally written in terms of angular scale, $\theta$, 
with the correspondence $r = \theta d_A$,
where $d_A$ is the comoving angular diameter distance. To calculate such a broad correlation function in redshift space,
we average over the galaxy redshift distribution associated with clustering measurements such that
\begin{equation}
w_p^{ij}(\theta|L,z) =\int dr n^2(r) \int \frac{k dk}{2\pi} P_{ij}(k|L,z) J_0(k d_A \theta) \, ,
\end{equation}
where $n(r)$ is the normalized radial distribution of galaxies with $\int dr n(r)=1$.

In our model predictions we calculate the
projected correlation function at the mid point of the redshift distribution of galaxies
used in that sample.  The measurements where the
exact redshift distribution plays an appreciable role is those
of the GOODS survey (Lee et al. 2005) and the Subaru  Deep Field (Ouchi et al. 2005)
since galaxies are broadly distributed over a wide range in redshift from 2.5 to 4.5 and
from 3.5 to 4.5, respectively. Fortunately, for both these surveys the expected
redshift distribution of galaxies is known either through Monte Carlo simulations, in the case of GOODS 
(Lee et al 2005), or, in the case of Subaru Deep Field, 
through a combination of spectroscopic redshift measurements
and Monte Carlo estimates  (Ouchi, private communication).
We take these distributions into account when model fitting GOODS and Subaru $w_p(\theta)$ measurements.

Another uncertainty in some of these measurements is the exact luminosity distribution of galaxies
in the sample. For surveys such as SDSS and DEEP2 galaxy luminosities are a priori known and samples are
binned in luminosity, while for surveys such as
GOODS and Subaru Deep Field, the exact luminosity distribution remains uncertain, though
statistics in terms of the apparent magnitude at the observed wavelength.
As appropriate, given the redshift information, we converted some of the suggested apparent magnitudes of galaxies
in the sample to rest-frame luminosities at the observed wavelength, usually in the rest B-band,
and used that information to establish the minimum luminosity of galaxies in the sample.
The minimum luminosity usually plays a larger role while the maximum luminosity of galaxies in
the sample does not due to the bright-end exponential cut-off in the galaxy LF.

In Fig.~7, we show the projected correlation function of SDSS galaxies with $M_r <-20$ from 
Zehavi et al. (2004)  and a comparison to model predictions based on the CLF 
using the fiducial description of model parameters. The corresponding CLFs of these model
fits are in Fig.~5. For reference,
Fig.~7 left panel illustrates the dependence of projected correlation 
function  when the power-law slope of the total luminosity-halo mass relation is varied with
$M_{\rm sat}$ fixed at the same fiducial value.  In general,
an increase in $\beta_s$ can be compensated by an increase in $M_{\rm sat}$.
This degeneracy will become clear when we study model fits to the data later.
In the right panel of Fig.~7, we illustrate the projected correlation function of galaxy types as well as the cross-correlation between two galaxy
types with $M_r <-21$. The presence of a non-linear part for the cross-correlation between galaxy types
can be easily understood based on the fact that both early- and late-type exists in similar mass halos (Zehavi et al. 2004).

In Fig.~8, we summarize the projected correlation function as a function of luminosity bins
 considered by Zehavi et al. (2004); For the faintest (-17 to -18) and the brightest (-22 to -23) bin,
only the total clustering correlation function is measured, though for comparison,
we continue to show the clustering correlation function for both early- and late-type galaxies.
Note that with our fiducial model parameters, measured projected correlation functions  in magnitude bins between
-19 to -20, -20 to -21, and -21 to -22 are generally well described, 
while fits are generally less than perfect in both the low luminosity and high luminosity bins. 
This is due to the fact that our fiducial model parameters are extracted from an overall fit to the whole sample assuming the
same underlying  description for the CLF for the whole galaxy sample.
When model fitting the data, since measurements in mid magnitude bins are better determined, 
the fits are weighted more for these bins than ones at the
two ends.  We did not attempt to weight different bins equally. At this initial stage of analysis we
are mostly interested in extracting a consistent model for the overall CLF of galaxies
from current measurements or trying to understand the extent to which
data can constrain parameters related to the CLF. 
The models considered in Zehavi et al. (2004) involved different occupation number descriptions for different luminosity bins.
The CLF approach avoids having to describe occupation numbers separately for different luminosity
bins, though it is likely and, guaranteed to be, that some parameters such as $M_{\rm sat}$ will
be luminosity dependent, though parameters such as $\beta_s$ should not be. We will also
show results where we model fitted the data separately based on divisions to luminosity bins.
While the overall fits are not strong, we do find certain limited evidence for variation in
$M_{\rm sat}$  as a function of the luminosity.
While the CLF description leads to
a certain reduction in the number of parameters to be determined from data, though we note that,
due to our introduction of new parameters involving galaxy types etc, there is in fact no reduction but rather an increase in parameters.
Later, in the discussion, we will propose additional measurements related to the same sample of galaxies
in SDSS  as Zehavi et al. (2004) and these measurements could further aid in improving model fits to determine
current parameters better.

To show that our models are consistent, in Fig.~9, we compare our prediction for the galaxy-mass cross-correlation function, in real space,
for a volume limited sample with $-21.5  > M_r > -23$, and in the redshift range between 0.1 and 0.174. This
galaxy sample has been used by Sheldon et al. (2004) to make a measurement of the galaxy-mass cross-correlation 
function via galaxy-galaxy weak lensing
measurements in SDSS. We find our predictions  agree well with measurements, and as a further application, in Fig.~9, we also
show the expected cross-correlation if the foreground galaxy sample of Sheldon et al. (2004) had been further divided to
blue- and red-galaxies, following essentially the same division to galaxy types as in Zehavi et al. (2004).

In Cooray \& Milosavljevi\'c (2005a) we made use of SDSS galaxy-galaxy weak lensing measurements in the z'-band to construct
$L_c(M)$ relation at higher wavelengths. These measurements are analyzed using the halo model in a variety of studies (e.g.,
Mandelbaum et al. 2004; Guzik \& Seljak 2002; Yang et al. 2003a; Sheldon et al. 2004)
and we do not make use of the galaxy-mass correlation function when model fitting parameters here.
This is due to the fact that we are primarily interested in understanding the extent to which CLFs can be
constructed from galaxy clustering measurements and to check our estimates, say, on the halo mass of galaxies
at a given luminosity from estimates made by prior studies using galaxy-galaxy lensing measurements.
We do this in the context of probability distribution for halo mass at a given galaxy luminosity
(Mandelbaum et al. 2004).

The approach based on CLFs easily allows us to model clustering statistics at high redshifts
within the same parameter description provided that redshift dependences are properly taken into account.
Given the results from Cooray (2005b) on the redshift evolution of the $L_c(M)$ relation,
here we take the redshift dependence of the central galaxy luminosity with redshift into account
with parameters $\alpha$ and $\eta$ in equation~5. For parameters in the satellite CLF, such as
$\beta_s$ and $M_{\rm sat}$, we do not attempt to include redshift variations given the lack of knowledge.
On the other hand, redshift dependences can be extracted by  analyzing clustering measurements as a function of redshift
and by looking for differences in parameters constrained at different redshifts.
This was the approach used in Cooray (2005b) to establish redshift variation in $L_c(M)$ relation.

In Fig.~10, we compare our predictions for projected galaxy
clustering at redshifts 0.4 to 0.8 as determined by  the
COMBO-17 survey (Phleps et al. 2005). These data involve rest B-band
magnitudes and we make use of the $L_c(M)$ relation as appropriate for rest B-band
from Cooray (2005b) including the redshift evolution with parameters
described with respect to equation~3.  While our fiducial parameters
describe the non-linear clustering part of early- and late-type galaxies
in this sample well, we find that large-scale clustering of late-type
galaxies is not modeled by our parameters. We find the same difference
between measurements and model fits based on the halo occupation number
appears in the analysis by Phleps et al. (2005) as well. 
We  use this data set to extract parameters
related to the satellite CLF and find that constraints on $\beta_s$
and $M_{\rm sat}$ allowed by COMBO-17 at a mean redshift of 0.6 is
in good agreement with SDSS suggesting that no strong evolution of
parameters such as $\beta_s$ and $M_{\rm sat}$ out to this redshift.

In  Fig.~11, we consider galaxy clustering at $z\sim 0.8$ to 1.3 from the DEEP2 survey (Coil et al. 2004).
Here, clustering measurements are divided to two-luminosity bins, in the rest B-band, and the combined sample to 
early- and late-types.  As shown in Fig.~11, our
fiducial model parameters related to central and satellite CLFs describe DEEP2 clustering measurements
at $z \sim 1$ reasonably well. Unfortunately,
DEEP2 data mostly probe large-scale linear clustering pattern rather than the non-linear 1-halo
part that is strongly sensitive to model parameters related to the satellite CLF. As we find later,
because of this reason, DEEP2 data only allow upper limits to be placed on
model parameters such as $\beta_s$ and $M_{\rm sat}$ at a redshift of unity.
Since measurements considered here only come from the first subset of the total DEEP2 galaxy sample,
the final clustering analysis should improve parameter estimates significantly.

Extending to higher-redshifts, we make use of
the rest B-band clustering measurements at $z \sim 3$ by Lee et al. (2005) in the
GOODS survey. Due to limited number statistics, measurements only exist for the
total galaxy sample though in Fig.~12 we also show the clustering of early- and late-type
galaxies as well. At $z \sim 4$, the recent angular clustering measurements by Ouchi et al. (2005) in the
Subaru/XMM-Newton Deep Field can also be modeled using the CLF approach. In Fig.~13 ({\it left panel}), we show the measurements
with $i$-band magnitudes brighter than 27.5. This magnitude limit roughly corresponds to the
rest $M_B <-18.5$, and this conversion is consistent with the galaxy number density expected from the rest B-band
galaxy LF at a redshift of 4 (with a number density of $5 \times 10^{-3}$ h$_{70}^3$ Mpc$^{-3}$
from Cooray 2005b) and the suggested number density of $5.8 \pm 1.4 \times 10^{-3}$ h$_{70}^3$ Mpc$^{-3}$ 
in Ouchi et al. (2005: see their Table~1) down to the same magnitude limit.
To describe non-linear clustering at these redshifts,
the satellite CLF must have distinctively different parameters for the slope $\beta_s$ and the low mass
cut-off $M_{\rm sat}$ for the appearance of satellites when compared to parameters. In Figure~1,
we show  the expected clustering level based on best-fit parameters that we will return to below.
For comparison, in the same figure, we also show expected clustering of galaxies down to the same magnitude
level at redshifts 5 and 6. At large angular scales, as the redshift is increased, clustering strength is
expected to increase due to redshift evolution of bias factor, which is in return associated with the decreasing
number density of halos that host the galaxies with the same luminosity when compared to the number density at a lower
redshift. At small scales corresponding to the 1-halo term, galaxy clustering should show a decrease in strength
as the number of galaxies that appear as satellites with the same luminosity is decreasing as the redshift
is increased.

In Fig.~13 ({\it right panel}), we consider clustering as a function of the galaxy luminosity at $z \sim 4$.
The measurements shown here now come from Kashikawa et al. (2005) based on clustering measured with the 
Subaru Deep Field\footnote{http://soaps.naoj.org/sdf}. To describe luminosity dependent galaxy clustering we use the
same CLFs as the ones used to describe galaxy clustering at $z \sim 4$ in
 the right panel, but divided the absolute luminosities of galaxies  following the division in
 Kashikawa et al. (2005)  based on apparent magnitudes. While galaxy clustering in the fainter bins are
adequately described, the non-linear clustering seen in the brighter bin is overestimated.
This is clearly due to a wrong choice of parameters related to the satellite CLF at $z \sim 4$
for these bright galaxies. Since our models here assume the best-fit parameters with $M_B <-18.5$,
the over prediction of non-linear clustering for galaxies with $M_B <-20$ clearly shows that galaxies that appear as 
luminous satellites are only present with a higher cut off for $M_{\rm sat}$. While the Kashikawa et al. (2005)
measurements only allow one estimate of clustering in the non-linear regime, we have begun a separate analysis
of luminosity dependent clustering from the same imaging data as those used in Ouchi et al. (2005).
Those measurements increase the signal-to-noise of non-linear clustering estimates as a function of redshift
allowing the mass scale associated with satellites, as a function of their luminosity, be established
better when compared to published measurements from Kashikawa et al. (2005) shown in Figure~13 ({\it right panel}).
We will present these results in an upcoming study (Cooray \& Ouchi 2005).

To study the extent to which model parameters related to the CLF can be constrained, we now
model fit LFs and clustering measurements by varying various parameters in our model.
From these model fits, we establish likelihoods to describe the data given model parameters.
In this analysis, we only make use of published variance measurements of both the LF and
clustering statistics. It is likely that the measurements are affected by a covariance
resulting in correlations between clustering measurements at different physical
or angular scales. The presence of a substantial covariance in angular projected
correlation function is well known due to non-Gaussianities and overlapping
 window functions (e.g., Eisenstein \& Zaldarriaga 1999; Cooray \& Hu 2001).
While the  model based on CLFs has a large number of free parameters ($\sim 20$),
various experimentation with the data showed that only a handful of
parameters are constrained while other parameters remain unconstrained for various reasons.
Thus, we only consider a subset of parameters to model fit while other parameters
are fixed based on various other arguments and observations.
For example, we fix parameters of the $L_c(M)$ relation and do not attempt
to establish them from galaxy clustering data.
As discussed in Cooray \& Milosavljevi\'c (2005a), such a relation is best determined with
galaxy-galaxy lensing data and we have reanalyzed $r$-band galaxy-galaxy lensing data from SDSS to
reestablish the $L_c(M)$ relation; The central galaxy mass estimates obtained
agrees well with estimates in Mandelbaum et al. (2004).
In the case of the central galaxy CLF we treat $\sigma_{\rm cen}$ and $M_{\rm cen-cut}$ as
free parameters, while for the satellite CLF, we take $M_{\rm sat}$ and
$\beta_s$ as free parameters, with the value of $\gamma$ fixed at $-1$ and equation~5 fixed
following the description below it. For description involving galaxy types,
we take $f_{\rm cen-E}$, $M_{\rm cen}$, $M_{\rm sat}$, $L_{\rm sat}$, $g_{\rm sat-E}$, and
$f_{\rm sat-E}$ as free parameters. 

While there are 10 free parameters, when model fitting the data we only consider a smaller subset of
these parameters for different datasets due to an important reason that some statistics are more
sensitive to certain parameters when compared to others. When considering the LF of the total galaxy sample,
we fit parameters $\sigma_{\rm cen}$, $M_{\rm cen-cut}$,  $M_{\rm sat}$, and
$\beta_s$, though there are no useful constraints on the latter two parameters from the LF.
This is clear from Fig.~2, where we show that
the LF is mostly determined by statistics of central galaxies; Another way to explain this is that,
at a given luminosity, the total density of galaxies is dominated by
a larger fraction of central galaxies in low mass halos, which have a higher density, than satellites 
of the same luminosity in more massive halos. 

In the case of LFs of galaxy types, with early- and late-type galaxies fitted
simultaneously given that parameters describing early-type galaxies also
describe late-type galaxies, we take $\sigma_{\rm cen}$, $M_{\rm cen-cut}$,
 $f_{\rm cen-E}$ and $M_{\rm cen}$ as free parameters.
Figs. 14 and 15 right panels  show constraints on two parameters from this parameter set
with likelihood of other parameters marginalized over. In the case of the
total correlation function, as a function of luminosity from SDSS, we fit
$\sigma_{\rm cen}$, $M_{\rm cen-cut}$,  $M_{\rm sat}$, and
$\beta_s$, and show parameter constraints on the central galaxy CLF in Figure~14 
to be compared with constraints for same parameters from the total galaxy LF.

We use the same set of parameters as the ones used to fit galaxy type LFs
to also fit the correlation functions divided to galaxy types from SDSS. The
constraints on $f_{\rm cen-E}$ and $M_{\rm cen}$ are shown in Fig.~15 can be compared with
constraints for same parameters from the galaxy LF. Since parameters related to satellite CLF
are better described with the correlation function, we expanded the parameter space and also fitted
parameters related to satellite galaxy types. The constraints on these parameters,
with ones related to central galaxies marginalized over based on LF
constraints, are shown in Fig.~16. Beyond SDSS, we also consider model fits separately to the total 
clustering data at different redshifts separated into galaxy luminosities when available.
Here, we treat  $\sigma_{\rm cen}$, $M_{\rm cen-cut}$, $M_{\rm sat}$, and
$\beta_s$ as free parameters as there is no information related to galaxy types
in the high redshift data except in DEEP2, though we do not use that information explicitly since
DEEP2 clustering measurements do not probe non-linear clustering in detail.
Fig.~17 summarizes these results for parameters  related to the satellite galaxy CLF.

As shown  in Figs. 14 and 15, the total LF and LFs galaxy types in SDSS
allow better estimates of parameters such as $\sigma_{\rm cen}$, the log-normal
scatter in the $L_c(M)$ relation at a given mass, and
$f_{\rm cen-E}$ the fraction of early-type galaxies that are in halo centers.
From the SDSS total LF data down to $M_r <-17$ (from Blanton et al. 2004), $\sigma_{\rm cen}$ is constrained to be
$0.17^{+0.01}_{-0.02}$ at the 68\% confidence.
In Cooray \& Milosavljevi\'c (2005b), we found $\sigma_{\rm cen} \sim 0.22$ 
to describe the field-galaxy luminosity function in the K-band
(Huang et al. 2003), while in Cooray (2005a), we suggested
a value for the dispersion of $\sim$ 0.17 $\pm 0.1$ 
in the 2dFGRS $b_J$ band. Unfortunately, the underlying reason
for a difference between the dispersion at K-band and lower wavelengths
is not understood. Our estimate for $\sigma_{\rm cen}$ is
in good agreement with the value of 0.168 found for the dispersion of
central galaxy luminosities by Yang et al. (2003b) where these authors used a
completely different parameterization for the CLF based on a priori assumed Schechter function shape.
When compared with the  Fig.~14 right panel clustering measurements do not allow
stronger constraints to be placed on these two parameters when compared to the constraint based on the galaxy LF.
This is due to the fact that the correlation function is more strongly sensitive
to satellite galaxies rather then central galaxies through the non-linear 1-halo term.

While $\sigma_{\rm cen}$ is well determined, we find no evidence for a general low-mass cut-off in the central
galaxy LF with a 95\% confidence level on the upper limit of $M_{\rm cen-cut} < 3 \times 10^{10}$
M$_{\sun}$. As discussed before this cut-off should not be interpreted as the M$_{\rm min}$ parameter in
halo occupation number models of Zehavi et al. (2004).
The cut-off suggested in models based on halo occupation number is present in
CLFs through the $L_c(M)$ relation as shown for central galaxy CLFs in  Fig.~5. We expect a global cut off in the
LF if effects such as reionization (Benson et al. 2002; Bullock et al. 2000;
Tully et al. 2002; see, review in Barkana \& Loeb 2001)  affect galaxy formation significantly.
As discussed in Cooray \& Cen (2005), the feedback effects may be
more complex than considered before and could depend on the time scale of formation relative to the reionization (e.g.,
Tully et al. 2002) and additional heating history of IGM by supernovae from first galaxies. 
The galaxy group LFs, down to magnitudes below -13, show
partial evidence for a cut off in the galaxy density corresponding to
central galaxies at a halo mass around 10$^{11}$ M$_{\sun}$ with a significant absence of dwarf galaxies.
On the other hand, dwarf galaxy statistics in massive clusters, hosted in dark matter halos with masses much below the cut-off
halo mass in galaxy groups, are consistent with the expectation based on the subhalo mass function.  It is not clear why we do not detect an overall
turn over given that such a cut-off is necessary to explain the low power-law slope of the $b_J$-band LF of 2dFGRS at the faint end
(Cooray 2005a) and that galaxies in our sample do probe mass scales down to
10$^{11}$ M$_{\sun}$. On the other hand, the phenomena leading to an absolute lower
cut off the halo mass hosting galaxies may be local rather then affecting the galaxy population as a whole,
though this does not explain the low-end difference between LFs of SDSS and 2dFGRS.
In a future study, we plan to analyze the faint-end LF of SDSS in detail to address if there is evidence
for a global cut off. We encourage extending clustering studies of galaxies to fainter magnitudes to obtain
a better handle on their properties and to extend CLFs down to fainter luminosities than possible so far,
though due to reasons that clustering statistics are not sensitive to central galaxies, it is unlikely
such measurements alone would be helpful.

As shown in Fig.~15, the galaxy LF also provides best constraints
on parameters related to galaxy types that appear in halo centers.
Marginalizing over other parameters, we constrain at 68\% confidence level
$f_{\rm cen-E}=0.62 \pm 0.19$, while the mass-scale $M_{\rm cen}$
describing the early-type fraction of central galaxies is
$M_{\rm cen}= (3.1^{+8.2}_{-2.7})\times 10^{11}$ M$_{\sun}$.
As in Fig.~14, constraints from SDSS galaxy clustering statistics
are lower when compared to  constraints from the LF for same parameters.
While parameters related to central galaxies are not well determined
by clustering statistics, certain parameters related to satellite galaxies are.
As shown in Fig.~16, while no useful constraint exist for
$M_{\rm sat-type}$, as well as $L_{\rm sat}$ though we do not show its constraint
here explicitly, $f_{\rm sat-E}=0.5\pm 0.15$ 
while $g_{\rm sat-E}=0.25 \pm 0.15$ at the 68\% confidence level from SDSS clustering data.

With clustering statistics, the best constraints are on parameters related to the total satellite CLF. 
In Fig.~17,  we summarize constraints on parameters $\beta_s$ and $M_{\rm sat}$
 as a function of redshift of the dataset.
Surveys such as SDSS and COMBO-17 allow these parameters to be determined in detail.
At high redshift, while Subaru data at $z\sim 4$ from Ouchi et al. (2005)
allow some constraints,
DEEP2 and COMBO-17 data only allow  an upper limit to be placed on say $M_{\rm sat}$ at a given value of $\beta_s$.
The contours show significant degeneracy between these two parameters even in the cases where 
these parameters can be separately measured from each other.
For example, with SDSS, we find
$M_{\rm sat}=(1.2_{-1.1}^{+2.9}) \times 10^{13}$ $h^{-1}$ M$_{\sun}$ with a 
power-law slope, $\beta_s$, of $(0.56^{+0.19}_{-0.17})$ for the total luminosity--halo mass relation, both 
the 68\% confidence level. 
The mass limit can be compared to other estimates from the literature. For example, based on numerical simulations
combined with semianalytic models, Zheng et al. (2004) finds  that even halos of mass $10^{12.4 \pm 0.1}$ M$_{\sun}$
host satellites with $M_r < -19$. Note that the one-sigma lower limit of the allowed range
we find from model fitting the data is $10^{12}$ M$_{\sun}$. Our results generally applies to galaxies with $M_r <-17$.
If we concentrate on galaxies with $M_r <-19$ only, we again find that the lower limit does not change significantly
suggesting that the appearance of satellite galaxies with $M_r <-19$ in Zheng et al. (2004; see their Figure~11)
in halos with mass above $10^{12.4 \pm 0.1}$ M$_{\sun}$ is not contradicted by SDSS clustering data.

At $z \sim 0.6$, COMBO-17 data allows these
parameters for  $M_B < -18$  galaxies
to be constrained as $M_{\rm sat}=(3.3_{-3.0}^{+4.9}) \times 10^{13}$ $h^{-1}$ M$_{\sun}$  and
 $\beta_s=(0.62^{+0.33}_{-0.27})$ at the 68\% confidence level, respectively, while
at $z \sim 4$, Subaru measurements constrain these parameters for $M_B < -18.5$ galaxies
as $(4.12_{-4.08}^{+5.90}) \times 10^{12}$ $h^{-1}$ M$_{\sun}$  and
 $(0.55^{+0.32}_{-0.35})$, respectively.
The large range allowed for  $M_{\rm sat}$, over an order of magnitude in mass at the
68\% confidence level, supports the suggestion in Zehavi et al. (2004)
that halo occupation models suggested there are not unique. This large  range also
shows that halo occupation number predicted here and the models in Zehavi et al. (2004) are likely to be
consistent with each other, but given that Zehavi et al. (2004)  did not present detailed
fits to data, a straight forward comparison is impossible.

The degeneracy patterns in Fig.~17, however, suggest that a certain combination of
$\beta_s$ and $M_{\rm sat}$ is better determined when compared to these two parameters separately.
The degeneracy direction is such that as M$_{\rm sat}$ is decreased,
$\beta_s$ is decreasing as well. Thus, the increase in
the total number of satellite galaxies, or more appropriately in the context of CLFs, the satellite luminosity,
is compensated by the decrease in $\beta_s$ so as to conserve the total satellite luminosity.
To understand this further, we calculate
the sample averaged total luminosity of satellites, over the luminosity distribution of the galaxy sample, as
\begin{eqnarray}
&&\langle L_{\rm sat}(z) \rangle=\\
&&\frac{\int dL \int dM \Phi(L|M,z) \frac{dn(z)}{dM} \left[ L_{\rm tot}(M,z)-L_c(M,z)\right]}{\int dL \Phi(L,z)} \, , \nonumber
\label{eqn:lsatave}
\end{eqnarray}
where $L_{\rm tot}(M,z)$ is given in equation~(\ref{eqn:ltot}). Since $L_{\rm tot}(M,z)$ is a function of parameters
$\beta_s$ and $M_{\rm sat}$, we calculate $\langle L_{\rm sat}(z) \rangle$ as a function of these two parameters.
In Fig.~18, we plot contours of constant $\langle L_{\rm sat}(z) \rangle$  at redshifts corresponding to
SDSS and GOODS (at $z\sim 3$), and, for comparison,
we also show constraints on this parameter plane from SDSS. The comparison reveals that the
single parameter best constrained by the combination of $\beta_s$ and $M_{\rm sat}$
is $\langle L_{\rm sat}(z) \rangle$, the sample-averaged
total luminosity of satellite galaxies. We find a similar behavior with other parameterization of
$L_{\rm sat}(M)$ relation as well.

In Fig.~19, we plot contours of constant $\langle L_{\rm sat}(z) \rangle$  at the redshift corresponding to
Subaru (at $z \sim 4$), and for comparison, in the right panel, contours of constant number density of galaxies
with $M_B < -18.5$ at $z=4$ (in units of $10^{-3}$ h$_{70}^3$ Mpc$^{-3}$) as a function of parameters related to the satellite CLF. Just as $\langle L_{\rm sat}(z) \rangle$ traces the degeneracy of the two parameters $\beta_s$ and $M_{\rm sat}$,
the same degeneracy is traced by contours of $\bar{n}(z=4)$ as well. The range allowed by  constraints
on $\beta_s$ and $M_{\rm sat}$ is consistent with the value of 
$5.8 \pm 1.4 \times 10^{-3}$ h$_{70}^3$ Mpc$^{-3}$ measured directly in the data by Ouchi et al. (2005).
As shown in Fig.~19, in fact, the density is better constrained by non-linear clustering pattern
when compared to a direct analysis related to the LF.

Using  $\langle L_{\rm sat}(z) \rangle$  parameter instead of $\beta_s$ and $M_{\rm sat}$, 
with SDSS, we find $\langle L_{\rm sat}(z<0.1)\rangle =(2.1^{+0.8}_{-0.4}) \times 10^{10}$ h$^{-2}$ L$_{\sun}$,  
while with COMBO-17, $\langle L_{\rm sat}(z\sim0.6)\rangle =(2.4^{+1.1}_{-0.6}) \times 10^{10}$ h$^{-2}$ L$_{\sun}$.
Moving to higher redshifts, with DEEP2, $\langle L_{\rm sat}(z\sim1)\rangle < 3.9 \times 10^{10}$ h$^{-2}$ L$_{\sun}$,  
for GOODS at $z \sim 3$, $\langle L_{\rm sat}(z\sim3)\rangle < 2 \times 10^{11}$ h$^{-2}$ L$_{\sun}$,
and for Subaru at $z \sim 4$,
$\langle L_{\rm sat}(z\sim 4)\rangle =(4.2^{+2.3}_{-3.1}) \times 10^{10}$ h$^{-2}$ L$_{\sun}$.  
Based on results from SDSS and COMBO-17, if $L_{\rm sat}(z)=L_{\rm sat}(z=0)(1+z)^\epsilon$, then 
we find that $\epsilon = 0.31 \pm 0.52$, while between COMBO-17 and Subaru at $z \sim 4$ is
$\epsilon = 0.49 \pm 0.74$. The difference between the two observational wavelength bands
between SDSS and COMBO-17, r- and B-band respectively, and galaxy luminosities in the two samples make
this comparison less useful. On the other hand, COMBO-17 sample is for galaxies with $M_B <-18$ while
for Subaru at $z \sim 4$ is for galaxies with $M_B <-18.5$. While there is  a small difference between the two samples,
given the large redshift difference, 0.6 and 4 for COMBO-17 and Subaru data respectively, it is safe to conclude that 
we find no evidence for redshift evolution in the sample-averaged total luminosity of satellites.

This conclusion is in agreement with Yan et al. (2003) who compared clustering of galaxies in 2dFGRS at low redshifts
and in DEEP2  and suggested no evidence for evolution between now and a redshift of unity in the
total CLF as parameterized by Yang et al. (2003b).
Either confirming or refuting the redshift evolution could help in understanding how satellite galaxies merge within halos
to form central galaxies, whose luminosities do evolve with redshift.
 Given that clustering measurements by Coil et al. (2004) involved
only a subset of the final DEEP2 galaxy sample, the complete analysis should improve the estimate of
$\langle L_{\rm sat}\rangle$ at a redshift of 1, which when combined with  SDSS and COMBO-17, should improve an estimate on the
 redshift evolution of $\langle L_{\rm sat}\rangle$  compared to the estimate here.

In Figure~20, we present the comparison between constraints on $\beta_s$ and $M_{\rm sat}$ parameters from
SDSS and Subaru and COMBO-17 and Subaru respectively. While constraints on $L_{\rm sat}(z)$ show no
evidence for evolution, at the 1$\sigma$ confidence level, we find that 
 $\beta_s$ and $M_{\rm sat}$ parameters at $z=0.6$ and $z=4$ differ each other suggesting that these
parameters in fact show some evolution. The fact that these parameters show differences (as in Figure~20), while
a parameter such as $\langle L_{\rm sat}\rangle$ remain the same may, at the first instance, contradictory.
The difference in parameters such as $\beta_s$ and $M_{\rm sat}$ between low and high redshifts
comes from the difference in halo mass functions between redshifts. As the halo mass function evolves,
there are no high mass halos, and satellites, if exist, should be appearing at a lower mass halo.
This is the general trend we see in Figure~20. If that's the case, one could argue that $\langle L_{\rm sat}\rangle$ 
should decrease as a function of increasing redshift. We do not find this behavior as halos at a high redshift
gets assigned brighter central galaxies than at a low redshift due to the redshift evolution in the $L_c(M)$ relation.
This anti-hierarchical behavior is consistent with what is generally referred to in the literature as ``down-sizing''
or mass-dependent luminosity evolution where brighter galaxies form first than less luminous galaxies.
Since small halos are assigned brighter central galaxies at high redshift, given our description of the CLF,
it is natural  that such halos end up with brighter satellites as well, relative to a same mass halo at a lower redshift.
Thus, while  $\beta_s$ and $M_{\rm sat}$  change with redshift,  $\langle L_{\rm sat}\rangle$ remains the same.
Note that in our models of the CLF, we have not a priori assumed this behavior. In fact, CLF parameters
may take any value, and we only note this behavior because of the  model fitting to the data. Thus, our
model fit results provide support for apparent brightening of galaxies at high redshifts both in the case of
central galaxies, as discussed in Cooray 2002b, and satellites, as discussed here in terms of the clustering statistics.

In Figure~21, we show constraints on $M_{\rm sat}$ and $\beta_s$ as a function of the galaxy luminosity.
For clarity, we only plot constraints when $-19 < M_r < -18$ and $-22 <M_r <-21$.
These constraints reveal, though not significant, a trend in 1$\sigma$ constraint on $M_{\rm sat}$
as a function of the luminosity bin such that as galaxy luminosities are increased, $M_{\rm sat}$
is also increased. Such an increase is heavily favored in halo occupation model fits
of Zehavi et al. (2004), though, we find that large uncertainties in our model parameters
related to CLFs do not allow us to establish the same dependence of $M_{\rm min}$, the
minimum mass for the appearance of a central galaxy in Zehavi et al. (2004), with galaxy luminosity
here for the appearance of satellites through our model parameter $M_{\rm sat}$
as a function of luminosity. As stated in Zehavi et al. (2004), the halo occupation models shown
there are not unique and it could be that the largely increasing values of $M_{\rm min}$
as a function of galaxy luminosity is partly accounted through unusually large power-law
slopes in the halo occupation number models suggested there. It is likely that this
result can be further improved with galaxy-galaxy lensing studies, and as we discuss later, 
more likely with cross-clustering between faint and bright galaxies.

Instead of using galaxy clustering data to establish the mass scale at which satellite galaxies begins to
appear, as a function of the luminosity, one can establish the same relation directly from the data
if galaxy sample can be divided to a distribution of galaxy groups and clusters, with some mechanism to estimate the halo mass
of that halo. Following this approach, we make use of a suggested catalog of galaxy groups in SDSS by Weinmann et al. (2005) and
use the luminosity distribution to study the minimum luminosity of galaxies in these halos 
as a function of the halo mass. In Figure~22, we summarize our results where we consider close to $\sim 10^4$
groups and clusters in SDSS. The halo masses are estimated based on the total luminosity of the halo,
though due to small number of galaxies when halo masses are below a few times $10^12$ M$_{\sun}$, the mass
estimates may become highly uncertain. The catalog may also be affected by uncertain galaxy assignments, especially when
a galaxy that is parter of a large mass halo such as cluster gets assigned systematically to a lower mass galaxy group.
Ignoring these complications, which affect the low mass end, we see a trend in minimum luminosity with halo mass.
This trend can be roughly described as $M_{r'}(Min) \approx -19.8 (M/10^{12}\; M_{\sun})^{0.03}$.
Galaxies with luminosities greater than -21 only begin to appear in dark matter halos with mass above $10^{13}$ M$_{\sun}$,
while the mass limit for galaxies with luminosities brighter than -17 in the $r'$-band is below 10$^{12}$ M$_{\sun}$.

The result we derived earlier where we suggest that mass limit is $(1.2^{+2.9}_{-1.1})\times 10^{13}$  M$_{\sun}$
is for the whole sample and is generally weighted by galaxies in magnitude bins between -19 to -21 (see, Figure~8 for example).
The result based on clustering analysis is thus generally consistent with the direct estimate from the cluster catalog.
In Figure~23, we plot the constraints on minimum halo mass and the power-law slope for individual bins in luminosity
between -18 and -22 (dashed lines). If we make use of the general result that $M_{r'}(Min) 
\approx -19.8 (M/10^{12}\; M_{\sun})^{0.03}$, then we find that $\beta_s > 0.4$ at the 3$\sigma$ confidence level.
Returning to Figure~21, we then see the clear trend between minimum luminosity and the halo mass even based on galaxy clustering.

As a further application of our results,
our CLFs can be easily used to estimate the average fraction of satellite galaxies in dark matter halos over a given
luminosity range, $\langle f_{\rm sat}(L) \rangle$:
\begin{eqnarray}
&&\langle f_{\rm sat}(L) \rangle=\frac{\int dM \Phi^{\rm sat}(L|M,z) \frac{dn(z)}{dM}}{\Phi(L,z)} \, .
\label{eqn:fsatave}
\end{eqnarray}
In Fig.~23, we show contours of constant $\langle f_{\rm sat}(L) \rangle$ for several luminosity bins
between -18 and -22 in $M_r$ as appropriate for SDSS as a function of $\beta_s$ and
M$_{\rm sat}$. For reference, we also plot
the constraint from SDSS clustering data on these two parameters as a function of the
luminosity bin.

To estimate the satellite fraction as a function of the luminosity bin, instead of $M_{\rm sat}$ and $\beta_s$ as
parameters, we determine the likelihood for the single parameter
$\langle f_{\rm sat}(L) \rangle$ directly from clustering data. These results are summarized in Figure~24.
The satellite fraction is in each of the bins is $0.13^{+0.03}_{-0.03}$,
$0.11^{+0.05}_{-0.02}$, $0.11^{+0.12}_{-0.03}$, and $0.12^{+0.33}_{-0.05}$ for
galaxies with $r'$-band luminosities of -22 to -21, -20 to -21, -19 to -20, and -18 to -19, respectively (see, also, Figure~22).
As we discussed with respect to Figure~21, there is an indication that $\beta > 0.4$ to be consistent with the
minimum luminosity of galaxies as a function of the halo mass. Thus,
if $\beta_s>0.4$, the satellite fractions are $0.105^{+0.035}_{-0.025}$,
$0.12^{+0.06}_{-0.05}$, $0.13^{+0.08}_{-0.05}$, and $0.13^{+0.10}_{-0.06}$, for galaxies
in luminosity bins of -21 to -22, -20 to -21, -19 to -20, and -18 to -19, respectively.

Though our fractions are slightly lower, considering the errors, these fractions are  fully consistent
with the values suggested in Mandelbaum et al. (2004) in three luminosity bins
based on an analysis of galaxy-galaxy lensing data with numerical simulations.
Given that current data allow a large range of satellite fractions, for most practical purposes, one can assume that the
satellite fraction is a constant with a value around 0.1 to 0.2 in luminosity bins
between -18 and -21 in $M_r$ for general prediction calculations (e.g.,
Slosar et al. 2005), though when estimating cosmological parameters or other parameter constraints,
it may be best to take into account suggested variations. Unlike calculations in Mandelbaum et al. (2004) or
Slosar et al. (2005), in the present description of galaxy statistics with CLFs, satellite fraction is not
an independent free parameter and is only determined to the extent that parameters related to the
satellite CLF are known. Thus, we need not establish the satellite fraction separately.

Involved with the above expression for
$\langle L_{\rm sat}(z) \rangle$, in equation~\ref{eqn:lsatave}, is
the probability of halo mass to a host a galaxy with luminosity $L$ at a redshift $z$ given by
\begin{equation}
P(M|L,z)dM=\frac{\Phi(L|M,z)}{\Phi(L,z)} \frac{dn(z)}{dM} \; dM \, .
\end{equation}
These probabilities are shown in left and right panels of Fig.~25 
for a faint and a bright sample of galaxies at three different redshifts,
respectively.
The two panels, when combined, show the mass-dependent redshift evolution  of
the galaxy luminosity following Cooray (2005b). Luminous galaxies at high redshifts
are found at lower mass halos than dark matter halo masses that corresponds to the same galaxy luminosity
today. At the faint-end, $-18 > M_r > -19$, regardless of the redshift, faint galaxies are
essentially found in dark matter halos with a factor of 2 less
range in mass, though at low redshifts, a 30\% or more fraction of
low-luminous galaxies could be satellites in more massive halos.

In Fig.~26, we show the same probabilities divided into three magnitude bins as a function of redshift in four panels.
When $-22 < M_r <-21$,  at $z \sim 3.5$, galaxies are primarily in dark matter halos of mass $\sim 3 \times 10^{12}$ M$_{\sun}$.
In comparison, such galaxies are central galaxies in groups and clusters today with masses above
10$^{14}$ M$_{\sun}$.  The same probabilities have been estimated based on galaxy-galaxy weak lensing
studies in SDSS by Mandelbaum et al. (2004). The mean mass estimates, at a given luminosity bin, and the
dispersion of the mean mass based on probabilities shown in Fig.~26 are in agreement with
estimates by Mandelbaum et al. (2004). 
For example,  probabilities shown in Fig.~26 suggest that the mean halo mass for the bin -21 to -22 in $M_r$
is about $8 \times 10^{12}$ h$^{-1}$ M$_{\sun}$ which agrees with 
the mass estimate of   $9.71 \times 10^{12}$ h$^{-1}$ M$_{\sun}$
 based on NFW fits  to galaxy-galaxy lensing data. 
Since galaxy-galaxy lensing measurements trace the
galaxy--dark matter correlation function while our estimates are based solely on galaxy-galaxy clustering,
these agreements suggest that to the extent probed by current data galaxy distribution traces dark matter
as assumed in these halo-based models. We will, however, test this assumption in detail in an upcoming analysis.

At high redshifts, the halo masses are again consistent with various prior estimates.
For example, in Conroy et al. (2004), the dark matter halo masses of
$-22 < M_B <-21$ galaxies are measured based on velocity profiles
with a halo mass estimate of $5.5^{+2.5}_{-2.0} \times 10^{12}$ h$^{-1}$ M$_{\sun}$.
Our probability distribution function for halo mass in this luminosity bin and
at a redshift of unity suggests a mean halo mass of $5 \times 10^{12}$ h$^{-1}$ M$_{\sun}$
in good agreement with this result. The agreement of halo masses based on
galaxy LFs and prior estimates based on clustering etc at higher redshifts, in the context of
LBGs, is discussed in Cooray (2005b). 

While certain parameters related to the satellite CLF are constrained well by current clustering data at low redshifts,
parameters related to satellite galaxy types are not. The measurements by Zehavi et al. (2004)
involve clustering of galaxies in the same luminosity bin, as well as the clustering of
galaxies in the same luminosity bin and the same type (except in Figure~7, the cross-clustering
between early- and late-type galaxies for $M_r  <-21$). These measurements, while useful, do not provide
all the information related to clustering for the same sample of galaxies. For example,
to probe the CLF of satellites better one can consider cross-correlating galaxies
that do not have a significant overlap in halo mass in terms of the central galaxy CLF.
The possibility exists when considering a faint and a bright subsample of galaxies.
As shown in Fig.~5, the central galaxy CLF for galaxies with $-18 < M_r < -17$
peaks at a halo mass of few times 10$^{11}$ M$_{\sun}$. The same CLF peaks at a halo
mass of few times 10$^{15}$ M$_{\sun}$ when galaxies with luminosities
$-23 < M_r <-22$ are considered. While the CLF of central galaxies do not overlap,
resulting in no contribution to the cross-correlation between these two samples,
there is a certain overlap in the satellite CLF and to a lower extent between the central galaxy
CLF of the brighter sample and the satellite CLF of the fainter sample.

The cross-correlation of galaxies between these two luminosity bins
will provide an additional handle on the luminosity distribution of satellites
in clusters of galaxies. In fact, one can consider cross-correlations between
different luminosity bins as well as different galaxy types; For example,
the cross-correlation between early type galaxies in the fainter sample
and late-type galaxies in the brighter sample. We illustrate the expected cross-correlation between
$-18 < M_r < -17$ and $-23 < M_r <-22$  galaxies in Fig.~26. For reference, in the same plot,
we also show the galaxy clustering correlation function of galaxies measured by SDSS
in each of the two luminosity bins. While the cross-correlation has not been measured in the data
yet, we propose these additional measurements for the whole sample. Such measurements, in addition to
clustering at each luminosity bin, would provide all the information related to galaxy clustering
at the two-point level from any given survey.
This information could in return help further constrain  CLF of satellite galaxies
as well as the fraction of galaxy types in the form of satellites.

While we have concentrated primarily on the use of galaxy LF and clustering measurements to
constrain parameters related to central and satellite galaxy CLFs here, a primary interest of these
statistical measurements is to establish global cosmological parameters. This has been achieved
mostly by combining CMB data, such as from WMAP, with large-scale linear clustering with surveys such
as SDSS (e.g., Tegmark et al. 2004) or with non-linear clustering part modeled based on a
simple parameterization of the halo occupation number (Abazajian et al. 2005).
The latter approach can be done with clustering measurements at different redshifts
and the combination, as a function of redshift, would provide additional constraints on the growth evolution of density
perturbations. The CLF approach suggested here may make this a possibility since  CLFs
provide estimates of galaxy bias,  both as a function  of luminosity and redshift, once
the galaxy sample used for clustering measurements at various redshifts is well defined.
While we have not considered cosmological parameters measurements here this is of
significant interest and we hope to return to this once several high-redshift surveys
provide more accurate clustering measurements for a well defined sample of galaxies.

\section{Summary and Conclusions}

To summarize our discussion involving model descriptions of
the galaxy LF and clustering statistics with CLFs and estimates of CLF parameters directly from the data,  our main results are:

(1) Instead of the halo occupation number, it may be useful to
describe galaxy properties through the CLF when describing the
galaxy LF and clustering statistics.
As discussed in Section~2, CLFs provide a consistent
way to compare, and understand, differences in measurements between different samples conditioned
in terms of galaxy properties. While occupation numbers have allowed
model fits to clustering statistics, their use is restricted to
clustering statistics alone as LFs cannot easily be described by occupation statistics that
treat all galaxies the same.

(2) We have outlined  a general procedure to describe CLFs of central and satellite galaxies
by improving prior descriptions of CLFs by a priori assumed Schechter function shapes (e.g., Yang et al. 2003b, 2005).
The log-normal distribution for central galaxies and the power-law assumption for
satellites combine to produce an overall Schechter function shape for galaxy LF (e.g., Cooray \& Milosavljevi\'c 2005b), but
at the same time, also explain why the cluster LF (e.g., Trentham \& Tully 2002) cannot be explained
with a single Schechter function.

(3) The combination of SDSS LF and clustering data at low redshifts and
clustering measurements at high redshifts allow certain model parameters related
to central and satellite galaxy CLFs be determined from the data.
For example, the appearance of satellites  with luminosities $M_r < -17$ at $z < 0.1$,
using a total  luminosity--halo mass relation of the form $L_c(M)(M/M_{\rm sat})^\beta_s$,
is constrained with SDSS to be at a halo mass of
$M_{\rm sat}=(1.2_{-1.1}^{+2.9}) \times 10^{13}$ $h^{-1}$ M$_{\sun}$ with a power-law slope $\beta_s$ of 
$(0.56^{+0.19}_{-0.17})$ at the 68\% confidence level. At $z \sim 0.6$, COMBO-17 data allows these
parameters for galaxies with $M_B < -18$ 
to be constrained as $(3.3_{-3.0}^{+4.9}) \times 10^{13}$ $h^{-1}$ M$_{\sun}$ with a power-law slope of $(0.62^{+0.33}_{-0.27})$ at the 68\% confidence level, while at higher redshifts,
Subaru measurements constrain these parameters for $M_B < -18.5$ galaxies
as $(4.12_{-4.08}^{+5.90}) \times 10^{12}$ $h^{-1}$ M$_{\sun}$  and
 $(0.55^{+0.32}_{-0.35})$, respectively at $z=4$.
DEEP2 and GOODS measurements only allow an upper limit on the power-law slope of total luminosity
at a given minimum halo mass for the appearance of satellites. 

(4)  The single parameter well constrained
by clustering measurements is the average of total satellite galaxy luminosity
corresponding to the dark matter halo distribution probed by the galaxy sample.
This parameter traces the degeneracy between $M_{\rm sat}$, the minimum halo mass in which satellites appear, and $\beta_s$.
For SDSS, $\langle L_{\rm sat}\rangle =(2.1^{+0.8}_{-0.4}) \times 10^{10}$ h$^{-2}$ L$_{\sun}$,
while for GOODS at $z \sim 3$, $\langle L_{\rm sat}\rangle < 2 \times 10^{11}$ h$^{-2}$ L$_{\sun}$.
While current data do not suggest any redshift variation in $\langle L_{\rm sat}\rangle$, consistent
with a prior suggestion (Yan et al. 2003) that CLFs do not evolve in redshift, at the
1$\sigma$ confidence level, we note that parameters related to satellite CLFs do change between $z \sim 0.6$ 
and $z=4$. Such a difference is expected given the redshift evolution of the halo mass function
and the difference in parameters are such that the halo masses where brighter satellites appear 
at high redshifts host fainter satellites at low redshifts.
Parameters such as the fraction of satellites at a given luminosity are not well determined by the data.
Such parameters are built into the CLF description and does not need to be specified separately
as in the halo models of Mandelbaum et al. (2004).

(5) In addition to constraints on central and satellite CLFs, we  also determine
model parameters of the analytical relations that describe the fraction of
early- and late-type galaxies in dark matter halos. We use our CLFs to establish probability distribution of halo mass
in which galaxies of a given luminosity could be found either at halo centers or as satellites and find good
agreement with prior estimates based on an analysis of galaxy-galaxy lensing and direct mass estimates based on
velocity profiles, among others.

(6) Finally, to help establish further properties of the galaxy distribution, we propose the measurement of
cross-clustering between galaxies divided into different luminosity bins.

{\it Acknowledgments:} 
The author thanks Alison Coil, Kyoungsoo Lee, Masami Ouchi, Stefanie Phleps, and Idit Zehavi 
for information and electronic data files of measurements from DEEP2, GOODS, Subaru/XMM-Newton Deep Field, COMBO-17,
and SDSS surveys, respectively, Uros Seljak for suggesting the
inclusion of Figures~23 and 24, and an anonymous referee for suggesting the analysis based on
the group catalog to determine minimum halo mass for a given luminosity (shown in Figure~22).  Author also thanks
members of Cosmology and Theoretical Astrophysics groups at Caltech and UC Irvine 
for useful discussions, and comments from the community
and anonymous referees on author's recent papers related to the CLF.
This study was initiated while the author was at the 
Aspen Center for Physics in Summer of 2005.

\end{document}